\documentclass[11pt,twoside]{article}

\usepackage{geometry}
\usepackage[linesnumbered, algoruled, boxed, lined]{algorithm2e}
\usepackage{amsmath,amssymb,fullpage,graphicx,mathtools,amsthm,enumitem,subfig,dsfont,xspace}
\usepackage{makecell}
\usepackage{tablefootnote}
\usepackage{hyperref,cleveref}
\usepackage[numbers,sort&compress]{natbib}
\usepackage{booktabs}
\usepackage{caption}
\usepackage{nicefrac}

\let\citet\cite
\let\citep\cite

\newtheorem{theorem}{Theorem}
\newtheorem{proposition}{Proposition}
\newtheorem{lemma}{Lemma}
\newtheorem{corollary}{Corollary}

\newtheorem{example}{Example}

\newcommand{\Expect}{\mathbb{E}}
\newcommand{\Prob}{\mathbb{P}}
\newcommand{\sample}{\sim}

\newcommand{\reals}{\mathbb{R}}

\newcommand{\given}{\mid}

\newcommand{\defn}{:=}
\newcommand{\defnright}{=:}

\DeclareMathOperator*{\argmax}{argmax}

\DeclareMathOperator*{\supp}{supp}

\DeclarePairedDelimiter\abs{\lvert}{\rvert}

\DeclarePairedDelimiter\ceil{\lceil}{\rceil}
\DeclarePairedDelimiter\floor{\lfloor}{\rfloor}

\newcommand{\indicator}{\mathds{1}}

\newcommand{\union}{\cup}

\newcommand{\stepone}{\text{(i)}\xspace}
\newcommand{\steptwo}{\text{(ii)}\xspace}
\newcommand{\stepthree}{\text{(iii)}\xspace}
\newcommand{\stepfour}{\text{(iv)}\xspace}

\newcommand{\textth}{\textrm{th}}

\newcommand{\lessorder}{\lesssim}
\newcommand{\gtorder}{\gtrsim}
\newcommand{\eqorder}{\asymp}

\newcommand{\binomial}{\text{Binom}}

\newcommand{\const}{c}

\newcommand{\naive}{na\"ive\xspace}

\newcommand{\numstudents}{n}
\newcommand{\idxstudent}{i}

\newcommand{\numschools}{n}
\newcommand{\idxschool}{j}
\newcommand{\idxtype}{t}
\newcommand{\numtypes}{T}
\newcommand{\vectypes}{{\boldsymbol{t}}}

\newcommand{\utility}{u}
\newcommand{\prior}{p}
\newcommand{\length}{L}
\newcommand{\lottery}{r}
\newcommand{\dist}{\mathcal{D}}

\newcommand{\strategy}{{\boldsymbol{\pi}}}
\newcommand{\partition}{\mathcal{P}}
\newcommand{\numpartitions}{K}
\newcommand{\idxpartition}{k}

\newcommand{\texth}{\textrm{H}}
\newcommand{\textr}{\textrm{R}}
\newcommand{\textp}{\textrm{P}}
\newcommand{\texti}{\textrm{I}}
\newcommand{\textj}{\textrm{J}}

\newcommand{\parenh}{{(\textrm{H})}}
\newcommand{\parenr}{{(\textrm{R})}}
\newcommand{\parenp}{{(\textrm{P})}}
\newcommand{\pareni}{{(\textrm{I})}}
\newcommand{\parenj}{{(\textrm{J})}}

\newcommand{\textmin}{\textrm{min}}
\newcommand{\textmax}{\textrm{max}}
\newcommand{\lotterymax}{{\lottery_\textmax}}
\newcommand{\lotterymin}{{\lottery_\textmin}}

\newcommand{\lnum}{lottery number\xspace}
\newcommand{\lnums}{lottery numbers\xspace}
\newcommand{\withlnum}{with \lnum}
\newcommand{\flow}{q}
\newcommand{\flowr}{\flow^\parenr}
\newcommand{\flowh}{\flow^\parenh}
\newcommand{\flowp}{\flow^\parenp}
\newcommand{\vecflow}{\mathbf{q}}

\newcommand{\sw}{\textrm{SW}}
\newcommand{\swh}{\sw^\parenh}
\newcommand{\swr}{\sw^\parenr}
\newcommand{\swp}{\sw^\parenp}

\newcommand{\mr}{\textrm{MR}}
\newcommand{\mn}{\textrm{MN}}

\newcommand{\mnh}{\mn^\parenh}
\newcommand{\mnr}{\mn^\parenr}
\newcommand{\mnp}{\mn^\parenp}

\newcommand{\utilunif}{\omega}
\newcommand{\ratioprob}{\rho}

\newcommand{\mrr}{\mr^\parenr}
\newcommand{\mrh}{\mr^\parenh}
\newcommand{\mrp}{\mr^\parenp}

\newcommand{\homo}{\textrm{hom}}

\newcommand{\zeropos}{0^+}

\newcommand{\ratiomrhr}{\mr_{\texth/\textr}^*}
\newcommand{\ratiomrrh}{\mr_{\textr/\texth}^*}
\newcommand{\ratiomrhrhomo}{\mr_{\texth/\textr}^{*,\homo}}
\newcommand{\ratiomrrhhomo}{\mr_{\textr/\texth}^{*,\homo}}

\newcommand{\ratioswhr}{\sw_{\texth/\textr}^*}
\newcommand{\ratioswrh}{\sw_{\textr/\texth}^*}
\newcommand{\diffswhr}{\sw_{\texth-\textr}^*}
\newcommand{\diffswrh}{\sw_{\textr-\texth}^*}

\newcommand{\ratioswhrhomo}{\sw_{\texth/\textr}^{*,\homo}}
\newcommand{\ratioswrhhomo}{\sw_{\textr/\texth}^{*,\homo}}
\newcommand{\diffswhrhomo}{\sw_{\texth-\textr}^{*,\homo}}
\newcommand{\diffswrhhomo}{\sw_{\textr-\texth}^{*,\homo}}

\newcommand{\pref}{v}

\newcommand{\probavail}{\alpha}

\newcommand{\probavailh}{\alpha^\parenh}
\newcommand{\probavailr}{\probavail^\parenr}
\newcommand{\probavailp}{\probavail^\parenp}
\newcommand{\textcond}{\textrm{cond}}

\newcommand{\probavailrmean}{\overline{\probavail}^\parenr}

\newcommand{\probavailbefore}{\beta}

\newcommand{\vecmatching}{\boldsymbol{\sigma}}
\newcommand{\matching}{\sigma}
\newcommand{\econ}{\mathcal{E}}

\newcommand{\numminority}{m}

\newcommand{\term}{T}

\newcommand{\thresh}{\tau}

\newcommand{\counthigh}{n_1}
\newcommand{\countlow}{n_2}
\newcommand{\countdiff}{\Delta}

\newcommand{\numtiers}{D}
\newcommand{\setschools}{S}
\newcommand{\idxtier}{d}
\newcommand{\countschools}{m}
\newcommand{\fractier}{\gamma}
\newcommand{\strategynum}{\boldsymbol{\sigma}}
\newcommand{\strategynumopt}{\strategynum^*}
\newcommand{\mass}{x}
\newcommand*\dd{\mathop{}\!\mathrm{d}}
\newcommand{\priorflexible}{\prior_{\textrm{F}}}
\newcommand{\distdelta}{\boldsymbol{\delta}}
\newcommand{\probassign}{z}
\newcommand{\probmix}{\pi}

\newcommand{\singletype}{flexible-only\xspace}
\newcommand{\Singletype}{Flexible-only\xspace}
\newcommand{\equalprior}{equal-prior\xspace}
\newcommand{\Equalprior}{Equal-prior\xspace}

\usepackage[dvipsnames]{xcolor}

\newenvironment{quotex}%
  {\list{}{\leftmargin=0.3in\rightmargin=0.3in}\item[]}%
  {\endlist}

\begin{document}

\begin{center}
{\bf{\LARGE{The Double-Edged Sword of Information:\\Revealed versus Hidden Lotteries in School Choice}}}
\vspace*{.2in}

\renewcommand*{\thefootnote}{\fnsymbol{footnote}}
{\large{
\begin{tabular}{cc}
Parinaz Naghizadeh$^{1}$, Jingyan Wang$^{2}$\footnotemark
\end{tabular}
}}

\vspace*{.2in}
\footnotetext[1]{Corresponding author: \texttt{jingyanw@ttic.edu}
}

\begin{tabular}{c}
$^{1}$University of California, San Diego\\
$^{2}$Toyota Technological Institute at Chicago
\end{tabular}
\renewcommand*{\thefootnote}{\arabic{footnote}}
\setcounter{footnote}{0}

\vspace*{.2in}

September 2026; \qquad Revised: November 2026

\vspace*{.2in}

\begin{abstract}
In school choice, a lottery number is often used by the matching mechanism to break ties when more students prefer the same school than the number of seats available. In practice, in recent years, the NYC Public Schools started revealing the lottery number to students. Theoretical findings from prior literature suggest that revealing the lottery number strictly improves the number of matches under the deferred acceptance algorithm. However, these theoretical results are based on the over-simplifying assumption that all students share the same preference ranking for schools. In this work, we relax this assumption and allow students to have heterogeneous preference rankings. Under a game-theoretic model where student strategies form a Bayesian Nash equilibrium, we characterize scenarios where revealing the lottery number can either improve or worsen the matching outcome, measured by two metrics of the match rate and social welfare, when students submit preference lists of length one. We further consider revealing partial information about the lottery, and demonstrate non-monotonic effects in the amount of information available to students. From these theoretical results, we identify a mechanism through which lottery revealing deteriorates the overall matching quality: Under the revealed lottery, students with weak preferences but high priorities slightly improve their matching outcome at the cost of increasing congestion and substantially worsening the outcome of students with strong preferences. Using a mean-field approximation, we show via simulations that this mechanism continues to operate when the length of preference lists is greater than one. These theoretical and simulation results together illustrate complex tradeoffs induced by the lottery revealing policy, and we identify students' preference heterogeneity and preference intensity as two key factors in shaping these tradeoffs.

\end{abstract}
\end{center}


\section{Introduction}

Centralized school-choice mechanisms, matching students to schools based on the students' indicated preferences, are widely used in the US and worldwide.\footnote{Such as in Boston, Chicago, New York City, Paris, and London. See~\citet[Table 1]{fack2019estimation} for a list of other countries and cities.} In New York City alone, over $120,000$ students~\citep{url_nyc_outcomes} participate in the admissions process run by the Department of Education (DOE), seeking admission to middle schools and high schools in New York City Public Schools (NYCPS).

The deferred acceptance algorithm~\citep{gale1962marriage} is widely adopted in school choice. At a high level, every student indicates their preferences as a ranked list of schools, and the deferred acceptance algorithm matches students to schools based on these reported preferences, subject to seat availability at each school. An important part of the process is to associate each student with a randomly-generated lottery number~\citep{abdulkadiroglu2005nyc}, such that when the number of students applying to a school exceeds the school's capacity, ties are broken based on the lottery number. More precisely, schools often set up priority criteria for students who, for example, are continuing students or siblings of current students, live within the designated zone,  come from low-income families, or satisfy other diversity criteria~\citep{url_nyc_diversity}. Then the lottery number is used to break ties for students within the same priority group.

This lottery number makes ``an outsized impact''~\citep{marian2023crowdsourcing} on students' matching outcome. Moreover, the importance of the lottery number has further grown due to post-COVID policy changes, where more schools have removed academic screenings or auditions. Even at schools that conduct screenings or auditions, sometimes the screening assessment is graded coarsely such that the lottery number remains important for breaking ties.\footnote{As an example, the Manhattan / Hunter Science High School assigned the maximum essay score of 400 to all students who submitted the essay~\citep{marian2023crowdsourcing}.} For transparency and compliance to New York State’s Freedom of Information Law (FOIL), since 2022, the NYC DOE has started to reveal the lottery number to all students, before their applications are due.

While the deferred acceptance algorithm is strategy-proof in theory when every student submits a total ranking of all schools, in practice, students are often constrained to list a small number of schools (e.g., 12 schools in NYC before 2024~\citep{url_nyc_mayor_announcement}, $12$ in Washington DC~\citep{url_dc}, and $20$ in Chicago~\citep{url_chicago}); it is also shown theoretically that a limited list length can improve the matching outcome~\citep{kanoria2025partners} in certain cases.  Even when students are allowed to list an unlimited number of schools (starting from the application cycle for Fall 2025 enrollment~\citep{url_nyc_mayor_announcement}), it is infeasible for a student to rank hundreds of schools, as gathering information about each school is time-consuming. Students thus make a strategic choice about which subset of schools to apply to.

Specifically, students strategize based on their lottery number when this information is available to them. As an anecdote:
\begin{quotex}
    \emph{[A] Manhattan mom noticed the [good] lottery number her daughter received [...] –- which she learned was likely to give her daughter first dibs on schools at the top of her list. The mom then decided her daughter should re-do her list by putting more desirable schools at the top.}~\citep{url_nyc_glitch}
\end{quotex}
Moreover, in the application process, students receive guidance from an official tool that takes into account the student's lottery number and predicts the student's chance of receiving an offer from a high school~\citep{url_nyc_tool}. Students' strategies are again based on their lottery number via these projected chances.

Revealing the lottery number to students has a number of benefits. On an individual level, it allows students to focus their research on a small subset of schools based on their lottery number, and manage their expectations of the matching outcome~\citep{marian2023crowdsourcing}. On a societal level, it increases transparency and accountability. However, theoretical foundations that compare the impact of revealed versus hidden lotteries remain very limited. As a notable exception, Huang and Zhang~\cite{huang2025lotteries} show that revealing the lottery number to students strictly improves the match rate. However, their conclusions rely on the key assumption that all students share the same preference ranking for schools, which is restrictive and unrealistic -- if all students share the same preference ranking, there is no need to collect students' preferences to start with.

In this work, we contest the assumption that students have an identical preference ranking for schools. Under a simplified model where students' priorities are solely determined by their lottery numbers, we analyze the matching outcome in terms of two standard metrics of match rate and social welfare. Beyond the hidden and revealed lotteries, we also consider a ``partial information lottery'' where students know that their lottery number lies within a certain range; revealed and hidden lotteries are thus special cases of the partial information lottery. We show the following theoretical results, focusing on the setting when students submit preference lists of length one: 
\begin{itemize}
    \item Under the hidden lottery, if there exist multiple equilibria, all of them lead to the same match rate and social welfare (Theorem~\ref{thm:equivalence}). The same result holds for the partial information lottery and revealed lottery under additional assumptions (to be defined precisely in Section~\ref{sec:invariance}).
    
    \item For the match rate, despite the fact that, as shown in prior work, the revealed lottery always does better when students have homogeneous preference rankings, we show that it can decrease the match rate when we allow students' preference rankings to be heterogeneous (Theorem~\ref{thm:MR-gaps}).

    \item For social welfare, there exist scenarios where revealing the lottery decreases the social welfare, even when students have homogeneous preference rankings. This decrease can be even larger under heterogeneous preference rankings (Theorems~\ref{thm:SW-gaps} and \ref{thm:SW-diff-gaps}).

    \item We show non-monotonic behavior for both the match rate and social welfare in the amount of information available to students, even under homogeneous preference rankings. Specifically, under homogeneous preference rankings, while the match rate is strictly higher under the revealed lottery than the hidden lottery, revealing partial information about the lottery may lead to a match rate that is strictly lower than both (Proposition~\ref{prop:partial-MR}). Under heterogeneous preference rankings, all 6 orderings among the revealed, hidden, and partial information lotteries are possible for both the match rate (\Cref{prop:partial-MR-hetero}) and social welfare (Proposition~\ref{prop:partial-SW}).
\end{itemize}
In establishing these theoretical results, we identify a mechanism through which lottery revealing deteriorates the matching outcome: When students differ in preference intensities, under the revealed lottery, students with weak preferences but high priorities slightly improve their matching outcome at the cost of increasing congestion and in turn substantially worsening the outcome of students with strong preferences. We conduct simulations to show that this mechanism generalizes to longer preference lists, using a mean-field approximation and assuming two tiers of schools (Section~\ref{sec:sim}). These theoretical and simulation results together reveal surprising intricacies in terms of the impact of information on the matching outcome. The high-level intuition is that information produces a double-edged effect: While individual students benefit from the lottery information by playing better strategies, they also face a more competitive market because other students also improve their strategies, and such competition may cause matching inefficiencies.
These results provide the following policy implications:
\begin{itemize}
    \item Given that multiple equilibria lead to the same match rate and social welfare, temporary interventions that aim to move from one equilibrium to another, without changing the information policy, cannot improve the matching outcome. This contrasts many classic games where it is beneficial to steer players from one equilibrium to another~\citep{balcan2013circumventing}.
    
    \item While the literature heavily focuses on the benefits of the revealed lottery, we show mixed effects by extending prior analysis in two directions: allowing heterogeneous preference rankings and examining social welfare in addition to the match rate. In other geographical regions outside New York City where a legal mandate is in absence, policy makers may face the decision of whether to reveal the lottery number to students or not. Our results encourage policy makers to scrutinize the impact of lottery revealing more carefully before making policy changes, especially when students differ in their preference rankings or preference intensities. 
    
    \item We provide two interpretations of our results on partial information. On the positive side, there are scenarios where partial information can be an effective intervention. Policy makers may consider a spectrum of policies with varying amounts of information, and choose the one that best balances transparency and matching efficiency. On the negative side, partial information may be an inevitable consequence due to the bounded rationality of students. In certain cases, the revealed lottery yields the best outcome, but students may not be able to accurately reason about their lottery number. Therefore, they apply suboptimally as if only partial information about the lottery number is released, which may deteriorate the matching outcome. Thus, revealing the lottery may cause unintended harm if students are not fully rational. 

\end{itemize}
Finally, we note that the insights from our results apply to other settings beyond lottery numbers in school choice. In test-based centralized admissions, student priorities are determined by their performance in an exam. A design choice concerns whether students submit their preferences before taking the exam, after taking the exam but before seeing their score, or after seeing their score. One way to view this process is that students have a prior belief about their performance, subsequently form a posterior after taking the exam, and this belief becomes exact when they see their scores. Our results imply that the matching outcome may not monotonically improve when students submit their preferences at a later time point. Another potential application arises in allocating computing resources in job scheduling. More generally, the matching process is a game where individual players' utilities are based on an inherent order that is stochastic, and we examine whether revealing this order leads to higher utilities in total, contributing to the broad area of information design in game theory and mechanism design. 

\section{Related work}

\paragraph{School choice.}

    Prior work has proposed various models to characterize how students strategize their school selection to hedge against uncertainty~\citep{chade2014portfolios}, for example, by mixing ``reach'', ``match'', and ``safety'' schools~\citep{beyhaghi2017selfish,ali2025hedging}. For the widely used deferred acceptance algorithm specifically~\citep{gale1962marriage}, it is only strategy-proof when students report a total ranking of all schools, and strategic behavior may arise when students are constrained to apply to a limited number of schools, as shown  theoretically~\citep{beyhaghi2017selfish,haeringer2009constrained} and empirically~\citep{calsamiglia2010constrained}. Beyond strategic behavior, students may also exhibit different types of cognitive biases~\citep{hakimov2010survey}, such as loss aversion, over-confidence~\citep{pan2019overconfident}, and lack of confidence~\citep{chen2019selection}.

    The role of lotteries has been studied in the context of comparing a single lottery versus independent lotteries~\citep{abdulkadiroglu2009indifferences,ashlagi2020tie,arnosti2023lottery}, where in the latter a lottery number is independently drawn to break ties at each individual school. To our knowledge, formal studies on lottery revealing policies remain limited with the exception of Huang and Zhang~\cite{huang2025lotteries}, drawing the conclusion that revealing the lottery number strictly improves the match rate under the assumption that all students share identical preference rankings. Another exception is Sato and Shirakawa~\citep{sato2025structures}, whose focus is on how to design the information structure to implement a given desired equilibrium, and for a problem setup that is significantly different from ours. Lastly, Marian~\cite{marian2023crowdsourcing} provides an empirical analysis of lotteries in the NYC school match through collecting crowdsourced data (including the lottery number) from individual families.

    Another line of research compares pre- versus post-exam preference submission~\citep{lien2017fairness,chen2023timing,pan2019overconfident}. Exam scores are only a noisy measurement of students' true abilities, rendering post-exam preferences suboptimal if the goal is to optimize with respect to students' unknown true abilities. In our lottery revealing problem, the lottery number is the ``true'' ordering that the matching mechanism wants to follow, and there is no such concern about noisy measurements.

\paragraph{Interventions in matching markets.}

    Beyond lottery revealing policies, interventions to reduce congestion in a matching market include limiting the number of applications~\citep{kanoria2025partners,arnosti2021congestion}, increasing the application cost~\citep{arnosti2021congestion,he2022costs}, or setting a market-wide wage~\citep{arnosti2021congestion}.

\paragraph{Information-based interventions.}

    Recommendation tools have been created to provide public or personalized information that helps students understand their own preferences, avoid common mistakes, and make informed decisions, in theory~\citep{besbes2025impact} and in practice~\citep{larroucau2025mistakes,chiang2026nudge}.
    In a different application concerning the labor market, it has been shown that from a school's perspective, only disclosing partial information (such as coarse grades) about the students leads to better job placement~\citep{ostrovsky2010disclosure}.

\paragraph{Game theory.} 

    Our problem can be written as an extensive-form game with incomplete information. While there is an inherent (unknown)  ordering of students determined by their lottery numbers, in contrast to sequential games, players do not observe the actions by previous players. In the domain of voting, a theoretical analysis comparing sequential versus simultaneous elections is a comparison between a sequential game and a simultaneous game~\citep{hummel2015elections}.
    
    Our problem is also a Bayesian game. It shares similarity in spirit to Bayesian persuasion where supplying partial information can be optimal~\citep{kamenica2011persuasion}.  However, the type of information considered in our problem does not involve injecting randomness and is independent from student types. In our theoretical results, we analyze the Bayesian Nash equilibria of students strategies, which can be viewed as generating a restricted class of Bayes correlated equilibria because the signal (i.e., information about the student's lottery number) sent to each individual student is restricted and does not allow for more general information structures such as the realized types of other students. 

    When lottery numbers are revealed, our analysis yields equilibrium strategies where every student iteratively computes the strategies of students with better lottery numbers, and then optimizes their expected utility accordingly. This computation is similar in spirit to the concept of cognitive hierarchy~\citep{camerer2004hierarchy}, a behavioral model where players have different levels of reasoning, and players of higher levels infer the strategies of players of lower levels and best-respond to them. Zhang~\cite{zhang2021level} applies this hierarchical reasoning to derive student strategies under the Boston mechanism. In contrast, in our model, the lottery number determines a student's priority in the matching, and students with worse lottery numbers incur a higher cognitive burden, because they have to reason about students with better lottery numbers.

\section{Problem formulation}

Assume that there are $n$ students and $n$ schools. Students have (potentially) heterogeneous preferences over schools. 
Specifically, we assume that there are $T$ distinct types of students. A student of type $t \in [T]$ has a positive valuation $v_{tj}\in \reals^+$ for school $j\in [\numschools]$. We assume that these valuations are normalized to have a unit sum, i.e., $\sum_{j=1}^n v_{tj}=1$ for every $t\in[T]$. Students have a valuation of $0$ for being unmatched to any school. A special case is when students have homogeneous preference rankings for schools.  In this case, without loss of generality, we index schools such that $v_{t1}\geq v_{t2} \geq \ldots \geq v_{tn}$ for every $t\in [\numtypes]$. For heterogeneous preference rankings, schools are indexed arbitrarily. 

Let $\mathcal{D}:=\{p_t\}_{t\in[T]}$ denote the distribution of types in the population, where $p_t\in (0,1]$ is the prior that any given student is of type $t\in [\numtypes]$. Denote by $t_\idxstudent$ the type of student $\idxstudent$ drawn i.i.d. from $\mathcal{D}$, and let $\vectypes = (t_1, \ldots, t_\numstudents)$. We assume that each student $\idxstudent$ knows their own type $t_\idxstudent$, but not that of other students. We assume that the prior $\mathcal{D}$ and valuations $\{\pref_{\idxtype\idxschool}\}_{t,\idxschool}$ are common knowledge. 

Schools have no preference over students, meaning that student priorities are solely determined by the lottery number. Specifically, students are assigned lottery numbers $1$ through $n$ uniformly at random, where $r_i\in [n]$ denotes the lottery number assigned to student $i\in [n]$. A smaller $r_i$ means that the student has a higher priority. Without loss of generality, we re-index the students so that $r_i = i$.  Students do not know their index $\idxstudent$ unless revealed to them by the information policy. 

Each student selects $L \ge 1$ schools and submits an ordering among these $L$ schools. Each school has a limited number of seats, and the matches are determined by the deferred acceptance algorithm, where all schools break ties according to the students' lottery numbers. In this case, the deferred acceptance algorithm reduces to serial dictatorship: The student with lottery number $1$ is assigned to their most preferred school; then the student with lottery number $2$ is assigned to their most preferred school among the ones that still remain available, and so on. If none of the $L$ schools selected by a student is available, then the student is unmatched. The algorithm continues until all students either receive a school assignment or become unmatched. We assume that each school has one seat available. In our theoretical results (\Cref{sec:results}), we focus on the case where each student chooses a single school (i.e., $L=1$). We also examine the general case of $L \ge 1$ via simulation (\Cref{sec:sim}).

We consider three policies that provide students with information about their lottery number: \emph{Revealed} (R), \emph{Hidden} (H), and \emph{Partial Information} (P). Under the \emph{revealed} lottery, students know their exact lottery number. Under the \emph{hidden} lottery, students have no information and hence assume being assigned any lottery number uniformly at random. Under the \emph{partial information} lottery, lottery numbers are partitioned into $\numpartitions$ blocks $\mathcal{P}_1\cup \mathcal{P}_2 \cup \ldots \cup \mathcal{P}_K =[n]$.
Students are informed about the partition structure and the specific block that their lottery number belongs to, but not their exact lottery number within the block.  We make a natural assumption that blocks are ordered and contiguous. That is, we have $r < r'$ for any $r\in \partition_\idxpartition$ and $r'\in \partition_{\idxpartition'}$ with $\idxpartition<\idxpartition'$. As an example, with $\partition_1 = \{1, \ldots, n/2\}$ and $\partition_2 = \{n/2 +1, \ldots, n\}$, students in block $1$ (resp. block 2) have lottery numbers better (resp. worse) than average. As another example, if there are ten blocks of equal size, then each block corresponds to a decile of lottery numbers. Note that this partial information lottery reduces to the revealed lottery when $\mathcal{P}_\idxpartition = \{\idxpartition\}$ for each $\idxpartition\in [n]$, and reduces to the hidden lottery when $\mathcal{P}_1 = [n]$.

Denote a matching outcome by $\vecmatching\in ([\numschools]\union \{\emptyset\})^\numstudents$, where student $\idxstudent$ is unmatched if $\matching_\idxstudent = \emptyset$, and otherwise is matched to school $\matching_\idxstudent$. We consider two common metrics to evaluate a matching outcome. The first metric is the \emph{match rate}, defined as the fraction of students who are matched: 
\begin{align*}
    \mr(\vecmatching) \defn \frac{1}{\numstudents} \sum_{\idxstudent=1}^\numstudents \indicator\{\matching_\idxstudent\ne \emptyset\}.
\end{align*}
The second metric is \emph{social welfare}, defined as the sum of valuations over all matched students:
\begin{align*}
    \sw_\vectypes(\vecmatching) \defn \sum_{\idxstudent\in [\numstudents]:\, \matching_\idxstudent\ne \emptyset} \pref_{\idxtype_\idxstudent, \matching_\idxstudent}.
\end{align*}
%

\subsection{Equilibrium strategies} 
We define student strategies at an equilibrium induced by the partial information policy, and then comment on the two special cases of the revealed and hidden lotteries. Consider students of type $\idxtype$ in block $\partition_\idxpartition$. We consider symmetric strategies where these students adopt the same (possibly mixed) strategy, denoted by $\strategy_{t,\partition_\idxpartition}$. We present the equilibrium definitions when the length of preference lists is $\length=1$, the setting considered in our theoretical analysis. In this setting, we have $\strategy_{t,\partition_\idxpartition}\in \Pi$, with $\Pi$ denoting the $(n-1)$-dimensional probability simplex; the $\idxschool$-th entry of $\strategy_{\idxtype, \partition_\idxpartition}$ specifies the probability of these students applying to school $\idxschool$. This definition extends naturally to preference lists of length $\length >1$, where $\strategy_{t,\partition_\idxpartition}$ is a distribution over ordered lists of $\length$ schools. We denote by $\strategy_{\idxtype i}\in \Pi$ the strategy adopted by the student who is assigned lottery number $\idxstudent$. We have 
\begin{align}
    \strategy_{ti} = \strategy_{t, \mathcal{P}_k}  \quad\text{for every }\idxstudent\in \partition_k \text{ and }t\in [\numtypes].\label{eq:symmetric_constraint_partial}
\end{align}
The utility of a student depends on their own lottery number $\idxstudent$ (which may be unknown to them), their type $t_\idxstudent$, and strategy $\strategy_{\idxtype_\idxstudent, \idxstudent}$; it also depends on the types $\vectypes_{[i-1]} \defn (t_1, \ldots, t_{i-1})$ and strategies $(\boldsymbol{\pi}_{t_1, 1},  \ldots, \boldsymbol{\pi}_{t_{i-1}, i-1})$ of all students with better lottery numbers $1$ through $(\idxstudent-1)$. Note that the matching outcome of student $\idxstudent$ is unaffected by students with lottery numbers worse than $\idxstudent$. We denote by $U_{\idxtype\idxstudent} \big(\boldsymbol{\pi};\vectypes_{[\idxstudent-1]},\{\boldsymbol{\pi}_{t_l, l}\}_{l\in [i-1]} \big)$ the expected utility of student $\idxstudent$ of type $\idxtype$ conditional on student types $\vectypes_{[\idxstudent-1]}$, with the expectation taken over the randomness in the mixed strategies $\{\strategy_{\idxtype_\ell, \ell}\}_{\ell\in [\idxstudent-1]}$ and $\strategy$.

Denote by $\strategy^\parenp= \{\strategy^\parenp_{t, \partition_k}\}_{t, k}$ the equilibrium strategies under the partial information lottery. At equilibrium, each student best-responds to maximize their expected utility. Formally, the best response of students of type $t$ in block $\partition_\idxpartition$ is given by:
\begin{align}
    \strategy^{\parenp}_{t,\mathcal{P}_k} \in \argmax_{\boldsymbol{\pi}\in \Pi}\; 
    \frac{1}{|\mathcal{P}_k|} \sum_{i \in\mathcal{P}_k}
    \Expect_{\vectypes_{[\idxstudent-1]}\sample \mathcal{D}} \, \Big[U_{\idxtype\idxstudent}\Big(\strategy; \vectypes_{[\idxstudent-1]}, \{\strategy^\parenp_{t_l, l}\}_{l\in [i-1]}\Big)\Big].
\label{eq:BR-partial}
\end{align}
In words, a student, upon knowing their block $\partition_\idxpartition$, maximizes their expected utility by taking an expectation over their lottery number within the block and the types of students with better lottery numbers. A Bayesian Nash equilibrium (BNE) is a solution to the best-response conditions~\eqref{eq:BR-partial} under constraints~\eqref{eq:symmetric_constraint_partial}.

These strategies can be computed iteratively from block $\partition_1$ to block $\partition_\numpartitions$. Note that in equation~\eqref{eq:BR-partial}, strategy $\strategy^\parenp_{t, \mathcal{P}_k}$ only depends on the strategies of students in preceding blocks and the same block. Hence, students in the first block $\partition_1$ solve their strategies $\{\strategy^\parenp_{t, \mathcal{P}_1}\}_t$ by solving the fixed-point equation~\eqref{eq:BR-partial}. 
Given the strategies for students in block $1$, we compute the strategies $\strategy_{t, \partition_2}$ for students in block $2$ by solving~\eqref{eq:BR-partial} and~\eqref{eq:symmetric_constraint_partial} again, and then iteratively solve the strategies for students in block $3$ up to block $\numpartitions$. Note that while it is unrealistic to assume that students are able to execute this iterative computation in practice, these Bayesian optimal strategies represent the maximum possible benefit that rational agents can extract from the knowledge of the block of their lottery number.

The revealed and hidden lotteries are special cases of the partial information lottery, with their equilibrium strategies\footnote{For brevity, we use ``equilibrium'' throughout to refer to a symmetric Bayesian Nash equilibrium.} denoted by $\strategy^\parenr$ and $\strategy^\parenh$. For the \emph{revealed} lottery, we have $\partition_k = \{k\}$ and constraint~\eqref{eq:symmetric_constraint_partial} becomes trivial. The fixed-point equation~\eqref{eq:BR-partial} reduces to
\begin{align*}
    \boldsymbol{\pi}^{\parenr}_{ti} \in \argmax_{\boldsymbol{\pi}\in \Pi}\; 
    \mathbb{E}_{\vectypes_{[i-1]}\sim \mathcal{D}} \, \Big[U_{\idxtype i} \Big(\boldsymbol{\pi}; \vectypes_{[\idxstudent-1]}, \{\boldsymbol{\pi}^{\parenr}_{t_l, l}\}_{l \in[i-1]}\Big)\Big],
\end{align*}
where again these strategies can be computed iteratively from lottery number $1$ through $n$. For the \emph{hidden} lottery, students are unaware of their lottery number (or equivalently, their index $\idxstudent$). Constraint~\eqref{eq:symmetric_constraint_partial} requires all students of the same type $t$ to adopt the same strategy, denoted by $\strategy_{t}\in \Pi$ as shorthand:
\begin{align*}
    \strategy_\idxtype\defn \strategy_{\idxtype, [n]}=\strategy_{\idxtype 1} =\cdots = \strategy_{\idxtype\numstudents},  \quad \text{ for every }\idxtype\in [\numtypes].
\end{align*} 
The fixed-point equation~\eqref{eq:BR-partial} becomes: 
\begin{align*}
    \boldsymbol{\pi}^{\parenh}_{t} \in  \argmax_{\boldsymbol{\pi}\in \Pi}\; \frac{1}{n}\sum_{i=1}^n \Expect_{\vectypes_{[\idxstudent-1]}\sim \mathcal{D}} \, \Big[U_{\idxtype\idxstudent}\Big(\boldsymbol{\pi}; \vectypes_{[\idxstudent-1]},\{\boldsymbol{\pi}^{\parenh}_{t_l}\}_{l\in [i-1]}\Big)\Big].
\end{align*}

\subsection{Matching outcomes under information policies}

Given a strategy profile $\{\strategy_{\idxtype, \partition_k}\}$, the matching outcome $\vecmatching$ produced by the deferred acceptance algorithm is determined by the realization of student types $\vectypes$ and the realization of (possibly mixed) strategies $\{\strategy_{\idxtype, \partition_k}\}$. We define the match rate and social welfare under this strategy profile by:
\begin{align*}
    \mr(\{\strategy_{\idxtype, \partition_k}\}) &= \Expect_{\vectypes\sample \dist}\, \Expect_{\vecmatching\given \vectypes,\{\strategy_{\idxtype, \partition_k}\}}\;\mr(\vecmatching)\\
    \sw(\{\strategy_{\idxtype, \partition_k}\}) &= \Expect_{\vectypes\sample \dist}\, \Expect_{\vecmatching\given \vectypes,\{\strategy_{\idxtype, \partition_k}\}}\;\sw_\vectypes(\vecmatching),
\end{align*}
where the expectation $\Expect_{\vectypes\sample \dist}$ is over student types $\vectypes$ and the expectation $\Expect_{\vecmatching\given \vectypes,\{\strategy_{\idxtype, \partition_k}\}}$ is over randomization in mixed strategies.
An economy $\econ\defn (\{v_{t\idxschool}\}, \dist)$ consists of valuations $\{v_{t\idxschool}\}$ and the type distribution $\dist$. For any economy $\econ$ and information policy $\texti\in\{\textr, \texth, \textp\}$, let $\strategy^\pareni(\econ)$ be a strategy profile at equilibrium. We define the expected match rate and expected social welfare attained under information policy $\texti$ by:
\begin{subequations}\label{eq:def_MR_SW}
\begin{align}
    \mr^\pareni &\defn  \mr(\strategy^\pareni) \label{eq:MR-def}\\
    \sw^\pareni &\defn \sw(\strategy^\pareni),\label{eq:SW-def}
\end{align}
\end{subequations}
where we omit the dependence on economy $\econ$ when it is clear from the context. We clarify definitions~\eqref{eq:def_MR_SW} in \Cref{sec:results} when there exist multiple equilibria $\strategy^\pareni$.

\subsection{Illustrative example} 

We begin with a numerical example to illustrate how different information policies affect the equilibria, match rates, and social welfare. 

\begin{example}\label{ex:equilibria}
Consider $3$ students, $3$ schools, and $2$ types of students. Consider an economy with valuations and prior of student types given by:
\begin{center}
    \begin{tabular}{c|c|c|c||c}
     &  school 1 & school 2 & school 3& prior\\\hline
    type 1 & $0.97$ & $0.02$  & $0.01$ & $0.8$\\
    type 2 & $0.65$ & $0.34$ & $0.01$ & $0.2$
\end{tabular}
\end{center}
Here students have homogeneous preference rankings, but different preference intensities. The common type-$1$ students are selective in their preferences, who strongly value school $1$ and have negligible valuations for other schools. The rare type-$2$ students are more flexible, who also prefer school $1$, but also non-trivially value school $2$. Both types have negligible valuation for school $3$. In what follows, we describe the Bayesian Nash equilibrium under each information policy for this economy. The complete derivations are provided in Appendix~\ref{app:derivations}.

\textbf{Revealed lottery.} Student $1$ (i.e., with lottery number $1$) is guaranteed admission to any school by the deferred acceptance algorithm, and hence applies to their most preferred school $1$ for both types. Student $2$ knows that school $1$ is taken by student $1$, and therefore best-responds by applying to their second choice of school $2$ for both types. The remaining student $3$ similarly best-responds, knowing that they are left only with school $3$. Therefore, it is an equilibrium that student $\idxstudent$ applies to school $\idxstudent$ regardless of their type, namely
\[\boldsymbol{\pi}^\parenr_{t,1} = (1,0,0),~~ \boldsymbol{\pi}^\parenr_{t,2} = (0,1,0),~~ \boldsymbol{\pi}^\parenr_{t,3} = (0,0,1)~~\quad \text{for } t\in \{1, 2\}.\] 
At this equilibrium, the matching outcome $\boldsymbol{\sigma}=(1,2,3)$ is deterministic. It is straightforward to show that 
\[\mrr = 1, \quad \swr = 1.\]

\textbf{Hidden lottery.} Type-$1$ students strongly prefer school $1$, and it can be verified that they always apply to school $1$ at equilibrium regardless of their lottery number. Type-$2$ students, on the other hand, apply to school $2$ regardless of their lottery number to avoid competition, because they know that school $1$ is likely taken by a common type-$1$ student with a better lottery number. 
The resulting equilibrium is:
\begin{align*}
    \strategy^\parenh_{1} = (1,0,0),\quad \boldsymbol{\pi}^\parenh_{2} = (0,1,0).
\end{align*}
At this equilibrium, school $3$ is always unfilled, while schools $1$ and $2$ are filled only if there exists a student of the respective type. The match rate and social welfare are:
\begin{align*}
    \mr^\parenh \approx 0.493, \quad 
\sw^\parenh \approx 1.128.
\end{align*}

\textbf{Partial information lottery.} Consider the partition $\mathcal{P}_1 =\{1,2\}$ and $\mathcal{P}_2=\{3\}$. In block $\partition_1$, a type-$1$ student always applies to school $1$ due to their strong preference. A type-$2$ student, on the other hand, uses a mixed strategy. They apply to school $1$ with probability approximately $0.85$, because being in block $\partition_1$ gives them a high priority, but also apply to school $2$ with probability approximately $0.15$ as they suspect that they may share the block with a type-$1$ student with a better lottery number. The student in block $\partition_2$ always applies to school $2$ regardless of their type, as they anticipate that school $1$ is taken with high probability by students in block $1$. The equilibrium is given by
\begin{align*}
    &\boldsymbol{\pi}^{(P)}_{1, \mathcal{P}_1} = (1,0,0), \quad \boldsymbol{\pi}^{(P)}_{2, \mathcal{P}_1} \approx (0.85,0.15,0)\\ &\boldsymbol{\pi}^{(P)}_{t, \mathcal{P}_2} = (0,1,0) \quad\text{ for } \idxtype\in \{1, 2\}.
\end{align*}
At this equilibrium, the match rate and social welfare are: 
\[\mr^\parenp \approx 0.666, \quad 
\sw^\parenp \approx 1.012.\]
\end{example}
\vspace{0.1in}

In this example, the most notable difference among information policies is the equilibrium strategy adopted by the type-$2$ students. Under the revealed lottery, a type-$2$ student chooses their preferred school $1$ when they have the best lottery number $1$. In contrast, under the hidden lottery, they deterministically yield to the common type-$1$ students by choosing their less preferred school $2$, therefore giving up school $1$ even if they have, unbeknownst to them, the best lottery number. Under the partial information lottery, if this type-$2$ student finds themselves in block $\partition_1$, this is a better situation than a uniform prior probability under the hidden lottery, but not as good as having lottery number $1$ under the revealed lottery. Therefore, they adopt a randomized strategy over schools $1$ and $2$. 

We also observe a tension created by the information policy in the match rate and social welfare. The revealed lottery leads to a perfect match rate as all seats are filled. In contrast, the hidden and partial information lotteries have lower match rate as no student applies to school $3$. On the other hand, when type-$2$ students yield and apply to school $2$ (which always happens under the hidden lottery and at times under the partial information lottery), they may allow the seat at school $1$ to be assigned to a type-$1$ student who values it more strongly, thereby improving social welfare. In the next section, we formalize such tradeoffs. 

\section{Theoretical results}\label{sec:results}

We next analytically compare the match rate and social welfare attainable under different information policies. To ascertain that these metrics are well-defined, we establish that multiple equilibria all lead to the same match rate and social welfare (Section~\ref{sec:invariance}). We proceed to contrast the revealed and hidden lotteries in terms of the match rate (\Cref{sec:MR-comp}) and social welfare (\Cref{sec:SW-comp}), followed by discussing the partial information lottery (\Cref{sec:partial-comp}). For all theoretical results, we consider preference lists of length $\length=1$.

We use the big-O notation for the match rate and social welfare in terms of the number of students $\numstudents$. The notation $f(\numstudents) = O(g(\numstudents))$ (resp. $f(\numstudents) = \Omega(g(\numstudents))$) means that there exists a positive constant $\const > 0$ such that $f(\numstudents) \le \const \cdot g(\numstudents)$ (resp. $f(\numstudents) \ge \const \cdot g(\numstudents)$) for all $\numstudents\ge 1$. We use the notation $f(\numstudents) = \Theta(g(\numstudents))$ when both $f(\numstudents) = O(g(\numstudents))$ and $f(\numstudents) = \Omega(g(\numstudents))$ hold.

\subsection{Invariance of multiple equilibria}\label{sec:invariance} 

Recall that we focus on symmetric Bayesian Nash equilibria, in which students of the same type with the same lottery information adopt the same strategy. Such a symmetric Bayesian Nash equilibrium always exists under all three information policies. 
\begin{lemma} \label{lem:exist}
    Let $n \geq 2$ and $T\geq 1$ be arbitrary. Consider any economy with $\numstudents$ students, $\numstudents$ schools, and $\numtypes$ types of students. There always exists a {symmetric} Bayesian Nash equilibrium under all information policies.
\end{lemma}
This lemma comes from a direct application of Kakutani's fixed point theorem for finite Bayesian games. While a symmetric Bayesian Nash equilibrium always exists, it may not necessarily be unique; an example of an economy with multiple equilibria is provided in Appendix~\ref{app:ex_multiple_equilibria}. That said, the following theorem establishes the conditions under which the match rate and social welfare remain the same across all equilibria. 
\begin{theorem}\label{thm:equivalence}
    Consider any ordered and contiguous partition $\partition_1\cup \mathcal{P}_2 \cup \ldots \cup \mathcal{P}_K = [n]$, and any two symmetric Bayesian Nash equilibria $\boldsymbol{\pi}$ and $\boldsymbol{\pi}'$ under the partial information lottery. Assume that for any singleton 
    block $\partition_k$ with $\abs*{\partition_k} = 1$, the two equilibria satisfy
    \begin{align}
        \strategy_{t, \partition_k} = \strategy'_{t, \partition_k}\quad \text{for every type } t\in [\numtypes].\label{eq:equivalence_assume_singleton}
    \end{align}
    Then we have
    \begin{align*}
        \mr^\parenp(\boldsymbol{\pi}) &= \mr^\parenp(\boldsymbol{\pi}')\\
        \sw^\parenp(\boldsymbol{\pi}) &= \sw^\parenp(\boldsymbol{\pi}').
    \end{align*}
\end{theorem}
The proof of this theorem is provided in Appendix~\ref{app:proof_equivalence}. The revealed and hidden lotteries are special cases. For the revealed lottery, every block is a singleton, and condition~\eqref{eq:equivalence_assume_singleton} specifies a unique equilibrium, rendering the theorem trivial.
Under the hidden lottery, condition~\eqref{eq:equivalence_assume_singleton} becomes trivially satisfied, because $\partition_1 = [\numstudents]$ and there exists no singleton block. All equilibria thus have the same match rate and likewise for social welfare.

We now outline the proof under the hidden lottery, and then generalize it to the partial information lottery as in \Cref{thm:equivalence}. Under the hidden lottery, a key quantity is the probability that a random student applies to school $\idxschool\in [\numstudents]$, defined by
\begin{align*}
    \flow_{\idxschool} = \sum_{\idxtype=1}^\numtypes \prior_\idxtype \pi_{\idxtype\idxschool},
\end{align*}
which we also call the ``flow'' to school $\idxschool$. The proof consists of two steps: First, we show that any two equilibria induce identical flows $\{\flow_\idxschool\}_\idxschool$. Second, we show the match rate and social welfare are solely determined by these flows.

The intuition for the first step is as follows. Consider two equilibria $\strategy$ and $\strategy'$, and assume for contradiction that they lead to different flows. Then there must exist a school $\idxschool$ such that $\flow_j > \flow_j'$, meaning that students are overall less likely to apply to school $\idxschool$ under the alternative equilibrium $\strategy'$. This decreases competition and leads to a higher admission probability of school $\idxschool$, attracting more students to apply to school $\idxschool$, which contradicts the assumption that students are more likely to apply to school $\idxschool$. Thus, this negative feedback stabilizes the flows under the two equilibria. This intuition is imprecise because while the admission probability of school $\idxschool$ increases under equilibrium $\strategy'$, it is possible that the admission probabilities of other schools increase even more, so school $\idxschool$  does not necessarily become more attractive in comparison. Therefore, in the proof, we formalize this intuition by identifying another school $\ell$ with $q_\ell \le q_\ell'$. Under equilibrium $\strategy'$, the admission probability of school $\idxschool$ increases and that of school $\ell$ decreases. We argue that it is impossible that school $\idxschool$ becomes less attractive and school $\ell$ becomes more attractive simultaneously.

To extend from the hidden lottery to general partial information lotteries, we proceed inductively from block $\partition_1$ through block $\partition_\numpartitions$. For every block $\partition_\idxpartition$, identical flows in all preceding blocks lead to identical probabilities for a school to be matched in preceding blocks under the two equilibria (or equivalently, the probability for a school to remain available when the matching process for block  $\partition_\idxpartition$ starts). If block $\partition_\idxpartition$ is a singleton, then identical flows arise directly from assumption~\eqref{eq:equivalence_assume_singleton}. If block $\partition_\idxpartition$ is not a singleton, then applying the argument of negative feedback on block $\partition_\idxpartition$ leads to identical flows in block $\partition_\idxpartition$.

We note that assumption~\eqref{eq:equivalence_assume_singleton} of identical strategies in singleton blocks is necessary for the theorem to hold. In singleton blocks, there are no other students to create the negative feedback that forces flows to be identical under different equilibria. When flows diverge, they may lead to different matching outcomes for students in subsequent blocks. For example, consider two types of students, with valuations $[0.5, 0.5]$ for type-$1$ students and $[0.9, 0.1]$ for type-$2$ students with any prior. Under the revealed lottery, if a type-$1$ student has lottery number $1$, they are indifferent between the two schools. Now if the student with lottery number $2$ is of type $2$, their matching outcome is affected by which choice was ultimately made by the student with lottery number $1$. 

Using the identical flows established in \Cref{thm:equivalence}, the following properties hold for the set of equilibria.
\begin{corollary}\label{cor:equilbria_convex}
    Consider any information policy. The number of symmetric Bayesian Nash equilibria is either one or infinite.
\end{corollary}
To see this result, consider the earliest block where two equilibria $\strategy$ and $\strategy'$ differ. If this block is a singleton, then at least two schools provide the same expected utility for a student type, and there is an infinite number of mixed strategies for this student type to break ties. If this block is not a singleton, then the two equilibria induce identical flows, so their linear combination also has identical flows and is also an equilibrium. The next corollary identifies a special case when the equilibrium is unique. This argument is formalized in Appendix~\ref{app:proof_cor_convevx}. 
\begin{corollary}\label{cor:unique_equil_two_schools}
    Assume that there are $2$ schools and any arbitrary $\numtypes\ge 1$ types of students. Under the hidden lottery, the symmetric Bayesian Nash equilibrium is unique. 
\end{corollary}
To see this result, we index the student types in the decreasing order of their valuation for school $1$. Then type-$\idxtype$ students find school $1$ utility maximizing only if type-$\ell$  students find school $1$ utility maximizing for every $\ell < \idxtype$. That is, $\pi_{t1} > 0$ implies $\pi_{\ell 1} = 1$ for every $\ell < \idxtype$. This creates a one-to-one correspondence between strategy profiles and the flows they induce, and the unique flows imply a unique equilibrium strategy profile. This argument is formalized in Appendix~\ref{app:proof_cor_two_schools}.

To substantiate assumption~\eqref{eq:equivalence_assume_singleton} on singletons, we formally define a tie-breaking rule as follows. Consider a type-$t$ student in a singleton block $\partition_\idxpartition$. Given equilibrium strategies of students from all preceding blocks, we denote  by $\utility_{\idxtype\idxpartition}(\idxschool)$ the expected utility of this type-$t$ student when applying to school $\idxschool\in [\numstudents]$. The set of equilibrium strategies is then characterized by all probability distributions over utility-maximizing schools, namely $\argmax_{\idxschool\in [\numschools]} \utility_{\idxtype\idxpartition}(\idxschool)$.
Let  $\omega_t: [n]\rightarrow [n]$ be a permutation over $\numstudents$ schools.\footnote{
More generally, a tie-breaking rule can also be allowed to depend on the block $\partition_k$ under the partial information lottery, and the lottery number $r$ under the revealed lottery. A tie-breaking rule is also allowed to depend on the valuations. For example, whenever multiple schools lead to the same expected utility, a risk-averse tie-breaking rule says that students always prefer the school with the lowest valuation.}
We say that a type-$\idxtype$ student follows the tie-breaking rule $\omega_t$, if the student, among all utility-maximizing schools, deterministically applies to the school $\idxschool$ with the lowest $\omega_t(\idxschool)$. A tie-breaking rule $\{\omega_t\}_t$ thus enforces assumption~\eqref{eq:equivalence_assume_singleton} in Theorem~\ref{thm:equivalence}. Under the revealed lottery, any tie-breaking rule $\{\omega_t\}_t$ leads to a pure equilibrium.

One natural tie-breaking rule arises under homogeneous preference rankings. Recall that we index schools such that $v_{t1}\geq v_{t2} \geq \ldots \geq v_{tn}$ for every type $t\in [\numtypes]$. The \emph{lexicographic tie-breaking rule} is defined by $\omega_t(i)=i$ for every type $\idxtype$. That is, ties are broken in the order of the homogeneous preference ranking. 
Under the lexicographic tie-breaking rule, we establish the following result for the revealed lottery.  
\begin{lemma}\label{lem:reveal_same_ranking_always_one}
    Assume students have homogeneous preference rankings and follow the lexicographic tie-breaking rule. Under the revealed lottery, a unique equilibrium is that student $\idxstudent$ applies to school $\idxstudent$ regardless of their type. At this equilibrium, the match rate is $\mrr = 1$, and the social welfare is $\swr =1$.
\end{lemma}
The proof of this lemma is provided in Appendix~\ref{app:proof_lem_reveal_homo_always_one}. The perfect match rate is shown in prior work~\cite[Lemma~1]{huang2025lotteries}.\footnote{Huang and Zhang~\cite{huang2025lotteries} assume a strict ordering of valuations $v_{t1}>v_{t2}>\ldots>v_{tn}$, which is similar to the lexicographic tie-breaking rule.}  To prove this result, note that the student with lottery number $1$, under homogeneous preference rankings, applies to their most preferred school $1$. Then the student with lottery number $2$, knowing that school $1$ is taken by student $1$, applies to their most preferred school that remains available, which is school $2$, and so on. This iterative argument proves the equilibrium strategy where every student $\idxstudent$ applies to school $\idxstudent$, yielding a match rate of $1$. For social welfare, an individual student has a uniform probability of being assigned each lottery number, and thus a uniform probability of being assigned to each school. Since their valuations over schools sum to one, their expected utility is $\frac{1}{n}$, yielding social welfare of $1$ by summing over all $\numstudents$ students. In our subsequent results, we use this equilibrium to benchmark the performance under the revealed lottery.

Finally, we note that these invariance results on the equilibria are in stark contrast with other classic games, such as coordination games or congestion games, where the quality of the best equilibrium and the worst equilibrium may differ significantly. Prior literature has explored using information to induce players' response dynamics to reach a more desired equilibrium~\citep{balcan2013circumventing}. Our findings here suggest that such interventions aimed at nudging students to move from one equilibrium to another are ineffective in improving either the match rate or social welfare. 

\subsection{Comparison of revealed versus hidden lotteries}

We next compare the matching outcome under revealed and hidden lotteries in terms of the match rate and social welfare.

\subsubsection{Ratio of match rate}\label{sec:MR-comp}
Recall that the match rate $\mr^\pareni$ defined in~\eqref{eq:MR-def} is the fraction of students who are matched under information policy $\texti$. To compare the attainable match rate under the revealed and hidden lotteries, we define the ratio:
\begin{align}
    \mr_{\texti/\textj}^* := \sup_{\econ} ~ \frac{\mr^\pareni}{\mr^\parenj}, \quad \text{for } \texti, \textj \in\{\textr,\texth\}.
    \label{eq:MR-ratio-def}
\end{align}
In words, ratio~\eqref{eq:MR-ratio-def} measures the \emph{best-case} multiplicative improvement in the match rate when moving from information policy $\textj$ to information policy $\texti$, over all possible economies
$\econ=(\{v_{t\idxschool}\}, \dist)$. More precisely, the set of all economies with $\numstudents$ students and $\numschools$ schools is defined by $$\econ_n \defn \union_{\numtypes\ge 1} \{(\{\pref_{tj}\}_{t\in [\numtypes], \idxschool\in [\numstudents]}, \{p_t\}_{t\in [\numtypes]})\}.$$
We define the ratio
\begin{align*}
     \mr_{\texti/\textj}^*(n) := \sup_{\econ_\numstudents} \inf_{\{\omega_t\}_t}~ \frac{\mr^\pareni}{\mr^\parenj},
\end{align*} 
where  $\{\omega_t\}_t$ is any tie-breaking rule, and we omit the dependence on $n$ when clear from the context.\footnote{We likewise define $\sw^*_{\texti/\textj}$ and $\sw^*_{\texti-\textj}$ in subsequent results.}
Since the match rate is at least $\frac{1}{n}$ and at most $1$ in any economy under any information policy, a \naive bound gives $\frac{1}{n}\le\frac{\mr^\pareni}{\mr^\parenj}\le n$. 

To analyze this ratio (both when moving from the revealed lottery to the hidden lottery, and vice versa), we distinguish between settings where all students share the same preference ranking of schools (as assumed in prior work~\cite{huang2025lotteries}) and settings where students have heterogeneous preference rankings. Specifically, when students have homogeneous preference rankings, the ratio is defined by:
\begin{align*}
    \mr^{*,\homo}_{\textrm{I}/\textrm{J}} := \sup_{\substack{\econ\text{ satisfying}\\v_{t1} \ge v_{t2} \ge \cdots \ge v_{t\numschools},\; \forall \idxtype\in [\numtypes]}} ~ \frac{\mr^\pareni}{\mr^\parenj}, \quad \text{for } \textrm{I}, \textrm{J} \in\{\textrm{R},\textrm{H}\}.
\end{align*}
The following result characterizes these ratios under homogeneous and heterogeneous preference rankings. 
\begin{theorem}\label{thm:MR-gaps}
    Assume that there are $n$ schools and $n$ students. 
    The ratios comparing the match rate under revealed and hidden lotteries are as follows.
    \begin{enumerate}[label=(\alph*)]
        \item Under homogeneous preference rankings, we have
        \begin{subequations}
        \begin{align}
            \ratiomrhrhomo &=1- \left(1-\frac{1}{n}\right)^n < 1\label{eq:ratio_mr_hr_homo}\\
            \ratiomrrhhomo &= n.\label{eq:ratio_mr_rh_homo}
        \end{align}

        \item Under heterogeneous preference rankings, we have
        \begin{align}
            \lim_{\numstudents\rightarrow \infty} \ratiomrhr >1.038, &\quad \text{and } \ratiomrhr = O(\log n)\label{eq:ratio_mr_hr_hetero}\\
            \ratiomrrh &= n.\label{eq:ratio_mr_rh_hetero}
        \end{align}
        \end{subequations}
    \end{enumerate}
\end{theorem}
The proof of this theorem is provided in Appendix~\ref{app:proof_MR-gaps}. When students have homogeneous preference rankings for schools, bound~\eqref{eq:ratio_mr_hr_homo} shows that the revealed lottery always has a higher match rate than the hidden lottery. This follows directly from \Cref{lem:reveal_same_ranking_always_one} as $\mrr = 1$, and is consistent with prior work~\citep{huang2025lotteries}. However, when students have heterogeneous preference rankings, the hidden lottery can have a strictly higher match rate than the revealed lottery according to~\eqref{eq:ratio_mr_hr_hetero}, though the advantage remains bounded by a log factor. In proving the lower bound in~\eqref{eq:ratio_mr_hr_hetero}, we construct an economy where students have different preference rankings of schools, and within students who share the same preference ranking, some students are more selective in their preferences and some are more flexible. Under the hidden lottery, flexible students avoid competition by applying to their less preferred schools. Under the revealed lottery, however, if flexible students have advantageous lottery numbers, they apply to their preferred schools, increasing congestion and decreasing the match rate. As a technical aside, this result emphasizes that the ratio~\eqref{eq:ratio_mr_hr_hetero} of the hidden lottery over the revealed lottery is greater than 1, though we did not intend to optimize this constant. 

On the other hand, as shown in bounds~\eqref{eq:ratio_mr_rh_homo} and~\eqref{eq:ratio_mr_rh_hetero}, the revealed lottery can be a linear factor of $n$ better than the hidden lottery, meaning that the revealed lottery has a perfect match rate $\mrr = 1$ and only one student is matched under the hidden lottery such that  $\mrh = \frac{1}{n}$. This is obtained when all students are selective and strongly prefer the same school, so all of them apply to this school under the hidden lottery, whereas under the revealed lottery, the lottery numbers signal students to apply to different schools, as outlined in \Cref{lem:reveal_same_ranking_always_one}.

\subsubsection{Ratio of social welfare }\label{sec:SW-comp}

Recall that the social welfare $\sw^\pareni$ defined in~\eqref{eq:SW-def} is the sum of valuations of all matched students under information policy $\texti$. Similar to our analysis of the match rate, we define the following ratio to compare the social welfare between revealed and hidden lotteries:
\begin{align}
    \sw_{\texti/\textj}^* := \sup_{\econ} ~ \frac{\sw^\pareni}{\sw^\parenj}, \quad \text{for } \texti, \textj \in\{\textr,\texth\}.
    \label{eq:SW-ratio-def}
\end{align}
We likewise define $\sw_{\texti/\textj}^{*,\homo}$ as the ratio under homogeneous preference rankings. For any information policy $\texti$, we have a \naive bound $\sw^\pareni \le n$.
Our analysis of this social welfare ratio is formalized below.
\begin{theorem}\label{thm:SW-gaps}
    Assume that there are $n$ schools and $n$ students.  The ratios comparing the social welfare under revealed and hidden lotteries are as follows.
    \begin{enumerate}[label=(\alph*)]
        \item \label{item:sw_ratio_homo}
        Under homogeneous preference rankings, we have
        \begin{subequations}
        \begin{align}
            \ratioswhrhomo &= \Theta(\log n)\label{eq:ratio_sw_hr_homo}\\
            \ratioswrhhomo &= 2-\frac{1}{n}.\label{eq:ratio_sw_rh_homo}
        \end{align}
    
        \item \label{item:sw_ratio_hetero}
        Under heterogeneous preference rankings, we have
        \begin{align}
            \ratioswhr&= \Theta(\sqrt{n})\label{eq:ratio_sw_hr_hetero}\\
            \ratioswrh &= O(\log^2 n).\label{eq:ratio_sw_rh_hetero}
        \end{align}
        \end{subequations}
    \end{enumerate}
\end{theorem}
The proof of this theorem is provided in Appendix~\ref{app:proof_SW-gaps}.
In contrast to the results on match rate, bound~\eqref{eq:ratio_sw_hr_homo} shows that even under homogeneous preference rankings, the social welfare under the hidden lottery can be better than the revealed lottery. That is, revealing the lottery number to students can strictly hurt social welfare. To prove the lower bound in~\eqref{eq:ratio_sw_hr_homo}, we construct a spectrum of student types with varying preference intensities, ranging from the most selective students who only value a single school to the most flexible students who value all schools equally. The same mechanism is at work: Under the revealed lottery, students with worse lottery numbers tend to apply to less preferred schools to avoid the competition at highly-demanded schools. Thus, the revealed lottery improves the match rate by avoiding congestion. On the other hand, however, students with better lottery numbers may strategically take advantage of their high priority in matching, which can give rise to matches with low utilities and thus hurt the social welfare. Under homogeneous preference rankings, we further note from bounds~\eqref{eq:ratio_sw_hr_homo} and~\eqref{eq:ratio_sw_rh_homo} that the hidden lottery has a higher best-case improvement. That is, the hidden lottery can be better than the revealed lottery by a multiplicative factor of $\Theta(\log n)$, whereas the revealed lottery can only be better than the hidden lottery by a constant factor of $\approx 2$.

Moving to heterogeneous preference rankings, we show that the improvement for the hidden lottery over the revealed lottery increases from $\Theta(\log n)$ to $\Theta(\sqrt{n})$, where the improvement for the revealed lottery remains small, from a constant factor of $\approx 2$ to a log factor of $O(\log^2 n)$. Combining parts~\ref{item:sw_ratio_homo} and~\ref{item:sw_ratio_hetero}, revealing the lottery not only does not guarantee an improvement over the hidden lottery, but also may cause a significant loss in social welfare especially under heterogeneous preference rankings. 
For the lower bound in~\eqref{eq:ratio_sw_hr_hetero}, we construct an economy (termed the ``common-rare'' economy formally defined in Appendix~\ref{app:prelim_common_rare}), where there is one ``common'' school that the majority of the students prefer. There are also a number of ``rare'' types of students where each rare type prefers a distinct ``rare'' school. In practice, these rare schools can represent specialized schools that attract a unique subgroup of students. Furthermore, we assume that the rare-type students are more selective, whereas common-type students are more flexible in that they also mildly value each rare school that a corresponding rare type of students strongly value. A lower social welfare under the revealed lottery arises under the same mechanism: Under the hidden lottery, all types of students apply to their respective preferred school. However, under the revealed lottery, some common-type students with worse lottery numbers, knowing that there likely exists a common-type student with a better lottery number who would have taken the seat at the common school, instead apply to rare schools. These common-type students block rare-type students who would benefit more from rare schools. 

Two technical remarks follow. First, to prove the upper bounds in~\eqref{eq:ratio_sw_hr_hetero} and~\eqref{eq:ratio_sw_rh_hetero}, we establish new connections between equilibrium strategies under the revealed and hidden lotteries. Second, the same economy construction in~\eqref{eq:ratio_sw_rh_homo} under homogeneous preference rankings trivially extends to heterogeneous preference rankings so that $\ratioswrh \ge 2-\frac{1}{n}$. The upper bound~\eqref{eq:ratio_sw_rh_hetero}  thus differs from the lower bound by a log factor.

\subsubsection{Difference of social welfare }

Our previous comparison of social welfare focuses on the best-case \emph{multiplicative} improvement. We now examine the \emph{additive} improvement, defined by:
\begin{align}
    \sw_{\texti-\textj}^* := \sup_{\econ} ~ \big(\sw^\pareni -\sw^\parenj\big), \quad \text{for } \texti, \textj \in\{\textrm{R},\textrm{H}\},
    \label{eq:SW-diff-def}
\end{align}
and likewise for $\sw_{\texti-\textj}^{*,\homo}$ under homogeneous preference rankings. The following result bounds these social welfare differences.\footnote{Note that the match rate is always between $0$ and $1$, so we have the trivial bound that the match rate difference is always $O(1)$.}
\begin{theorem}\label{thm:SW-diff-gaps}
    Consider a setting with $n$ schools and $n$ students. The relative performance of revealed and hidden lotteries in terms of the difference in attainable social welfare \eqref{eq:SW-diff-def} is as follows.
    \begin{enumerate}[label=(\alph*)]
        \item Under homogeneous preference rankings, we have
        \begin{subequations}
        \begin{align}
            \diffswhrhomo &= \Theta(\log n)\label{eq:diff_sw_hr_homo}\\
            \diffswrhhomo &= \frac{n-1}{2n-1}.\label{eq:diff_sw_rh_homo}
        \end{align}

        \item Under heterogeneous preference rankings, we have
        \begin{align}
            \diffswhr&= \Theta(n)\label{eq:diff_sw_hr_hetero}\\
            \diffswrh &= \Theta(n).\label{eq:diff_sw_rh_hetero}
        \end{align}
        \end{subequations}
    \end{enumerate}
\end{theorem}
The proof of this theorem is provided in Appendix~\ref{app:proof_SW-diff-gaps}. 
Under homogeneous preference rankings, the social welfare under the revealed lottery is always $\swr = 1$ by \Cref{lem:reveal_same_ranking_always_one}. Therefore, optimizing the difference is equivalent to optimizing the ratio, and the same economy constructions are used as those from \Cref{thm:SW-gaps}. Under heterogeneous preference rankings, we identify economies where both the hidden and revealed lotteries yield a social welfare linear in $n$, but still cause a difference as large as $\Theta(n)$ in either direction. The economy constructed for proving the lower bound in~\eqref{eq:diff_sw_hr_hetero} is a common-rare economy, similar to that for proving the lower bound in~\eqref{eq:ratio_sw_hr_hetero} from \Cref{thm:SW-gaps} but with more rare types.

\subsection{Partial information lottery}\label{sec:partial-comp}

With our understanding of the revealed and hidden lotteries, we now turn to partial information lotteries, which include revealed and hidden lotteries as extreme cases with complete and no information disclosed, respectively. One may initially expect that the performance under the partial information lottery lies between the two extreme cases. However, we show that this is not the case for either the match rate or social welfare. That is, the quality of the matching outcome is not monotonic in the amount of information.

We begin with the match rate. Recall from Lemma~\ref{lem:reveal_same_ranking_always_one} that when students have homogeneous preference rankings, the revealed lottery always has match rate $\mrr=1$ and thus outperforms the hidden lottery. Nonetheless, we show that the information provided by the partial information lottery may not increase the match rate compared to the hidden lottery. 
\begin{proposition}\label{prop:partial-MR}
    Assume students have homogeneous preference rankings for schools. There exist $\numstudents_0$, economy $\econ_{\numstudents_0}$, and partition for the partial information lottery satisfying
    \begin{subequations}
    \begin{align}
       \mrh < \mrp < \mrr =1,\label{eq:ranking_partial_mr_RPH}
    \end{align}
    as well as $\numstudents_0$, economy $\econ_{\numstudents_0}$, and partition for the partial information lottery satisfying
    \begin{align}
        \mrp < \mrh < \mrr =1.\label{eq:ranking_partial_mr_RHP}
    \end{align}
    \end{subequations}
\end{proposition}
The proof of this proposition is by construction and is provided in Appendix~\ref{app:proof_partial-MR}. The following result further establishes the non-monotonic role of information under heterogeneous preference rankings.
\begin{proposition}\label{prop:partial-MR-hetero} 
    Assume students have heterogeneous preference rankings for schools. For any of the six orderings among the match rate  $\{\mrr, \mrp, \mrh\}$, there exist $\numstudents_0$, economy $\econ_{\numstudents_0}$, and partition for the partial information lottery, such that the match rate in economy $\econ_{\numstudents_0}$ satisfies that given ordering.
\end{proposition}
The proof of this proposition is provided in Appendix~\ref{app:proof_partial_MR_hetero}. 
The following result establishes a similar non-monotonic effect in terms of social welfare, even under homogeneous preference rankings.
\begin{proposition}\label{prop:partial-SW} 
    Assume students have homogeneous preference rankings for schools.  For any of the six orderings among the social welfare $\{\swr, \swp, \swh\}$, there exist $\numstudents_0$, economy $\econ_{\numstudents_0}$, and partition for the partial information lottery, such that the social welfare in economy $\econ_{\numstudents_0}$ satisfies that given ordering.
\end{proposition}
The proof of this proposition is provided in Appendix~\ref{app:proof_partial-SW}.
The non-monotonic effect of information presents both opportunities and challenges. On the positive side, there exist cases where revealing the exact lottery number hurts the matching outcome, but revealing partial information improves it, suggesting that partial information can sometimes be effective in improving efficiency without sacrificing too much transparency. On the negative side, there also exist cases where revealing exact information improves the matching outcome, but partial information hurts. Under the revealed lottery, students' cognitive biases may prevent them from optimally processing the lottery information~\citep{miller1956seven}. If students only reason about their lottery number coarsely, they behave as if only partial information is available, so the information intervention that aims to improve the matching outcome may end up bringing unintended harm.

\section{Simulation}\label{sec:sim}

Our theoretical results so far have assumed that each student applies to $\length=1$ school. We now provide simulations for general $\length\ge 1$. In this case, the size of the action space is of order $\numschools^\length$, which consists of rankings of all subsets of $\length$ out of $\numschools$ schools; a mixed strategy is a probability distribution over this action space, which can take exponential time in $\numschools$ to find. To make the problem computationally tractable, we introduce a generalization of our previous model, under which we identify a simple case where student strategies can be computed numerically. 

In this general model, we assume that schools are partitioned into $\numtiers$ tiers by $[\numschools]=\setschools_1\union \cdots \union \setschools_\numtiers$, where the set $\setschools_\idxtier$ denotes schools that belong to tier $\idxtier$. For every student type $\idxtype\in [\numtypes]$, the valuation is identical for schools within the same tier. That is, for every tier $\idxtier\in [\numtiers]$, we have $\pref_{\idxtype\idxschool} = \pref_{\idxtype\idxschool'}$ for every $\idxschool, \idxschool'\in \setschools_\idxtier$.
We consider the following action space of students, parameterized by $(\countschools_1, \ldots, \countschools_\numtiers)$ with $\sum_{\idxtier=1}^\numtiers \countschools_\idxtier = \length$. Here, $\countschools_\idxtier$ denotes the number of schools from tier $\idxtier\in [\numtiers]$ that the student applies to, so that $0 \le \countschools_\idxtier \le \abs*{\setschools_\idxtier}$. A student forms a school ranking of length $\length$ as follows:
\begin{itemize}
    \item The student selects $\countschools_\idxtier$ schools uniformly at random from each tier $\idxtier\in [\numtiers]$.

    \item The student ranks all $\length$ selected schools according to their valuations, with ties broken arbitrarily. That is, the student ranks schools in the order of their tiers, with ties broken within each tier.
\end{itemize}
We consider symmetric Bayesian Nash equilibria where students of each type follow a (possibly mixed) strategy over $(\countschools_1, \ldots, \countschools_\numtiers)$. By standard arguments as in \Cref{lem:exist}, a symmetric Bayesian Nash equilibrium always exists. 
Note that our previous model is a special case of this general model, where each school forms its own tier and $\length=1$.

For our simulations, we consider $\numtiers=2$ school tiers and two types of students. We focus on homogeneous preference rankings where both student types prefer schools in tier 1 over tier 2. One student type is selective, with valuations $1$ and $0$ for the two school tiers, respectively.\footnote{We apply the argument in Appendix~\ref{app:proofs_prelims} to make the valuations strictly positive.} The other student type is flexible, with valuations $\pref_{F1}$ and $\pref_{F2}$ for the two schools tiers, where $\frac{1}{2} < \pref_{F1} < 1$ and $\pref_{F1} + \pref_{F2}=1$ (so we have $\pref_{F1} > \pref_{F2})$. Assume that a student has prior $\prior_{F}\in [0, 1]$ to be the flexible type, and thus prior $(1-\prior_F)$ to be the selective type. Selective students always apply to $\length$ schools in tier $1$ and none in tier $2$. For flexible students, we use $\countschools$ to denote the number of tier-1 schools they apply to. The strategy of flexible students is thus a distribution over $\countschools\in \{0, 1, \ldots, \min\{\length, \abs*{\setschools_1}\}\}$. We parameterize this mixed strategy by
\begin{align*}
    \strategynum \defn (\sigma_0, \ldots, \sigma_{\min\{\length, \abs*{\setschools_1}\}}),
\end{align*}
where $\sigma_\countschools$ denotes the probability that a flexible student applies to $\countschools$ schools of tier 1.

\subsection{Equilibrium computation}

We consider a mean-field approximation where students and schools both form a continuum of mass 1, and students' lottery numbers $\lottery\in [0, 1]$ are uniform. We define the fraction of schools of the two tiers by
\begin{align*}
    \fractier_\idxtier \defn \frac{\abs*{\setschools_\idxtier}}{\numschools},\quad \text{for }\idxtier\in \{1, 2\},
\end{align*}
and consider the equilibrium where the fractions $\fractier_\idxtier$ are assumed to be fixed constants as $\numschools\rightarrow \infty$.

Appendix~\ref{app:equil_comp} describes the computation of the equilibrium under the revealed and hidden lottery; we provide a brief summary here. 
Under the revealed lottery, we divide lottery numbers into $n$ intervals $(0, \frac{1}{n}), (\frac{1}{n}, \frac{2}{n})$, up to $(\frac{n-1}{n}, 1)$ for some large value of $n$, and iteratively compute the strategy for students in each interval like in our previous model. In each interval, the probabilities of admission to the two school tiers  are computed under a mean-field approximation, assuming that these probabilities only depend on students from earlier intervals but not those within the same interval. Under the hidden lottery, the following proposition significantly simplifies the space of mixed strategies to search over, at the mean-field equilibrium formally defined in Appendix~\ref{app:equil_comp}.
\begin{proposition}\label{prop:mf_equilibrium_consecutive}
Consider any list length $\length \ge 1$. Consider any symmetric mean-field equilibrium $\strategynum^*$ under the hidden lottery. Then $\strategynum^*$ is either a pure strategy, or a mixed strategy supported on two consecutive integers. That is, one of the following must be true:
\begin{itemize}
    \item There exists an integer $0 \le m\le \length$, such that
        \begin{align*}
            \sigma_m =1, \quad \text{and }\quad\sigma_{m'} = 0\quad\text{ for every }m' \ne m,
        \end{align*}

    \item There exists an integer $0\le m\le \length-1$, such that 
        \begin{align*}
            \sigma_m >0, \quad\sigma_{m+1} > 0,\quad \text{and }\quad \sigma_{m'} = 0\quad \text{ for every } m' \not\in \{m, m+1\}.
        \end{align*}
\end{itemize}
\end{proposition}
The proof of this proposition is provided in~\ref{app:proof_prop_mf_equilibrium_consecutive}. This result reduces the search of an equilibrium to finding the value of $\countschools$, and, for mixed strategies, also the probabilities assigned to $\countschools$ and $(\countschools+1)$. Finding the value of $\countschools$ can be done efficiently by a binary search, as described in Appendix~\ref{app:equil_comp}.

\subsection{Comparison of hidden and revealed lotteries}

We assume that each tier consists of half of the schools, so that $\fractier_{1} = \fractier_{2} = 0.5$. We examine the matching outcome as a function of the list length $\length$, the preference intensity $\pref_{F1}$ of flexible students, and the fraction $\priorflexible$ of flexible students.

\paragraph{Effect of list length $\length$.} We consider the following two economies:
\begin{itemize}
    \item \textbf{\Singletype economy:} We set $\prior_F = 1$ with $(v_{F1}, v_{F2}) = (0.75, 0.25)$. That is, there exists a single type of students who are flexible. 
    
    \item \textbf{\Equalprior economy:} We set $\prior_F=0.5$ with $(v_{F1}, v_{F2}) = (0.55, 0.45)$.  
\end{itemize}

\begin{figure}
\centering
    \subfloat[\Singletype economy]{\includegraphics[width=0.42\linewidth]{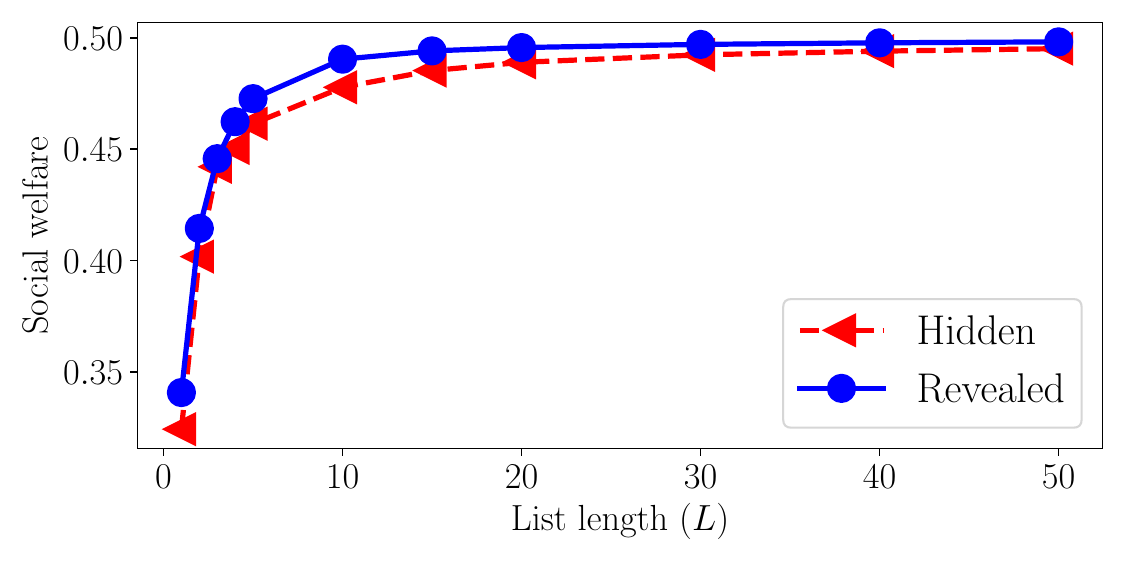}~~~~~~~
        \includegraphics[width=0.42\linewidth]{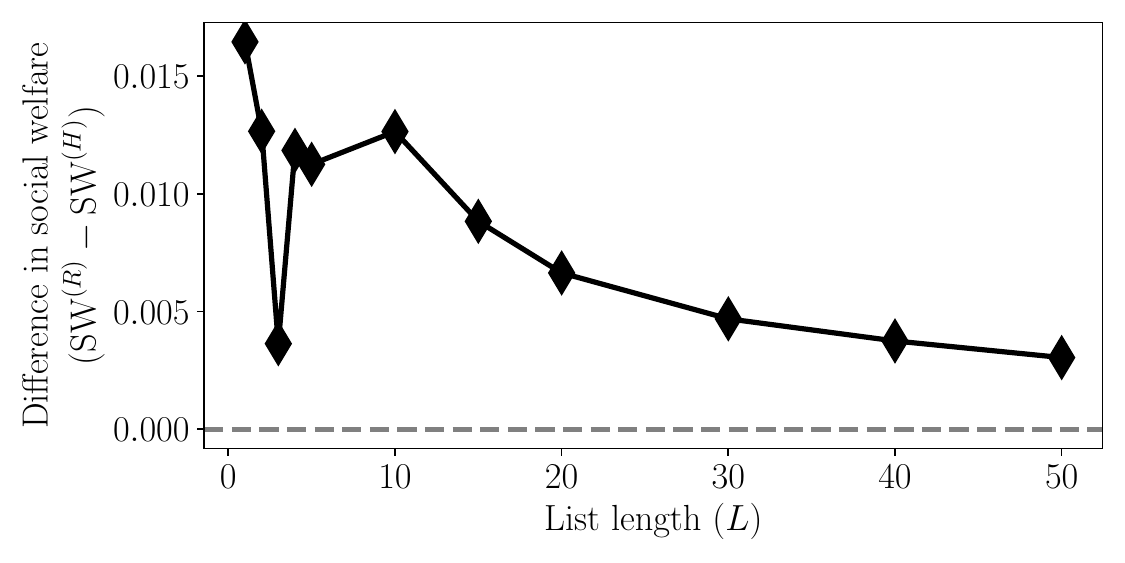}\label{float:sim_sw_one_type}}
    
    \subfloat[\Equalprior economy]{\includegraphics[width=0.42\linewidth]{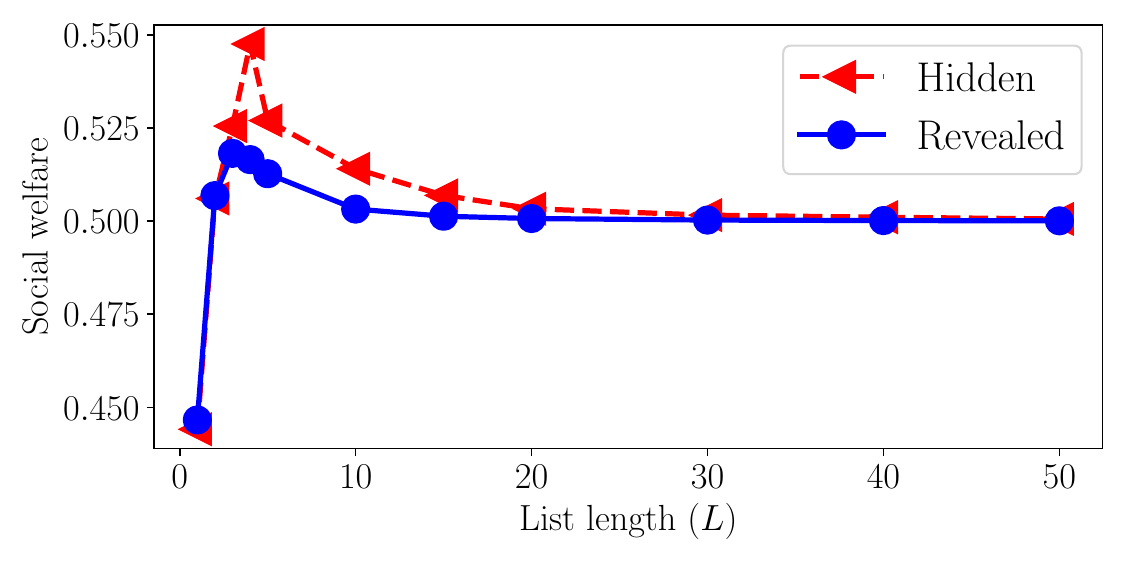}~~~~~~~
        \includegraphics[width=0.42\linewidth]{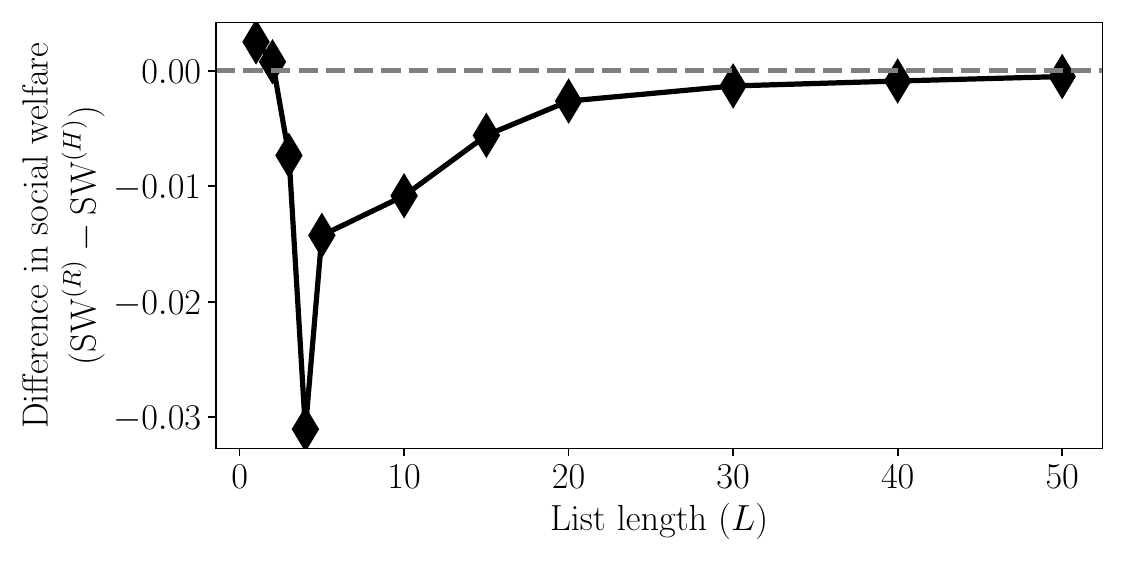}\label{float:sim_sw_two_types}}
    \caption{Social welfare under the revealed and hidden lotteries (left) and their difference (right).\label{fig:sim_sw}
    }
\end{figure}

We visualize the social welfare under hidden and revealed lotteries, along with their difference in \Cref{fig:sim_sw}. The difference in social welfare between the revealed and hidden lotteries is positive in the flexible-only economy (\Cref{fig:sim_sw}\subref{float:sim_sw_one_type}) and primarily negative in the equal-prior economy (\Cref{fig:sim_sw}\subref{float:sim_sw_two_types}). The difference has abrupt changes for a small list length $\length$, but the general trend is that its magnitude decreases as the list length $\length$ increases. This is expected because in the extreme case where $\length=\numschools$, all students report a total ranking truthfully and the difference in social welfare is known to be zero. In practice, however, these lists have limited lengths, e.g., $12$ in New York City before 2024, $12$ in Washington DC~\citep{url_dc}, and $20$ in Chicago~\citep{url_chicago}. With such a relatively small $\length$, the difference in social welfare remains visible. 

\begin{figure}
\centering
    \subfloat[]{\includegraphics[width=0.95\linewidth]{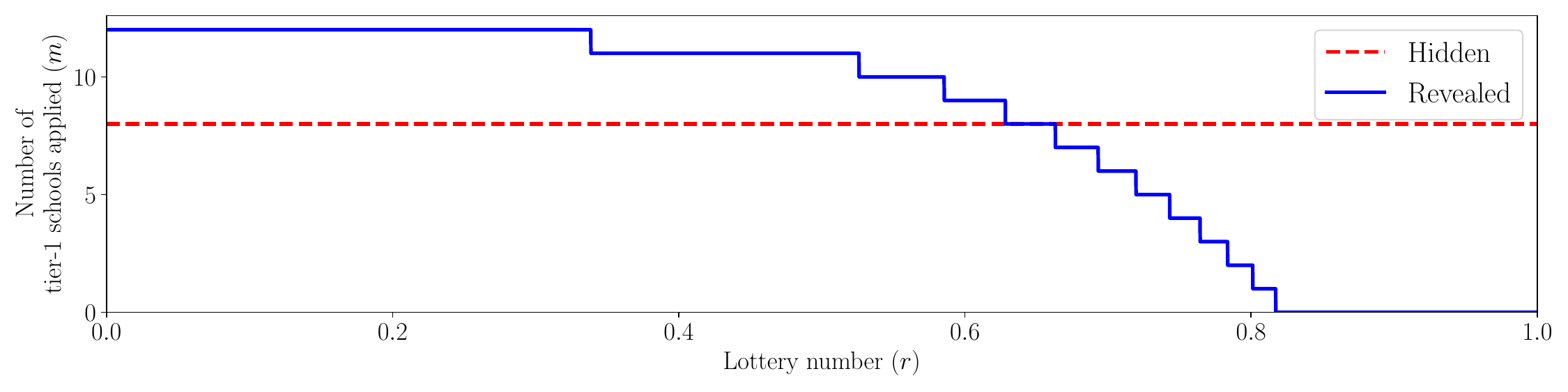}\label{float:sim_single_type_equil}}

    \subfloat[]{\includegraphics[width=0.4\linewidth]{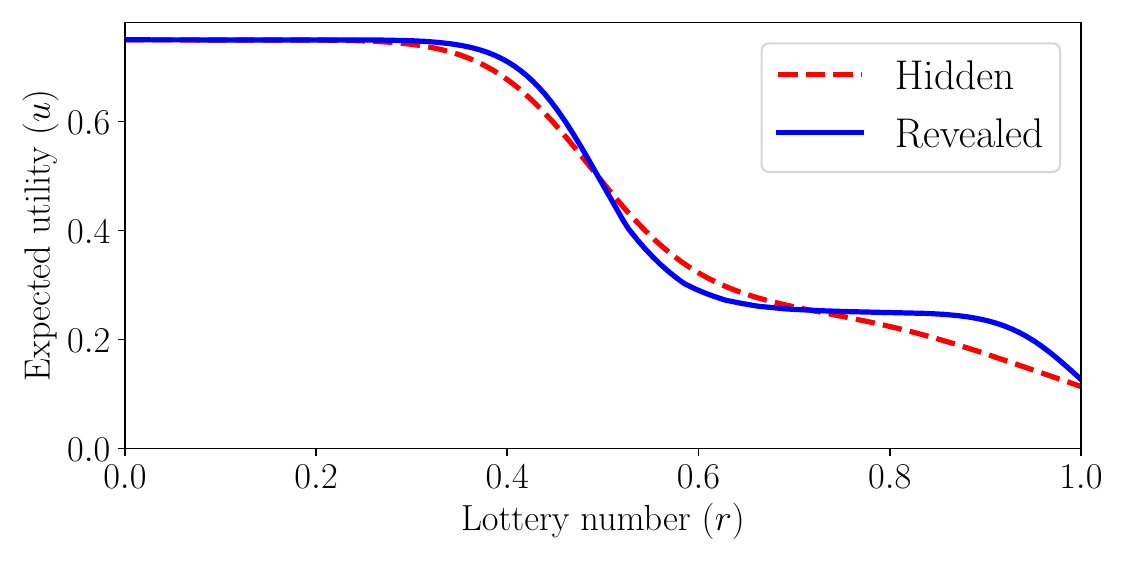}\label{float:sim_single_type_util}}
    ~~~~~~~
    \subfloat[]{\includegraphics[width=0.4\linewidth]{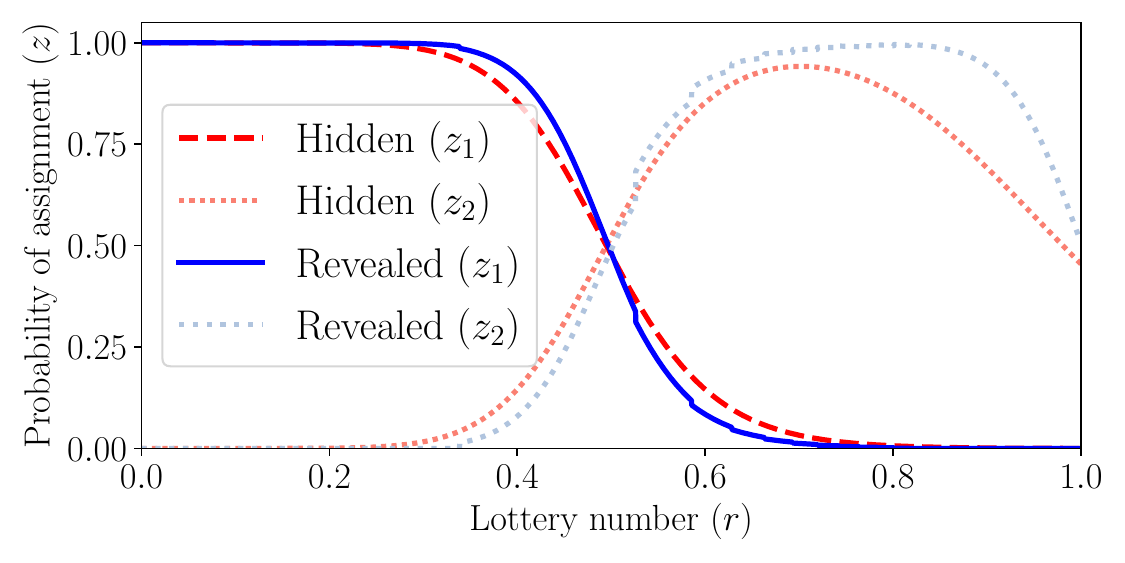}\label{float:sim_single_type_prob}}
    \caption{Visualization of equilibria in the \singletype economy. The subplots show, at each lottery number: (a) the number of tier-1 schools applied to; (b) the expected utility under the hidden and revealed lotteries; and (c) the probability of matching to a school in each tier. \label{fig:sim_one_type}
    }
\end{figure}

We now examine the equilibria in the two economies more closely. \Cref{fig:sim_one_type} visualizes the equilibria under the \singletype economy with list length $\length=12$. As shown in \Cref{fig:sim_one_type}\subref{float:sim_single_type_equil}, students under the hidden lottery apply to $8$ schools from tier 1. Under the revealed lottery, as the lottery number increases, students start from applying only to tier-1 schools (for students  with the best lottery numbers whose admission to tier-1 schools is almost guaranteed) to applying only to tier-2 schools (for students with later lottery numbers whose admission to tier-1 schools is so low that it is more worthwhile to focus on tier-2 schools). In \Cref{fig:sim_one_type}\subref{float:sim_single_type_util}, comparing the revealed lottery to the hidden lottery, students with the best lottery numbers (roughly in $[0, 0.2]$) have similar utilities, because admission to a tier-1 school is almost guaranteed. Students with lottery numbers roughly in $[0.2, 0.5]$ have higher utilities under the revealed lottery, because knowing their relatively good lottery numbers makes the students apply ambitiously to more tier-1 schools. Students with lottery numbers roughly in $[0.5, 0.7]$ have lower utilities under the revealed lottery, because they are negatively impacted by the students with $r\in [0.2, 0.5]$ who have taken more seats at tier-1 schools. Finally, students with lottery numbers roughly in $[0.7, 1]$ have notably higher utilities under the revealed lottery, by applying conservatively to more tier-2 schools. In \Cref{fig:sim_one_type}\subref{float:sim_single_type_prob}, we visualize a student's probability of being assigned a tier-1 school or a tier-2 school, denoted by $\probassign_1$ and $\probassign_2$ which are formally defined in Appendix~\ref{app:equil_comp}. We observe that for students with lottery numbers in $[0.7, 1]$, their probability of being assigned a tier-2 school is much higher under the revealed lottery.

\begin{figure}[tb]
\centering
    \subfloat[]{\includegraphics[width=0.95\linewidth]{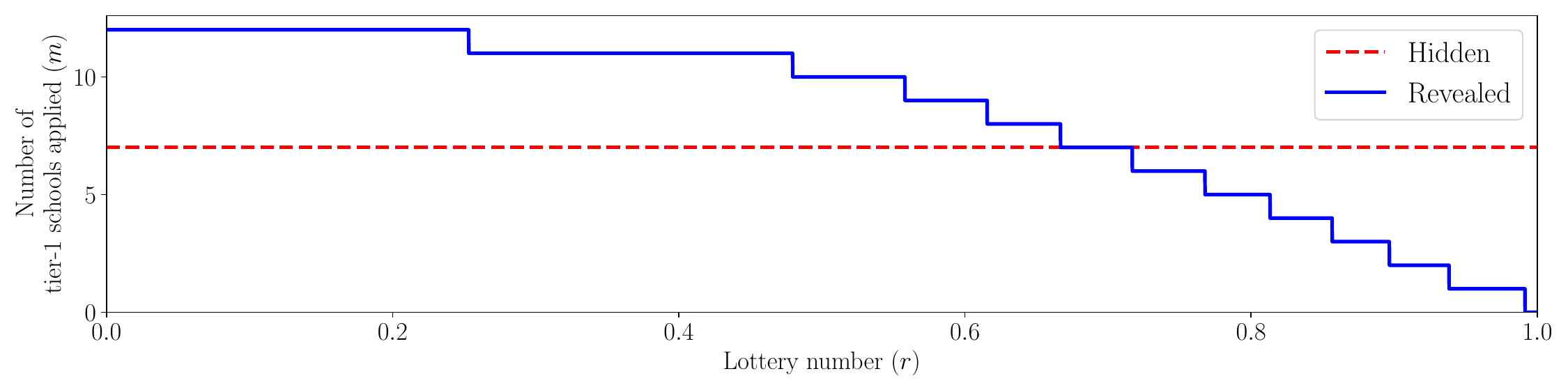}\label{float:sim_two_types_equil}}~~~~~~~

    \subfloat[]{\includegraphics[width=0.4\linewidth]{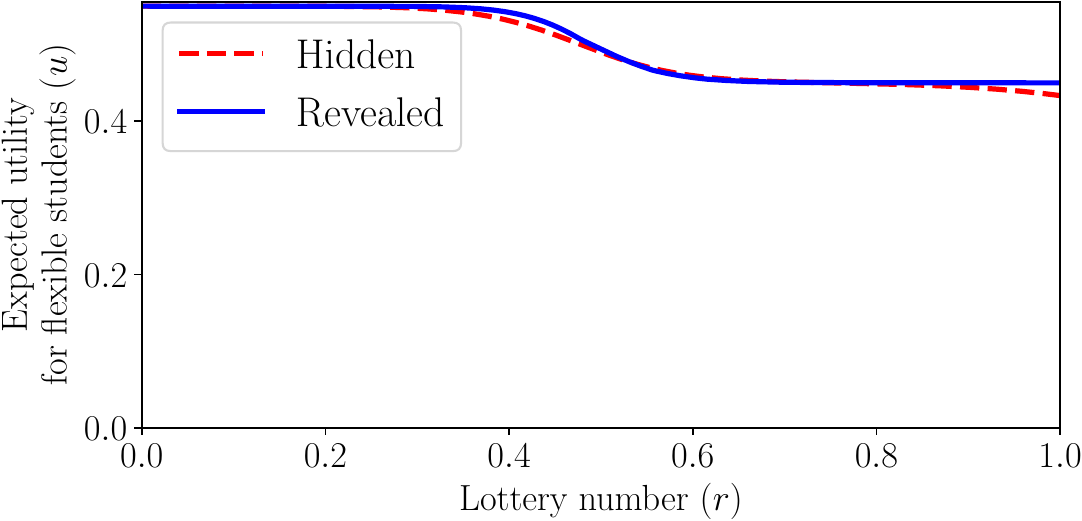}\label{float:two_types_util_flexible}}~~~~~~~
    \subfloat[]{\includegraphics[width=0.4\linewidth]{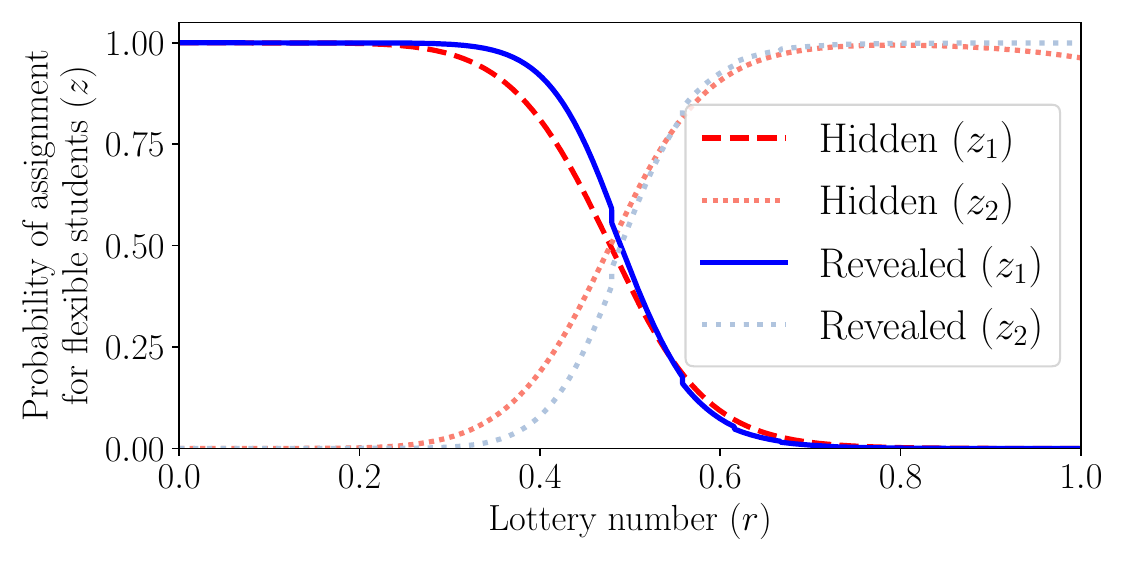}\label{float:two_types_prob_flexible}}

    \subfloat[]{\includegraphics[width=0.4\linewidth]{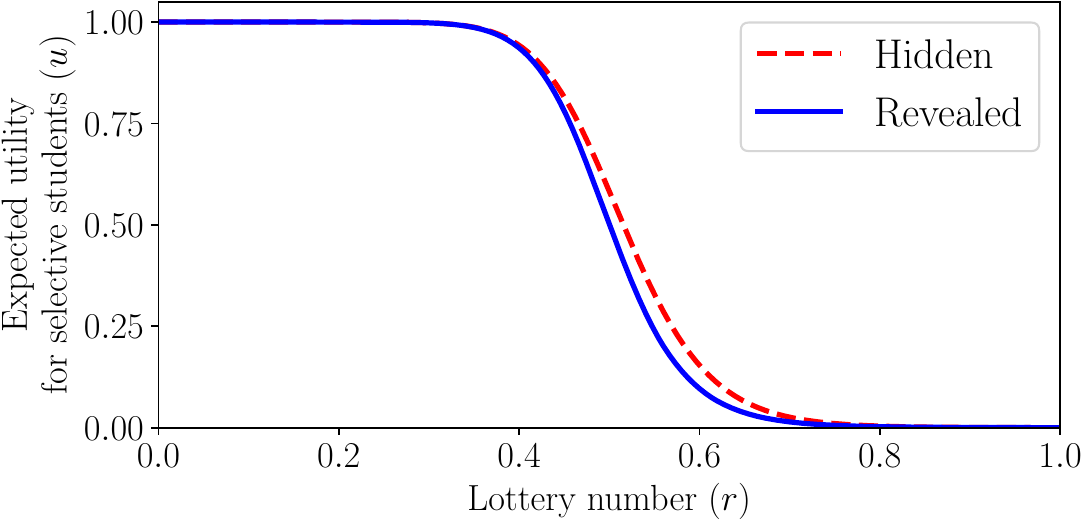}\label{float:two_types_util_selective}}~~~~~~~
    \subfloat[]{\includegraphics[width=0.4\linewidth]{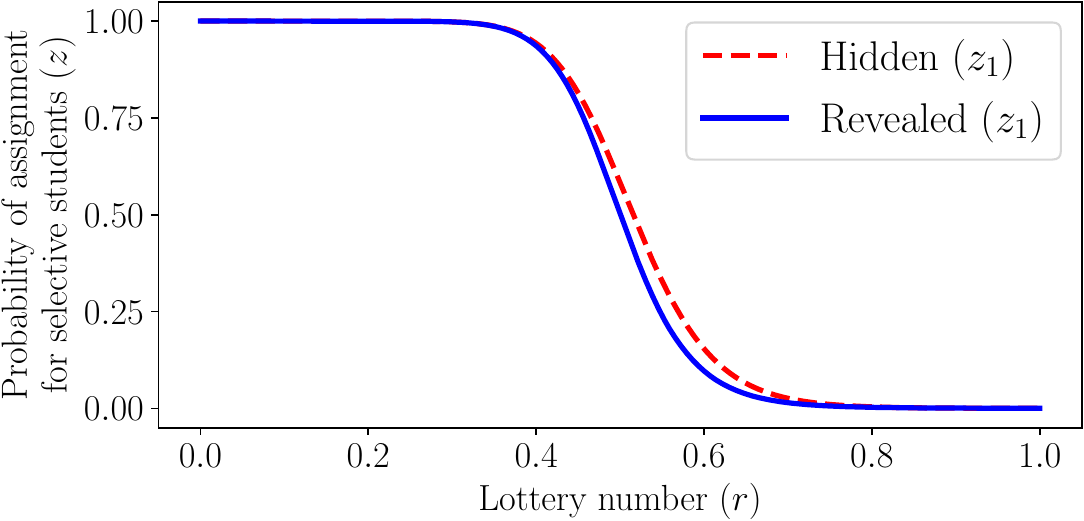}\label{float:two_types_prob_selective}}

    \caption{Visualization of equilibria in the \equalprior economy. The subplots show, at each lottery number: (a) the number of tier-1 schools applied to by flexible students; (b)/(d) the expected utility under the hidden and revealed lotteries for flexible/selective students; and (c)/(e) the probability of matching to a school in each tier for flexible/selective students. \label{fig:sim_two_types}
    }
\end{figure}

In Figure~\ref{fig:sim_two_types}, we visualize the equilibria in the \equalprior economy. Since selective students only apply to tier-1 schools, we visualize the equilibrium strategy of flexible students in \Cref{fig:sim_two_types}\subref{float:sim_two_types_equil}. We again observe that under the revealed lottery, flexible students apply more conservatively as their lottery number increases. We visualize the students' expected utility and probability of tier assignment, separately for flexible students (\Cref{fig:sim_two_types}(\subref*{float:two_types_util_flexible}-\subref*{float:two_types_prob_flexible})) and selective students (\Cref{fig:sim_two_types}(\subref*{float:two_types_util_selective}-\subref*{float:two_types_prob_selective})). We observe that lottery revealing benefits flexible students slightly, but hurts selective students significantly in terms of both the expected utility and the probability of admissions. This again illustrates the mechanism that we have identified: Under the revealed lottery, flexible students with good lottery numbers apply to more tier-1 schools, taking seats that would be taken by selective students under the hidden lottery, leading to a decrease in social welfare. 

\begin{figure}[tb]
\centering
    \includegraphics[width=0.42\linewidth]{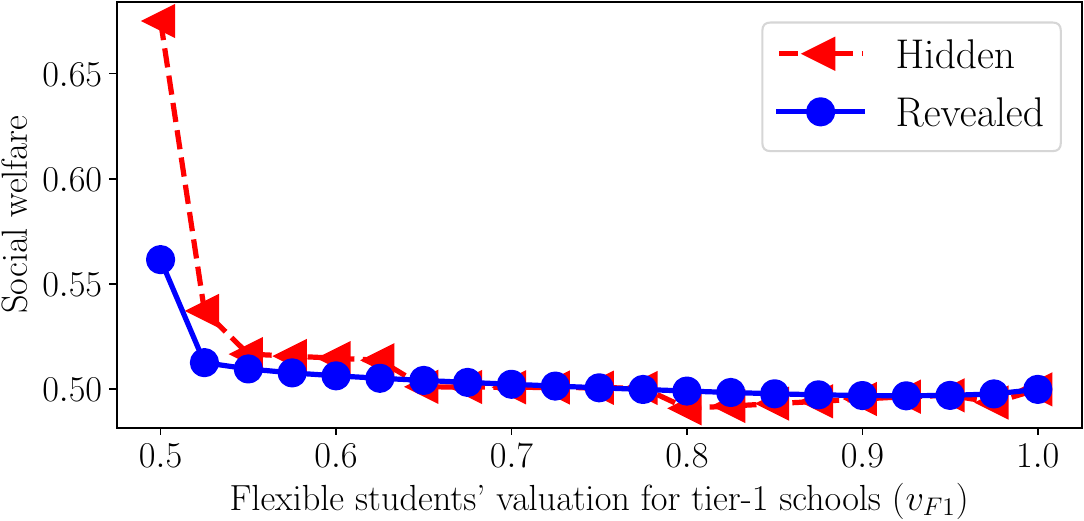}~~~~~~~
    \includegraphics[width=0.42\linewidth]{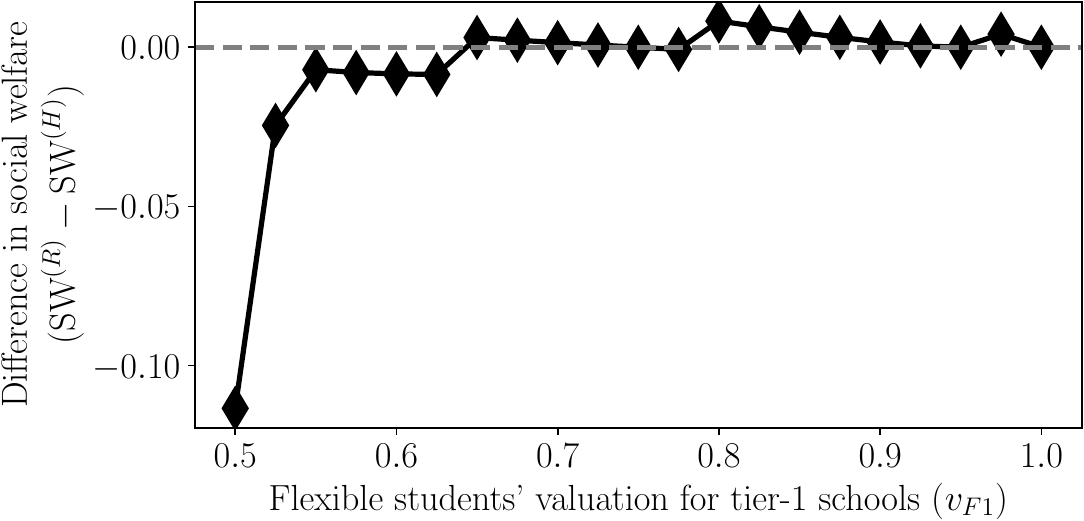}
    \caption{Social welfare under the revealed and hidden lotteries (left) and their difference (right), when varying the preference intensity of flexible students via the value of $v_{F1}$.\label{fig:sim_sw_vary_v}}
\end{figure}

\paragraph{Effect of preference intensity $\pref_{F1}$.} We examine how the preference intensity of flexible students affects the social welfare. We set the fraction of flexible students to be $\priorflexible=0.5$. We observe from Figure~\ref{fig:sim_sw} that the difference between the two lotteries is more prominent for a small list length. Therefore, we set the list length $\length=6$. Results are qualitatively similar for $\length=12$ (not shown). In \Cref{fig:sim_sw_vary_v}, we vary the value of $\pref_{F1}$ (and $\pref_{F2} = 1 - \pref_{F1}$) and plot the social welfare under the revealed and hidden lotteries as well as their difference. We observe that lottery revealing decreases the social welfare when flexible students have weak preferences for tier-1 schools over tier-2 schools. At the extreme, when $\pref_{F1} = 0.5$, lottery revealing decreases the social welfare the most. In this case, flexible students are indifferent from the two school tiers. Under the revealed lottery, some flexible students get admitted to tier-1 schools, which does not improve social welfare of flexible students yet hurts that of selective students. 

\begin{figure}[tb]
\centering
    \includegraphics[width=0.42\linewidth]{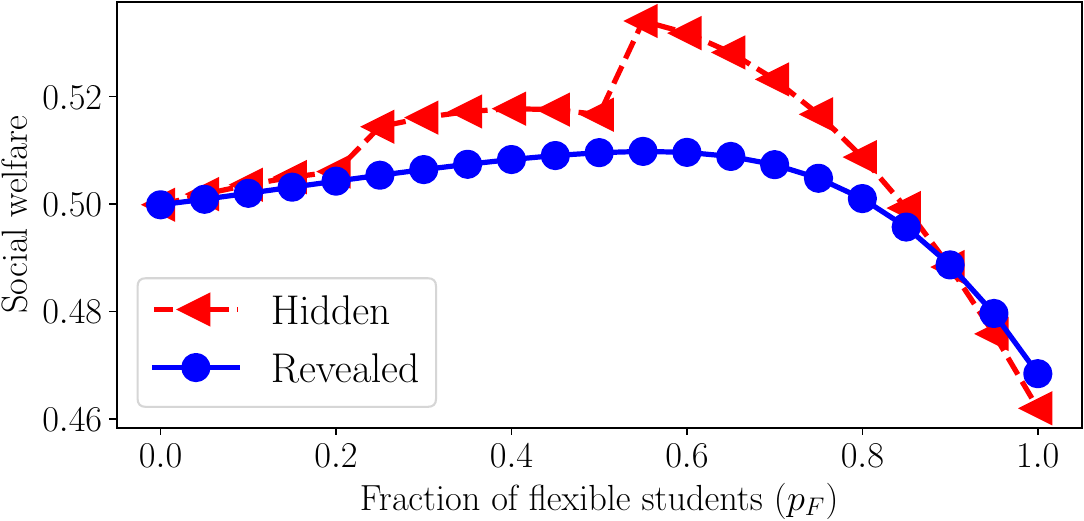}~~~~~~~
    \includegraphics[width=0.42\linewidth]{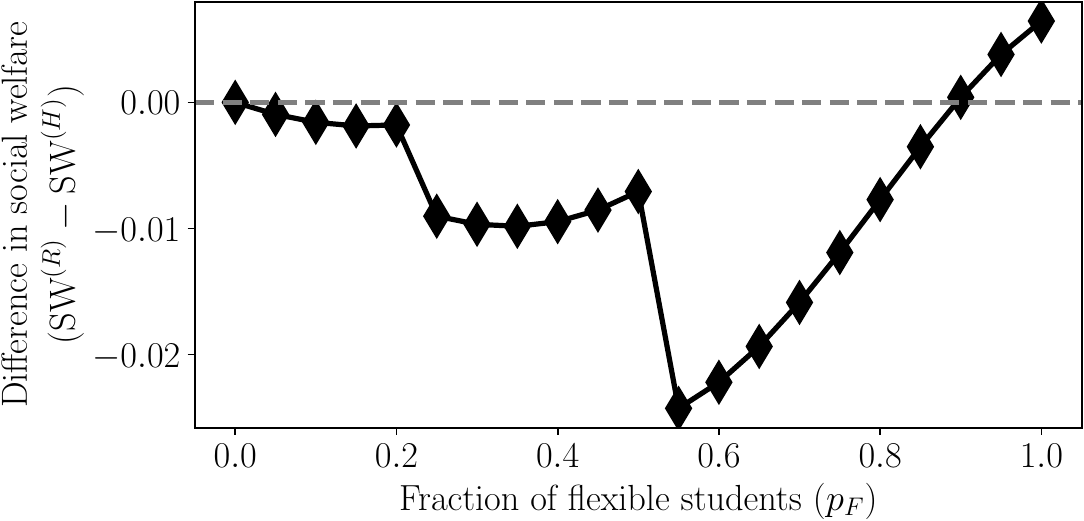}
    
    \caption{Social welfare under the revealed and hidden lotteries (left) and their difference (right) when varying the fraction $\priorflexible$ of flexible students.\label{fig:sim_sw_vary_frac}
    }
\end{figure}

\paragraph{Effect of the fraction $\priorflexible$ of flexible students.} We set $(\pref_{F1}, \pref_{F2}) = (0.55, 0.45)$ and set $\length = 6$. The results are again qualitatively similar when $\length =12$ (not shown). \Cref{fig:sim_sw_vary_frac} shows the social welfare under the revealed and hidden lotteries as well as their difference. At one extreme when $\priorflexible=0$, all students are selective and apply only to tier-1 schools regardless of the lottery information, and hence the difference in social welfare is exactly $0$. At the other extreme when $\priorflexible=1$, all students are flexible. Revealing the lottery information effectively signals students to apply to more (resp. fewer) tier-1 schools if they have good (resp. bad) lottery numbers, improving the social welfare. The difference is negative when a moderate fraction of students are flexible. This is the case when there are enough flexible students to increase congestion and block selective students, and there are also enough selective students who are blocked, which leads to a lower social welfare under the revealed lottery.

\section{Conclusion and discussion}

In studying different information policies in school choice lotteries, we present three main theoretical results when the length of preference lists is $\length=1$. First, we show that all equilibria, under certain assumptions, lead to the same matching outcome in terms of access (match rate) and efficiency (social welfare), suggesting that temporary interventions that aim to move students from one equilibrium to another do not work. Second, we compare the match rate and social welfare under revealed and hidden lotteries, and show that contrary to prior literature, revealing the lottery may worsen both metrics. Third, we consider partial information, and demonstrate the non-monotonic effect of information. A mechanism that reduces the match rate and social welfare under the revealed lottery is that students with weak preferences but good lottery numbers cause congestion and worsen the matching outcome of students with strong preferences. We show via simulation that this mechanism generalizes to list length $\length \ge 1$. Our findings highlight preference heterogeneity and preference intensity as two key factors in shaping the tradeoffs induced by lottery revealing policies, informing policy decisions that balance legal compliance, transparency, and utility.

The intervention to reveal partial information is of practical relevance. In contrast to Bayesian persuasion where signals are strategically designed, this partial information is a simple, intuitive coarsening of the true lottery numbers, which is critical in protecting public trust in high-stakes applications such as school admissions. Partial information is also consistent with the current practice in NYC. In more detail, the lottery number of a student is generated independently as a $32$-character hexadecimal number~\citep{url_nyc_random_number}. Students are often advised to look at the first two digits of their lottery number, which provide a rough percentile estimate by the law of large numbers.

\paragraph{Open problems.}

Immediate extensions of our problem formulation include theoretical analysis of preference lists of length $L > 1$, school-specific priority groups, metric structures for student valuations, and alternative behavioral models for students' responses to lottery numbers. In terms of our theoretical results, we conjecture that the log factors in \Cref{thm:MR-gaps} and \Cref{thm:SW-gaps} may be an artifact of the proofs, and it is an open question to tighten these bounds. 
Beyond the match rate and social welfare, it is also important to account for fairness criteria, for example, if certain sub-populations are not very skilled in interpreting their lottery number or strategizing actions.
More generally, given the mixed effects of lottery revealing policies, it would be valuable to establish practical guidelines based on which policy makers can determine whether it is beneficial to reveal the lottery for their specific use case. For example, simulation based on historical data may be useful. It would also be valuable to experiment with personalized recommendation tools that, for example, suggest schools or provide feedback on students' tentative preference lists based on their lottery number. 

Finally, the school choice problem is a simultaneous game, where students do not observe each others' submitted preferences. For other resource allocation problems that involve players taking actions sequentially, it may be permissible to disclose (partial) information about the preferences that have been submitted by previous players. One potential application is the allocation of permits to national parks, and it is an open direction to construct richer information policies that incorporate both the lottery number and previous players' actions.

\subsection*{Acknowledgments}

We thank Nikhil Garg for pointing us to relevant references, and Thanh Nguyen and David Kempe for helpful discussions.
{\small
    \bibliographystyle{plain}
    \bibliography{references}
}

\appendix

\section{Notation and preliminaries}\label{app:proofs_prelims}
In this appendix, we present additional notation and preliminary results that are used in subsequent proofs. Recall that under homogeneous preference rankings, we assume without loss of generality that schools are indexed so that $v_{t1}\geq v_{t2} \geq \ldots \geq v_{tn}$ for every type $t\in [\numtypes]$, and under the lexicographic tie breaking, if students are indifferent between two or more schools, they deterministically select the one with the lowest index.

Recall that we use the notation $\strategy^\pareni$ for the equilibrium strategy under information policy $\texti$.
We define the match number by $\mn^\pareni = n\cdot \mr^\pareni$. We omit the superscript $\texti$ for all quantities when the information policy is clear from the context.

To simplify the notation, in the proofs we construct economies where valuations are allowed to be zero. Whenever an economy $\econ$ has non-negative valuations $\{\pref_{\idxtype\idxschool}\}$, we construct an alternative economy $\econ'$ with strictly positive valuations
\begin{align*}
    \pref'_{\idxtype\idxschool} = \frac{1}{1+n\epsilon}(\pref_{\idxtype\idxschool}+\epsilon),
\end{align*}
where we take $\epsilon = o\big(\frac{1}{n}\big)$. Note that the equilibria under economy $\econ'$ are identical to the equilibria under economy $\econ$ with a valuation of $-\epsilon$ assigned to being unmatched. In subsequent proofs, we consider strict equilibria of economy $\econ$ that hold regardless of the tie-breaking rule, so our analysis also holds in economy $\econ'$ with minor modifications. For clarity, we use the notation $\zeropos$ to denote a valuation of zero in economy $\econ$.

In the proofs, we use the inequality
\begin{align}\label{eq:inequality}
    \frac{1}{4}\le \left(1-\frac{1}{n}\right)^n < \frac{1}{e}\quad \text{ for every }n\ge 2.
\end{align}
We also use the following property.
\begin{lemma}\label{lem:prod_sum}
Consider arbitrary $\const\in (0, 1]$. Consider any real-valued $z_1, \ldots, z_n \in [0, 1]$ satisfying
\begin{align*}
    \prod_{i=1}^n (1-z_i) \ge c.
\end{align*}
Then we have
\begin{align*}
    \sum_{i=1}^n z_i\le \log \left(\frac{1}{c}\right).
\end{align*}
\end{lemma}

\paragraph{Proof of \Cref{lem:prod_sum}.}

With $ 1-u\le e^{-u}$ for all $u\in \reals$ and the assumption that $z_i\in [0, 1]$ for every $i$, we have
\begin{align*}
    e^{-\sum_i z_i} &\ge \prod_i (1-z_i) \ge c\\
    \sum_i z_i &\le \log \left(\frac{1}{c}\right),
\end{align*}
completing the proof of the lemma.

\subsection{Partial information lottery}\label{app:prelim_partial}

Denote by $\flowp_{\partition_k, \idxschool} = \sum_{\idxtype=1}^\numtypes \prior_\idxtype\pi^\parenp_{t, \partition_k, \idxschool}$ the probability that a random student in block $\partition_\idxpartition$ applies to school $\idxschool$ under partial information equilibrium $\strategy^\parenp$, also termed the ``flow'' from block $\partition_\idxpartition$ to school $\idxschool$. For notational simplicity, we also write this probability as  $\flowp_{k\idxschool}$ when it is clear from the context that $k$ is the index for a block, and likewise for other quantities. For every block $\partition_\idxpartition$, we define the following probabilities concerning whether school $\idxschool\in [\numstudents]$ is available:
\begin{itemize}
    \item Denote by $\probavailbefore^\parenp_{k\idxschool}$ the probability that school $\idxschool$ is available after all students in previous blocks are assigned, that is, no student in block $\partition_1$ through $\partition_{\idxpartition-1}$ applies to school $\idxschool$. We have
    \begin{align}
        \probavailbefore^\parenp_{k\idxschool} &=\prod_{\ell=1}^{\idxpartition-1}\left(1-\flowp_{\ell\idxschool}\right)^{\abs*{\partition_\ell}}.\label{eq:partial_probavail_before}
    \end{align}
    For block $\partition_1$, we have $\probavailbefore^\parenp_{11}=\cdots = \probavailbefore^\parenp_{1\numstudents} = 1$ by definition.

    \item Denote by $\probavail^\parenp_{\idxpartition\idxschool, \textcond}$ the probability that no student in block $\partition_\idxpartition$ with a better lottery number, from the perspective of an individual student in block $\partition_\idxpartition$, applies to school $\idxschool$. We have
\begin{align}
    \probavail^\parenp_{k\idxschool, \textcond} &= \frac{1}{\abs*{\partition_\idxpartition}}\sum_{\lottery=1}^{\abs*{\partition_\idxpartition}} \left(1-\flowp_{\idxpartition\idxschool}\right)^{r-1} \nonumber\\
    &= \begin{cases}
            1 & \text{if } \flowp_{\idxpartition, \idxschool} = 0\\
            \frac{1-\big(1-\flowp_{\idxpartition\idxschool}\big)^{\abs*{\partition_\idxpartition}}}{\abs*{\partition_\idxpartition}\cdot \flowp_{\idxpartition\idxschool}} & \text{ if } 0 < \flowp_{\idxpartition, \idxschool} \le 1,
    \end{cases}\label{eq:prob_avail_partial}
\end{align}
where we have $\probavail^\parenp_{k\idxschool, \textcond} = 1$ for any singleton block  with $\abs*{\partition_\idxpartition} =1$, and strictly decreasing in $\flowp_{\idxpartition_\idxschool}$ when $\abs*{\partition_\idxpartition} >1$.

\item Denote by $\probavail^\parenp_{\idxpartition \idxschool}$ the admission probability that school $\idxschool$ is available to an individual student in block $\partition_\idxpartition$.
Since the randomness in mixed strategies is independent across students, we have
\begin{align}
    \probavail^\parenp_{\idxpartition \idxschool} = \probavail^\parenp_{\idxpartition\idxschool,\textcond} \cdot \probavailbefore^\parenp_{\idxpartition\idxschool}.\label{eq:partial_cond_prob_chain}
\end{align}
\end{itemize}
The expected utility of a type-$\idxtype$ student in block $\partition_\idxpartition$ is
\begin{align*}
    \utility^\parenp_{\idxtype, \partition_\idxpartition} \defn \max_{j\in [\numstudents]}~ \pref_{\idxtype\idxschool} \probavail^\parenp_{\idxpartition\idxschool}.
\end{align*}
The match number is
\begin{align}
    \mn^\parenp = \sum_{\idxschool=1}^\numschools \left(1 - \prod_{\idxpartition=1}^\numpartitions (1-\flowp_{\idxpartition\idxschool})^{\abs*{\partition_\idxpartition}}\right).\label{eq:mn_partial}
\end{align}
The social welfare is
\begin{align}
    \swp = \sum_{\idxpartition=1}^\numpartitions \abs*{\partition_\idxpartition} \cdot  \sum_{\idxtype=1}^\numtypes \prior_\idxtype\utility^\parenp_{\idxtype, \partition_\idxpartition}.\label{eq:sw_partial}
\end{align}
We now simplify these derivations in the special cases of revealed and hidden lotteries.

\subsection{Revealed lottery}\label{app:prelim_revealed}

With partition $\partition_\idxpartition = \{\idxpartition\}$ for $\idxpartition\in [\numstudents]$ under the revealed lottery, we denote by $q^\parenr_{r\idxschool}=\sum_{\idxtype=1}^{\numtypes}\prior_\idxtype \pi^\parenr_{\idxtype\lottery \idxschool}$ the probability, namely the flow, that the student with lottery number $r$ applies to school $j$.
We use the notation $\big(\probavailbefore^\parenr_{\lottery\idxschool}, \probavailr_{\lottery\idxschool, \textcond}, \probavailr_{\lottery\idxschool}\big)$ for $\big(\probavailbefore^\parenp_{\idxpartition\idxschool}, \probavail^\parenp_{\idxpartition\idxschool, \textcond}, \probavailp_{\idxpartition\idxschool}\big)$ from~\eqref{eq:partial_probavail_before}-\eqref{eq:partial_cond_prob_chain} under the revealed equilibrium. Given that $\probavail^\parenr_{\lottery\idxschool, \textcond} = 1$ trivially for every $\lottery$ and $\idxschool$, we have 
\begin{align}
    \probavailr_{\lottery\idxschool} = \probavailbefore^\parenr_{\lottery\idxschool} = \prod_{\idxstudent=1}^{\lottery-1} \big(1-\flowr_{\idxstudent\idxschool}\big).\label{eq:admission_prob_revealed}
\end{align}
with admission probabilities for the student with \lnum $\lottery=1$:
\begin{align*}
    \probavail_{1 1}^\parenr= \cdots =\probavail_{1\numschools}^\parenr = 1.
\end{align*}
From~\eqref{eq:admission_prob_revealed}, we also have the recursive relation
\begin{align}
    \probavail^\parenr_{r+1,j} = \probavail^\parenr_{rj} \cdot (1-q^\parenr_{rj}).\label{eq:prob_reveal_recursive}
\end{align}
We slightly abuse the notation and denote by $\utility^\parenr_{\idxtype\lottery}(\idxschool)$ the expected utility for a type-$\idxtype$ student with lottery number $\lottery$ to apply to school $\idxschool$, while all other students follow their equilibrium strategies $\strategy^{(R)}$. We have
\begin{align*}
    \utility^\parenr_{\idxtype\lottery}(\idxschool)= v_{tj}\probavailr_{rj}.
\end{align*}
We denote by $\utility_{tr}^\parenr$ the expected utility of a type-$\idxtype$ student with lottery number $\lottery$ under the revealed equilibrium:
\begin{align}\label{eq:util_revealed_max}
    \utility_{t\lottery}^\parenr = \max_{\idxschool \in [\numschools]}~\utility^\parenr_{\idxtype\lottery}(\idxschool)
    = \max_{\idxschool\in [\numschools]} ~ v_{\idxtype\idxschool}\probavail^\parenr_{r\idxschool}.
\end{align}
The match number is
\begin{align}
    \mn^\parenr = \sum_{\idxschool=1}^\numschools \left(1 - \prod_{r=1}^\numstudents \left(1-\flowr_{rj}\right)\right) 
    = \sum_{\idxschool=1}^\numschools\left(1-\probavailr_{\numstudents+1, \idxschool}\right).\label{eq:mn_reveal}
\end{align}
The social welfare is
\begin{align}
    \swr = \sum_{\lottery=1}^\numstudents\sum_{\idxtype=1}^\numtypes \prior_\idxtype \utility^\parenr_{\idxtype r}.\label{eq:sw_reveal}
\end{align}

\subsection{Hidden lottery}\label{app:prelim_hidden}

With a single block $\partition_1 = [\numstudents]$, we denote by $q_j^\parenh \defn \sum_{t=1}^T p_t\pi^{\parenh}_{tj}$ the probability, namely the flow, that a student applies to school $\idxschool$ under the hidden equilibrium. We use the notation $(\probavailbefore^\parenh_{\idxschool}, \probavailh_{\idxschool, \textcond}, \probavailh_{\idxschool})$ for $(\probavailbefore^\parenp_{\idxpartition\idxschool}, \probavail^\parenp_{\idxpartition\idxschool, \textcond}, \probavailp_{\idxpartition\idxschool})$ from~\eqref{eq:partial_probavail_before}-\eqref{eq:partial_cond_prob_chain} under the hidden equilibrium. 
Given that $\probavailbefore_{\idxschool}=1$ trivially for every $\idxschool$, we have 
\begin{align}
        \probavailh_\idxschool= \begin{cases}
        1 & \text{ if }\flowh_j = 0\\
        \frac{1-\big(1-\flowh_j\big)^n}{n\flowh_j}& \text{ if }0 < \flowh_j\le 1,
        \end{cases}\label{eq:prob_avail_expression}
\end{align}
where $\probavailh_\idxschool$ decreases in $\flowh_\idxschool$ on $[0, 1]$.
Moreover, we provide a bound on $\probavailh_\idxschool$ as follows. From the perspective of an individual student, let $N_j\sim \binomial(n-1, \flowh_j)$ be the number of \emph{other} students applying to school $j$. This student is admitted to school $\idxschool$ if and only if they have the best lottery number among all students applying to school $j$, which happens with probability $\frac{1}{N_j+1}$. 
By Jensen's inequality, we have
\begin{align}
    \probavailh_j = \Expect\left[\frac{1}{N_j+1}\right] \ge \frac{1}{\Expect[N_j]+1} = \frac{1}{(n-1)\flowh_j + 1} \ge \frac{1}{n},\label{eq:prob_avail_single_lb}
\end{align}
and thus
\begin{align}
    \sum_{j\in [\numschools]} \frac{1}{\probavailh_\idxschool} \le \sum_{j\in [\numschools]} \big((n-1)\flowh_j + 1\big) = n-1+n = 2n-1.\label{eq:prelim_sum_inv_alpha_ub}
\end{align}
We denote by $\utility_t^\parenh(\idxschool)$ the expected utility for a type-$\idxtype$ student to apply to school $\idxschool$, assuming that all other students (including those of type $\idxtype$) follow their strategies at the hidden equilibrium  $\strategy^{(H)}$:
\begin{align}
    \utility^\parenh_t(\idxschool) &= v_{\idxtype\idxschool}\probavailh_\idxschool.\label{eq:utility_hidden_deviate_j}
\end{align}
We denote by $u_t^\parenh$ the expected utility of a type-$\idxtype$ student at the hidden equilibrium: 
\begin{align*}
    \utility^\parenh_t &= \max_{j\in [\numschools]} u_t^\parenh(\idxschool) = \max_{\idxschool\in [\numschools]} v_{\idxtype\idxschool}\probavailh_\idxschool.
\end{align*}
The match number is
\begin{align}
    \mnh = \sum_{\idxschool=1}^\numschools \left(1 - \big(1-\flowh_j\big)^\numschools\right) = n\sum_{\idxschool=1}^\numschools \flowh_\idxschool\probavailh_\idxschool.\label{eq:mn_hidden}
\end{align}
The social welfare is
\begin{align}
    \sw^\parenh = \numstudents\sum_{\idxtype\in [\numtypes]}\prior_\idxtype\utility_\idxtype^\parenh. \label{eq:sw_hidden}
\end{align}

\subsection{Connection between revealed and hidden lotteries}

The following result establishes a relation between the flows under the revealed and hidden lotteries.
\begin{lemma}\label{lem:flow_set_compare}
    Fix any lottery number $\lottery\in [\numstudents]$. Let $\const >0$ be any positive constant. Let
    \begin{align}
        J_r(c) \defn \left\{\idxschool: \frac{\probavailr_{\lottery\idxschool}}{\probavailh_\idxschool} > \const\right\}\label{eq:def_set_J_r_c}
    \end{align}
    be the set of schools whose admission probability under the revealed lottery is more than the hidden lottery by at least a factor of $\const$.
    Then we have
    \begin{align*}
        \sum_{\idxschool\in J_r(c)} \flowr_{\lottery\idxschool} \ge \sum_{\idxschool\in J_r(c)} \flowh_\idxschool.
    \end{align*}
\end{lemma}
This lemma is used for establishing upper bounds when comparing the match rate (bound~\eqref{eq:ratio_mr_hr_hetero} in \Cref{thm:MR-gaps}) and social welfare (bound~\eqref{eq:ratio_sw_rh_hetero} in \Cref{thm:SW-gaps}) under heterogeneous preference rankings. 

\paragraph{Proof of Lemma~\ref{lem:flow_set_compare}.}

Consider any type $\idxtype\in [\numtypes]$. Consider any school  $h\in J_r(c)$ such that type-$\idxtype$ students have a positive probability to apply to school $h$ under the hidden equilibrium, i.e., $\pi^\parenh_{\idxtype h} > 0$. Consider any school $\idxschool$ that a type-$\idxtype$ student \withlnum $\lottery$ has a positive probability to apply to under the revealed equilibrium, i.e., $\pi^\parenr_{\idxtype\lottery\idxschool} > 0$. By the definition of the hidden and revealed equilibria, we have
\begin{subequations}\label{eq:def_two_equilibria}
\begin{align}
    v_{th} \probavailh_h &\ge v_{tj}\probavailh_j\\
    v_{tj}\probavailr_{\lottery\idxschool} &\ge v_{th}\probavailr_{\lottery h}.
\end{align}
\end{subequations}
Rearranging~\eqref{eq:def_two_equilibria}, we have
\begin{align*}
    \frac{\probavailh_h}{\probavailh_j}\ge \frac{v_{tj}}{v_{th}}\ge \frac{\probavailr_{\lottery h}}{\probavailr_{\lottery\idxschool}}\\
    \frac{\probavailr_{\lottery\idxschool}}{\probavailh_j} \ge \frac{\probavailr_{\lottery h}}{\probavailh_h} \stackrel{\stepone}{>} c,
\end{align*}
where step~\stepone is true due to the definition~\eqref{eq:def_set_J_r_c} of set $J_r(c)$ and the assumption that $h\in J_r$. Hence, we have $j\in J_r(c)$. That is, if type-$\idxtype$ students apply to some schools in $J_r(c)$ with non-zero probability under the hidden equilibrium, then the student with lottery number $\lottery$ only applies to schools in $J_r(c)$ under the revealed equilibrium. Equivalently, for every type $\idxtype$, 
\begin{align*}
    \sum_{j\in J_r(c)}\pi^\parenh_{tj} > 0 \quad \text{ implies }\quad \sum_{j\in J_r(c)} \pi_{t\lottery\idxschool}^\parenr = 1,
\end{align*}  
and hence
\begin{align}
    \sum_{j\in J_r(c)} \pi_{tr\idxschool}^\parenr \ge \sum_{j\in J_r(c)}\pi^\parenh_{tj}.\label{eq:flow_sum_over_Jr}
\end{align}
We have
\begin{align*}
    \sum_{j\in J_r(c)} \flowr_{\lottery\idxschool} &=\sum_{j\in J_r(c)} \sum_{t} p_t \pi_{trj}^\parenr \\
    &=  \sum_{t} p_t \sum_{j\in J_r(c)}\pi_{trj}^\parenr \\
    &\stackrel{\stepone}{\ge} \sum_{t} p_t \sum_{j\in J_r(c)}\pi_{tj}^\parenh \\
    &= \sum_{j\in J_r(c)} q_{j}^\parenh,
\end{align*}
where step~\stepone is true due to~\eqref{eq:flow_sum_over_Jr}.

\subsection{Construction of economies }\label{app:prelim_common_rare}

In this section, we construct two economies that achieve better matching outcome under the hidden lottery than the revealed lottery.

\paragraph{Uniform-prefix economy.}

Consider $\numtypes=\numstudents$ types. The uniform-prefix economy captures a spectrum of preference intensities under homonogeous preference rankings. The valuations for type-$t$ students are:
\begin{align*}
    v_{tj}=\begin{cases}
        \frac{1}{\idxtype} & \text{if }1\le j\le \idxtype\\
        \zeropos & \text{otherwise}.
    \end{cases}
\end{align*} 
Equivalently, we write $\{v_{tj}\}$ as a matrix:
\begin{align*}
    \begin{bmatrix}
        1 & \zeropos & \zeropos &  \cdots & \zeropos\\
        \nicefrac{1}{2} & \nicefrac{1}{2} & \zeropos &\cdots & \zeropos\\
        \vdots & \vdots & \vdots & \ddots & \vdots\\
        \nicefrac{1}{n} & \nicefrac{1}{n} & \nicefrac{1}{n} &\cdots & \nicefrac{1}{n}
    \end{bmatrix}.
\end{align*}
The prior is uniform, i.e., $p_1= p_2= \ldots = p_n =\frac{1}{n}$. The following lemma formalizes an equilibrium of this economy under the hidden lottery.

\begin{lemma}\label{lem:uniform_prefix_equilibrium}
    Consider a uniform-prefix economy. Under the hidden lottery, it is an equilibrium that type-$t$ students apply to school $t$ for every $\idxtype\in [\numstudents]$. 
\end{lemma}

\paragraph{Proof of Lemma~\ref{lem:uniform_prefix_equilibrium}.}

Consider the strategy profile that type-$\idxtype$ students apply to school $\idxtype$. Under this strategy profile, the flows are $q_t = p_\idxtype$.
The  expected utility for type-$\idxtype$ students to apply to school $\idxtype$ is:
\begin{align*}
    u_t^\parenh (\idxtype) = v_{tt}\alpha_\idxtype^\parenh &\stackrel{\stepone}{=}  v_{tt}\cdot\frac{1-(1-p_t)^n}{np_t}\\
    &\stackrel{\steptwo}{\ge} v_{tj}\cdot \frac{1-(1-p_j)^n}{np_j} = u_t^\parenh(j),
\end{align*}
where step~\stepone is true by plugging in~\eqref{eq:prob_avail_expression}, and step~\steptwo is true because $v_{tt} \ge v_{tj}$ in the uniform-prefix economy.
Hence, applying to school $\idxtype$ is utility maximizing, proving the equilibrium. 

\paragraph{Common-rare economy.}
We construct an economy that consists of one common type of students and $\numminority$ rare types with $\numminority < \numschools-1$. For notational convenience, we re-index the student types such that the common type is type $0$, and rare types are types $1$ through $\numminority$. Likewise, we re-index the schools to be $0, 1, \ldots, \numschools-1$ such that common-type students prefer school $0$ (termed the ``common school''), and also mildly value the other $\numminority$ schools (termed ``rare schools''):
\begin{subequations}\label{eq:construction_common_rare}
    \begin{align}\label{eq:construction_common_rare_common}
    \pref_{0\idxschool} = \begin{cases}
        1-\frac{\numminority}{\numschools} & \text{ if }\idxschool=0\\
        \frac{1}{\numschools} & \text{ if } 1\le \idxschool\le \numminority\\
        \zeropos & \text{ if }\numminority < \idxschool \le \numschools-1 .
    \end{cases}
\end{align}
Each rare type $\idxtype \in [\numminority]$ prefers a distinct rare school $\idxtype$, and has a valuation of zero if they are matched elsewhere:
\begin{align}
    \pref_{\idxtype\idxschool} = \begin{cases}
        1 & \text{ if } \idxschool = t\\
        \zeropos & \text{ if }\idxschool \ne t.
    \end{cases}    
\end{align}
The prior is
\begin{align}\label{eq:construction_common_rare_prior}
    \prior_0 = 1 - \frac{\numminority}{\numschools},\quad \text{ and }\quad \prior_\idxtype = \frac{1}{\numschools} \quad\text{ for every } \idxtype\in [\numminority].
\end{align}
\end{subequations}
The following lemma formalizes an equilibrium and the social welfare under the hidden lottery.
\begin{lemma}\label{lem:common_rare_equil_hidden}
    Consider a common-rare economy~\eqref{eq:construction_common_rare} with $\numminority < \numschools-1$ rare types. Under the hidden lottery, it is an equilibrium that common-type students apply to the common school 0, and students of rare type $\idxtype\in [\numminority]$ apply to their preferred rare school $\idxtype$. The social welfare is:
    \begin{align}\label{eq:sw_common_rare}
        \swh = v_{00} \cdot \left(1 - (1-p_0)^n\right) + m\cdot \left(1 - \Big(1-\frac{1}{n}\Big)^n\right) > \Big(1-\frac{1}{e}\Big) m.
    \end{align}
\end{lemma}
The proof of this lemma is provided at the end of this section. 
On the other hand, under the revealed lottery, common-type students with worse lottery numbers know that they are at a disadvantage to compete with other common-type students with better lottery numbers. Hence, they apply to rare schools to avoid competition, which leads to social welfare inefficiency because these seats could have been taken by rare-type students who have higher valuations for rare schools. We invoke this lemma to analyze the social welfare ratio $\ratioswhr$ with $\numminority = \Theta(\sqrt{n})$ (in Appendix~\ref{app:proof_sw_ratio_hr_different_rankings}), and the social welfare difference $\diffswhr$ with $\numminority = \Theta(\numschools)$ (in Appendix~\ref{app:proof_sw_diff_hr_different_rankings}). 

\paragraph{Proof of Lemma~\ref{lem:common_rare_equil_hidden}.}
    
We verify that it is a hidden equilibrium that type-$\idxtype$ students apply to school $\idxtype$ for each $0 \le \idxtype\le \numminority$. First, note that it is optimal for students of every rare type $\idxtype$  to apply to school $\idxtype$, because they have a valuation of $0$ for every other school. For common-type students, we first compute the flows
\begin{align*}
    q_\idxschool^\parenh = \begin{cases}
        p_\idxschool & \text{for }0 \le \idxschool\le \numminority\\
        0 & \text{for }\numminority < \idxschool < \numstudents.
    \end{cases}
\end{align*}
The expected utility for a common-type student to apply to the common school $0$ is:
\begin{align*}
    \utility^\parenh_{0}(0) = \pref_{00}\cdot \probavail_0^\parenh &\stackrel{\stepone}{=} \left(1-\frac{m}{n}\right)\cdot\frac{1 - \big(1-q_0^\parenh\big)^n}{n q_0^\parenh}\\
    &=\left(1-\frac{m}{n}\right)\cdot\frac{1 - (1-p_0)^n}{n p_0}\\
    & \stackrel{\steptwo}{=} \frac{1 - (\frac{m}{n})^n}{n},
\end{align*}
where step~\stepone is true by plugging in the valuation $v_{00} = 1-\frac{\numminority}{\numstudents}$ from~\eqref{eq:construction_common_rare_common} and the expression of $\probavail_0$ from~\eqref{eq:prob_avail_expression}; step~\steptwo is true by plugging in $p_0=1-\frac{\numminority}{\numstudents}$ from~\eqref{eq:construction_common_rare_prior}.

The expected utility for a common-type student to apply to any school $\idxtype > \numminority$ is $\utility^\parenh_0(\idxtype) = 0$. The expected utility for a common-type student to apply to school $\idxtype \in [\numminority]$ is:
\begin{align*}
    \utility^\parenh_{0}(\idxtype)= \pref_{0\idxtype}\cdot \probavail^\parenh_\idxtype &= \frac{1}{\numschools}\cdot \frac{1 - (1 - p_t)^n}{n p_t} \\
    &= \frac{1- \left(1-\frac{1}{n}\right)^n}{n}\\
    &\stackrel{\stepone}{<} \frac{1 - (\frac{m}{n})^n}{n} = \utility^\parenh_{0}(0),
\end{align*}
where step~\stepone is true because $\numminority < \numschools-1$ by assumption. 
We thus have $\utility^\parenh_{0}(0) > \utility^\parenh_{0}(\idxtype)$ for every $t \ge  1$, verifying the equilibrium. To compute the social welfare, note that the school $0$ is matched with probability  $1 - (1-p_0)^n$, and any school $t\in [\numminority]$ is matched with probability $1-(1-\frac{1}{n})^n$. We have
\begin{align*}
    \swh = v_{00} \cdot (1 - (1-p_0)^n) + m\cdot \left(1 - \Big(1-\frac{1}{n}\Big)^n\right) \stackrel{\stepone}{>} \Big(1-\frac{1}{e}\Big) m,
\end{align*}
where step~\stepone is true by inequality~\eqref{eq:inequality}.

\section{Derivations for Example~\ref{ex:equilibria}}\label{app:derivations}
We separately analyze the three information policies.

\paragraph{Revealed Lottery.} 
Since students have homogeneous preference rankings, invoking \Cref{lem:reveal_same_ranking_always_one} yields the desired equilibrium, match rate, and social welfare.

\paragraph{Hidden Lottery.} We verify that it is an equilibrium for students of type $\idxtype\in \{1, 2\}$ to apply to school $\idxtype$. 
Under this strategy profile, the flows are:
\begin{align*}
    \flowh_1 = 0.8,\quad, \flowh_2= 0.2, \quad \flowh_3=0.
\end{align*} 
The admission probabilities~\eqref{eq:prob_avail_expression} are:
\begin{align*}
    \probavailh_1 \approx 0.413,\quad \probavailh_2 \approx 0.813, \quad\probavailh_3 = 1.
\end{align*}
By~\eqref{eq:utility_hidden_deviate_j}, the expected utility of an individual type-$\idxtype$ student to apply to school $\idxschool$, when all other students follow the given strategy profile, is $\utility^\parenh_t(\idxschool) = v_{\idxtype\idxschool}\probavailh_\idxschool$. It can be verified that
\begin{align*}
    \argmax_{j} v_{tj} \probavailh_\idxschool = \idxtype \quad \text{for } \idxtype\in \{1, 2\},
\end{align*}
verifying the equilibrium. The match rate~\eqref{eq:mn_hidden} is
\begin{align*}
    \mr^\parenh &= \frac{1}{3}  \sum_{\idxschool=1}^3 \left(1 - \big(1-\flowh_j\big)^3 \right) \approx 0.493,
\end{align*}
The social welfare~\eqref{eq:sw_hidden} is
\begin{align*}
    \swh = 3\cdot \left(p_1 \cdot \pref_{11}\probavailh_1 + p_2\cdot \pref_{22}\probavailh_2\right) \approx 1.128.
\end{align*}

\paragraph{Partial Information Lottery.}

For block $\partition_1$, it can be verified that type-$1$ students apply to school $1$, because their utility of applying to school $1$ is at least $v_{11}\probavailp_{\partition_1, 1}\ge 0.97\cdot 0.5=0.485$, exceeding their valuation for schools $2$ and $3$.
For type-$2$ students, their utility of applying to school $1$ is at least $v_{21}\probavailp_{\partition_1, 1} \ge 0.65\cdot 0.5 = 0.325$, which is strictly higher than their valuation of school $3$, so they never apply to school $3$. It can be verified that deterministically applying to school 1 or school 2 alone is not optimal for type-$2$ students in block $\partition_1$.
We thus write their equilibrium strategy by  $\strategy^{\parenp}_{2,\partition_1} = (x, 1-x, 0)$ with $0<x<1$, which leads to flows
\begin{align}
    q^\parenp_{\partition_1 1}& =0.8+0.2x, \quad q^\parenp_{\partition_1 2} =0.2(1-x), \quad 
    q^\parenp_{\partition_1 3}=0.\label{eq:ex_partial_flow_x}
\end{align}
By~\eqref{eq:prob_avail_partial} and~\eqref{eq:partial_cond_prob_chain}, the admission probabilities are
\begin{align}
    \probavailp_{\partition_1,j} = \frac{1-(1-\flowp_{\idxpartition \idxschool})^{\abs*{\partition_1}}}{\abs*{\partition_1}\cdot \flowp_{\partition_1 \idxschool}}= \frac{2-\flowp_{\partition_1 \idxschool}}{2} \quad \text{for } \idxschool\in \{1, 2\}.\label{eq:ex_partial_admission_prob_x}
\end{align}
Since applying to schools 1 and 2 are both utility maximizing, we have
\begin{align}
    \pref_{21}\cdot \probavailp_{\partition_1 1} &= \pref_{22}\cdot \probavailp_{\partition_1 2}.\label{eq:ex_partial_equal_utility_x}
\end{align}
Plugging~\eqref{eq:ex_partial_flow_x} and~\eqref{eq:ex_partial_admission_prob_x} into~\eqref{eq:ex_partial_equal_utility_x} yields $x=\frac{28}{33}$, so the optimal strategy is $\strategy^\parenp_{2, \mathcal{P}_1} = (\frac{28}{33}, \frac{5}{33},0 )$, which leads to flows $\flowp_{\partition_1,1} = \frac{32}{33}$ and $\flowp_{\partition_1, 2} = \frac{1}{33}$.

For block $\mathcal{P}_2$, the admission probability is $\probavailp_{\partition_2,\idxschool}=\probavailbefore^\parenp_{\partition_2,\idxschool} =(1-\flowp_{\partition_1, j})^2$ for each school $\idxschool$, so we have
\begin{align*}
    \probavailp_{\partition_2,1} = \left(\frac{1}{33}\right)^2, \quad \probavailp_{\partition_2,2} =\left(\frac{32}{33}\right)^2, \quad \probavailp_{\partition_2,3} = 1.
\end{align*}
It can be verified that the optimal strategy for all students is to apply to school $2$ regardless of their type.

The match rate~\eqref{eq:mn_partial} is
\begin{align*}
    \mrp = \frac{1}{3}\left[1- \left(1-\flowp_{\partition_1, 1}\right)^2 + 1+0\right] \approx 0.666.
\end{align*}
The social welfare~\eqref{eq:sw_partial} is
\begin{align*}
    \swp = 2 \cdot (\prior_1 \pref_{11}+\prior_2\pref_{21} ) \cdot \probavailp_{\partition_1, 1}+ (p_1\pref_{12} + \prior_2\pref_{22})\cdot \probavailp_{\partition_2, 2} \approx 1.012.
\end{align*}

\section{Example with multiple equilibria}\label{app:ex_multiple_equilibria}
Consider the following economy with $n=3$ students of $T=2$ types:
\begin{center}
    \begin{tabular}{c|c|c|c||c}
     & school 1 & school 2 & school 3 & prior \\\hline
    type 1 & $\nicefrac{37}{60}$ & $\nicefrac{21}{60}$ & $\nicefrac{2}{60}$ & $0.5$\\\hline
    type 2 & $\nicefrac{37}{70}$ & $\nicefrac{21}{70}$ & $\nicefrac{12}{70}$ & $0.5$
\end{tabular}
\end{center}
Under the hidden lottery, it can be verified that both
\begin{align*}
    \strategy_1 = (1, 0, 0), \quad \strategy_2 = \left(\frac{1}{2}, \frac{1}{2}, 0\right)
\end{align*}
and
\begin{align*}
    \strategy_1' = \left(\frac{1}{2}, \frac{1}{2}, 0\right), \quad \strategy_2' = (1, 0, 0)
\end{align*}
are equilibria, generating flows
\begin{align*}
    \vecflow = \left(\frac{3}{4}, \frac{1}{4}, 0\right).
\end{align*}
The set of all equilibria is characterized by a linear combination between equilibria $\strategy$ and $\strategy'$:
\begin{align*}
    \lambda\strategy + (1-\lambda)\strategy' \quad \text{for every }\lambda\in [0, 1].
\end{align*}

\section{Equilibria for list length \texorpdfstring{$\length > 1$}{L>1}}\label{app:equil_comp}
In this appendix, we consider the equilibria under the revealed and hidden lotteries using a mean-field approximation, and compute these equilibria numerically. We consider the model in \Cref{sec:sim} that consists of $\numtiers=2$ school tiers.

\subsection{Mean-field approximation}

We consider the regime of $\numstudents\rightarrow \infty$, where the fraction $\fractier_\idxtier=\frac{\abs*{\setschools_\idxtier}}{\numstudents}$ of schools in each tier $\idxtier\in \{1, 2\}$ remains a constant. We use a mean-field approximation where students and schools are both a continuum of mass $1$; students' lottery numbers are uniform with  $\lottery\in [0, 1]$. 
For each school tier $\idxtier\in \{1, 2\}$, we denote by $\mass_\idxtier(\lottery; \strategynum)$ the fraction of tier-$\idxtier$ schools that are already filled when the algorithm assigns the student with lottery number $\lottery$, when all flexible students follow strategy $\strategynum$.

We assume that the list length satisfies $\length \le \min\{\abs*{\setschools_1}, \abs*{\setschools_2}\}$. This is the regime when in practice students are only able to meaningfully select and rank a small constant number of schools. Under the mean-field approximation, we assume that the event of each school on the student's list being filled is independent, with probability $\mass_\idxtier(\lottery; \strategynum)$ for schools in tier $\idxtier$. Thus, if the student with lottery number $\lottery$ applies to $m$ schools of tier 1 (and hence $(\length-m)$ schools of tier 2), their probability of being assigned a tier-1 or tier-2 school is respectively:
\begin{subequations}\label{eq:mf_prob_assignment}
    \begin{align}
    z_1(\countschools; \lottery, \strategynum) &\defn 1 - x_1(\lottery;\strategynum)^\countschools\\
    z_2(\countschools; \lottery,\strategynum) &\defn x_1(\lottery)^\countschools \cdot [1-\mass_2(\lottery;\strategynum)^{\length-\countschools}],
\end{align}
\end{subequations}
and the student is unmatched with probability $1-\probassign_1(\countschools; \lottery, \strategynum)-\probassign_2(\countschools; \lottery, \strategynum)$. Let us first consider the simple case when $\priorflexible=1$, so there is a single type of students who are flexible. We have mean-field trajectories
\begin{align*}
    \frac{\dd \mass_1(\lottery; \strategynum)}{\dd \lottery} &= \frac{1}{\gamma_1} \sum_{0\le \countschools\le \length} \sigma_\countschools \cdot \probassign_1(\countschools; \lottery, \strategynum)\\
    \frac{\dd x_2(\lottery; \strategynum)}{\dd \lottery}  &= \frac{1}{\gamma_2} \sum_{0\le \countschools \le \length} \sigma_\countschools \cdot \probassign_2(\countschools; \lottery, \strategynum),
\end{align*}
with initial conditions
\begin{align*}
    x_1(0; \strategynum) = \mass_2(0; \strategynum) = 0.
\end{align*}
In the general case when the prior of flexible students is $\priorflexible\in [0, 1]$, the $(1-\priorflexible)$ fraction of selective students apply to $\length$ schools of tier 1. The mean-field trajectories become
\begin{subequations}\label{eq:ode_general}
    \begin{align}
        \frac{\dd x_1(\lottery; \strategynum)}{\dd \lottery} &= \frac{1}{\gamma_1} \left( \priorflexible \cdot \sum_{0\le \countschools\le \length} \sigma_\countschools \cdot \probassign_1(\countschools; \lottery,\strategynum) + (1-\priorflexible) \cdot \probassign_1(\length; \lottery,\strategynum) \right)\\
        \frac{\dd \mass_2(\lottery; \strategynum)}{\dd \lottery}  &= \frac{1}{\gamma_2} \cdot \priorflexible\cdot \sum_{0\le \countschools \le \length} \sigma_\countschools \cdot \probassign_2(\countschools; \lottery, \strategynum),
\end{align}
\end{subequations}
The expected utility of a flexible student applying to $\countschools$ schools of tier 1, while all other flexible students follow strategy $\strategynum$, is:
\begin{align}
    \utility(\countschools; \strategynum) \defn \int_0^1 \left[ \pref_{F1}\cdot \probassign_1 (\countschools; \lottery, \strategynum) + \pref_{F2} \probassign_2(\countschools; \lottery, \strategynum) \right]\dd \lottery.\label{eq:mf_expected_util_flexible}
\end{align}
We define strategy  $\strategynum^*$ to be at the mean-field equilibrium under the hidden lottery, if it maximizes the expected utility~\eqref{eq:mf_expected_util_flexible} of flexible students:
\begin{align*}
    \supp(\strategynum^*) \subseteq \argmax_{0\le \countschools \le \length} \utility(\countschools; \strategynum^*).
\end{align*}
A symmetric mean-field equilibrium always exists, formalized in the following lemma.
\begin{lemma}
Under the hidden lottery, a symmetric mean-field equilibrium always exists.
\end{lemma}
Similar to \Cref{lem:exist}, the existence of a mean-field equilibrium comes from Kakutani's fixed point theorem, given that fractions $\mass_1(\cdot\;; \strategynum), \mass_2(\cdot\;; \strategynum)$, and utility $\utility(\cdot\;; \strategynum)$ are all continuous in $\strategynum$.

\subsection{Equilibrium computation}

We describe the computation of mean-field equilibria under the hidden and revealed lotteries.

\subsubsection{Hidden lottery}
Denote by $\distdelta_\countschools$ the pure strategy to apply to $\countschools$ schools of tier 1. That is, we have $(\distdelta_\countschools)_\countschools=1$ and $(\distdelta_\countschools)_{\countschools'}=0$ for every $\countschools' \ne \countschools$.
\Cref{alg:mf} computes a mean-field equilibrium under the hidden lottery, where we define 
\begin{align*}
    A_\countschools &\defn \utility(\countschools; \distdelta_\countschools) - \utility(\countschools-1; \distdelta_\countschools)\\
    B_\countschools &\defn \utility(\countschools+1; \distdelta_\countschools) - \utility(\countschools; \distdelta_\countschools).
\end{align*}
\begin{algorithm}[tb]
    \DontPrintSemicolon
    \uIf{$A_\length\ge 0$}{Return $\distdelta_\length$\label{line:return_max}\;}
    \uIf{$B_0 \le 0$}{Return $\distdelta_0$\label{line:return_zero}\;}
    Set $\ell = 0$ and $h=\length$ (satisfying $B_\ell > 0$ and $A_h < 0$)\label{line:search}\;
    
    \While{$h-\ell > 1$}{
        $m = \floor{\frac{\ell +h}{2}}$\;
        \uIf{$A_m \ge 0 \ge B_m$}{
            return $\distdelta_m$\label{line:return_m}
        }
        \uElseIf{$A_m < 0$}{
            Set $h=m$\;
        }
        \Else({(}satisfying $B_m > 0${)}){
            Set $\ell = m$\;
        }
    }({(}Now we have $h-\ell = 1${)})
    Find a root $\probmix^*$ of \begin{align*}
        \utility(\ell;\, \probmix\distdelta_{\ell} + (1-\probmix)\distdelta_{\ell+1}) = \utility(\ell+1;\, \probmix\distdelta_{\ell} + (1-\probmix)\distdelta_{\ell+1}).
    \end{align*}\;\label{line:root}
    Return $\probmix^*\distdelta_{\ell} + (1-\probmix^*)\distdelta_{\ell+1}$\;
    
    \caption{Computation of the mean-field equilibrium under the hidden lottery.\label{alg:mf}}
\end{algorithm}

\Cref{alg:mf} conducts a binary search by leveraging the concavity of the expected utility $\utility(\countschools, \strategynum)$ in $\countschools$, as shown in the following lemma.

\begin{lemma}\label{lem:mf_concavity}
    Consider any $\length \ge 2$. Consider any strategy $\strategynum$. The expected utility $\utility(\countschools; \strategynum)$ is concave in $\countschools$. That is, for every $0\le \countschools\le \length-2$, we have
    \begin{align}\label{eq:mf_concavity}
    \utility(\countschools+2; \strategynum) + \utility(\countschools; \strategynum) \le 2\utility(\countschools+1; \strategynum).
\end{align}
Moreover, strict concavity holds if \begin{align}
    \priorflexible(1-\sigma_0) + (1-\priorflexible) > 0,\label{eq:strict_concavity_cond}
\end{align}
which is satisfied by every mean-field equilibrium $\strategynum^*$.
\end{lemma}
The proof of this lemma is provided in Appendix~\ref{app:proof_lem_mf_concavity}. By~\eqref{eq:mf_concavity}, we have $A_\countschools \ge B_\countschools$. By definition, the pure strategy $\distdelta_\countschools$ is an equilibrium if it achieves a higher or equal expected utility than any other pure strategy $\distdelta_{\countschools'}$. The concavity property allows us to reduce the equilibrium conditions to only checking that $\distdelta_\countschools$ achieves a higher or equal expected utility than neighboring pure strategies $\distdelta_{\countschools-1}$ and $\distdelta_{\countschools+1}$. In words, \Cref{alg:mf} keeps track of an interval $[\ell, h]$ within which there exists an equilibrium, and this interval is initialized to be the entire interval $\{0, \ldots, \length\}$. Then the algorithm examines if the midpoint $m=\floor{\frac{\ell + h}{2}}$ is a pure equilibrium. If it is an equilibrium, the algorithm returns this solution. If it is not an equilibrium, the algorithm identifies that the support of an equilibrium lies in either $[\ell, m]$ or $[m, h]$, and iteratively shrinks this interval. At the end, the interval only consists of two consecutive integers $\ell$ and $(\ell+1)$, at which point the algorithm solves a single parameter $\probmix$ denoting the probability assigned to integer $\ell$ (Line~\ref{line:root}). Specifically, for mixed strategy $\strategynum_\probmix \defn \probmix\distdelta_{\ell} + (1-\probmix)\distdelta_{\ell+1}$, define
\begin{align*}
    g(\probmix)\defn \utility(\ell+1; \strategynum_\probmix) - \utility(\ell; \strategynum_\probmix).
\end{align*}
It can be verified that $g(\probmix)$ is non-positive when $\probmix=0$ and non-negative when $\probmix=1$. Hence, there exists a root $\probmix^*\in [0, 1]$ which can be found by bisection. Specifically, to evaluate $g(\probmix)$ for any $\probmix\in [0, 1]$, we solve ODEs~\eqref{eq:ode_general} with $\strategynum_\probmix = \probmix\distdelta_{\ell} + (1-\probmix)\distdelta_{\ell+1}$, with initial conditions $x_1(0) = x_2(0) = 0$.
Then the value of $g(\probmix)$ can be computed by integrating
\begin{align*}
    \int_{0}^1\Big[ \pref_{F1}\cdot \probassign_1 (\ell+1; \lottery, \strategynum_p) + \pref_{F2}\cdot \probassign_2(\ell+1; \lottery, \strategynum_p) \Big]- \Big[ \pref_{F1}\cdot \probassign_1 (\ell; \lottery, \strategynum_p) + \pref_{F2}\cdot \probassign_2(\ell; \lottery, \strategynum_p) \Big]\dd r.
\end{align*}
The following proposition formally establishes the correctness of \Cref{alg:mf}.
\begin{proposition}\label{prop:algo_search_correct}
    \Cref{alg:mf} returns a mean-field equilibrium under the hidden lottery.
\end{proposition}
The proof of this proposition is provided in Appendix~\ref{app:prop_algo_search_correct}.

\subsubsection{Revealed lottery}
Under the revealed lottery, we compute an approximate equilibrium by dividing students into $\numstudents$ intervals. Each interval has a mass of $\frac{1}{\numstudents}$, and we make the approximation that students within the same interval follow the same strategy, and do not affect the admission probability of each other. Denote by $\mass_\idxtier(\idxstudent)$ the fraction of schools that are already filled when the algorithm assigns students in the interval $\idxstudent$. For students in interval $\idxstudent\in [\numstudents]$, the probability of being assigned a tier-1 or tier-2 school is respectively computed by:
\begin{subequations}\label{eq:mf_discrete_prob}
\begin{align}
    \probassign_1(\countschools; \idxstudent) &\defn 1 - \mass_1(\idxstudent)^\countschools\\
    \probassign_2(\countschools; \idxstudent) &\defn \mass_1(\idxstudent)^\countschools\cdot [1-\mass_2(\idxstudent)^{\length-\countschools}].
\end{align}
\end{subequations}
Students in interval $\idxstudent$ follow strategy
\begin{align*}
    m^* \in \argmax_{0\le m \le \length} \; \Big[\pref_{F1}\cdot \probassign_1(\countschools; \idxstudent) + \pref_{F2}\cdot \probassign_2(\countschools; \idxstudent)\Big],
\end{align*}
where the ties are broken arbitrarily. In our simulation, we break ties in favor of the largest value of $m^*$. We observe empirically that the matching outcome does not appear to be sensitive to the choice of this tie-breaking strategy.
When there are only flexible students so that $\priorflexible=1$, we have
\begin{align*}
    \mass_1(\idxstudent+1) &= \mass_1(\idxstudent) + \frac{1}{\fractier_1}\left(\frac{\probassign_1(\countschools^*; i)}{\numstudents}\right)\\
    \mass_2(\idxstudent+1) &= \mass_2(\idxstudent) + \frac{1}{\fractier_2}\left(\frac{\probassign_2(\countschools^*; \idxstudent)}{\numstudents}\right),
\end{align*}
with the initial condition $\mass_1(1) = \mass_2(1) = 0$. In the general case with $\priorflexible \in [0, 1]$, we have
\begin{subequations}\label{eq:mf_discrete_update}
\begin{align}
    \mass_1(\idxstudent+1) &= \mass_1(\idxstudent) + \frac{1}{\fractier_1}\left(\frac{\priorflexible\cdot\probassign_1(\countschools^*; i) + (1-\priorflexible)\cdot\probassign_1(\length; i)}{\numstudents}\right)\\
    \mass_2(\idxstudent+1) &= \mass_2(\idxstudent) + \frac{1}{\fractier_2}\left(\frac{\priorflexible\cdot\probassign_2(\countschools^*; \idxstudent)}{\numstudents}\right).
\end{align}
\end{subequations}
Combining~\eqref{eq:mf_discrete_prob} and~\eqref{eq:mf_discrete_update}, the equilibrium strategies can be computed iteratively for intervals from $1$ to $\numstudents$.

\subsection{Proofs for the mean-field equilibria}

In this section, we provide proofs of the results on mean-field equilibria for the model with two school tiers.

\subsubsection{Proof of Lemma~\ref{lem:mf_concavity}}\label{app:proof_lem_mf_concavity}

Fix any strategy $\strategynum$. We denote $a\defn x_1(\lottery; \strategynum)$ and $b \defn x_2(\lottery; \strategynum)$. For any lottery number $\lottery\in [0, 1]$, the utility of a flexible student with lottery number $\lottery$ applying to $\countschools$ tier-1  schools is:
\begin{align*}
    \utility_\lottery(\countschools) \defn \utility_\lottery(\countschools; \strategynum) &=  \pref_{F1}\cdot \probassign_1 (\countschools; \lottery, \strategynum) + \pref_{F2}\cdot \probassign_2(\countschools; \lottery, \strategynum)\\
    &\stackrel{\stepone}{=} v_{F1} (1-a^\countschools) + v_{F2}\cdot a^\countschools (1-b^{\length-\countschools})\\
    &=  v_{F1} - (v_{F1} - v_{F2}) a^\countschools - v_{F2} a^\countschools b^{\length-\countschools},
\end{align*}
where step~\stepone is true by plugging in~\eqref{eq:mf_prob_assignment}.
We thus have
\begin{align}
    \utility_\lottery({\countschools+2}) + u_\lottery(\countschools) - 2&\utility_\lottery(\countschools+1) \nonumber\\
    &= -(v_{F1} - v_{F2})(a^\countschools + a^{\countschools+2} - 2a^{\countschools+1}) \nonumber\\
    &\quad\quad - v_{F2}(a^\countschools b^{\length -\countschools} + a^{\countschools+2} b^{\length-\countschools-2} - 2a^{\countschools+1}b^{\length-\countschools-1}) \nonumber\\
    &= -(v_{F1} - v_{F2})a^{\countschools} (a^2-2a+1) - v_{F2} a^\countschools b^{\length-\countschools-2}(b^2 + a^2 - 2ab)\label{eq:concavity_integrand_before} \\
    &\le 0.\label{eq:concavity_integrand}
\end{align}
Integrating~\eqref{eq:concavity_integrand} over $\lottery\in [0, 1]$ proves weak concavity. 

We now show strict concavity for any strategy $\strategynum$ satisfying~\eqref{eq:strict_concavity_cond}. To do so, we show that  $a=x_1(\lottery; \strategynum)$ is bounded away from $0$ and $1$ on a set with positive measure. 
At $\lottery=0$, we have
\begin{align*}
    \left.\frac{\dd \mass_1(\lottery)}{\dd \lottery}\right|_{\lottery=0} = \frac{1}{\fractier_1}\left(\priorflexible\sum_{1\le \countschools\le \length} \sigma_\countschools + (1-\priorflexible)\right) >0.
\end{align*}
Since $\mass_1(0; \strategynum) = 0$, by continuity, there exists $\delta>0$ such that
\begin{align*}
    0 < \mass_1(\lottery; \strategynum) < 1, \quad \text{for every }\lottery\in (0, \delta),
\end{align*}
Since $\pref_{F1} > \pref_{F2}$, then~\eqref{eq:concavity_integrand_before} is strictly negative on $(0, \delta)$. Integrating~\eqref{eq:concavity_integrand_before} over $\lottery\in [0, 1]$ yields strict concavity.

Finally, to show that~\eqref{eq:strict_concavity_cond} holds for any mean-field equilibrium $\strategynum^*$, we assume for contradiction that there exists an equilibrium $\strategynum^*$ such that \eqref{eq:strict_concavity_cond} does not hold. Then we must have  $\priorflexible=1$ and $\sigma^*_0=1$. That is, flexible students only apply to tier-2 schools, and we have $\mass_1(\lottery) = 0$ for every $\lottery\in [0, 1]$. If a flexible student instead applies to a tier-1 school, they are guaranteed admission and have a utility of $v_{F1}$, which is strictly greater than the utility for only applying to tier-2 schools, contradicting the fact that $\strategynum^*$ is an equilibrium. Hence, condition~\eqref{eq:strict_concavity_cond} holds for any mean-field equilibrium $\strategynum^*$.

\subsubsection{Proof of Proposition~\ref{prop:mf_equilibrium_consecutive}}\label{app:proof_prop_mf_equilibrium_consecutive}

Fix any mean-field equilibrium $\strategynumopt$. 
Given strict concavity in~\eqref{eq:mf_concavity} from Lemma~\ref{lem:mf_concavity}, we have
\begin{align*}
    \utility(\countschools+2; \strategynumopt) - \utility(\countschools+1; \strategynumopt) < \utility(\countschools+1; \strategynumopt) - \utility(\countschools; \strategynumopt).
\end{align*}
The difference in utility between two consecutive values of $\countschools$, namely $\utility(\countschools+1; \strategynumopt) - \utility(\countschools; \strategynumopt)$, is strictly decreasing in $\countschools$. Hence, there exists at most one value of $\countschools$ such that $\utility(\countschools+1; \strategynumopt) - \utility(\countschools; \strategynumopt)= 0$. 

\begin{itemize}
    \item When there does not exist $\countschools$ such that $\utility(\countschools+1; \strategynumopt)-\utility(\countschools; \strategynumopt) = 0$, it can be verified that there exists a unique $0\le \countschools \le \length$  that maximizes $\utility(\countschools; \strategynumopt)$. Thus, if $\strategynumopt$ is an equilibrium, it must be a pure equilibrium.

    \item When there exists $\countschools$ such that $\utility(\countschools+1; \strategynumopt)-\utility(\countschools; \strategynumopt) = 0$, it can be verified that
    \begin{align*}
        \utility(0; \strategynumopt) < \cdots < \utility(\countschools; \strategynumopt) = \utility(\countschools+1; \strategynumopt) > \cdots > \utility(\length; \strategynumopt).
    \end{align*}
    If $\strategynumopt$ is an equilibrium, it can only include values $\countschools$ and $(\countschools+1)$.
\end{itemize}
In both cases, if $\strategynumopt$ is an equilibrium, its support contains at most two consecutive values, completing the proof.

\subsubsection{Proof of Proposition~\ref{prop:algo_search_correct}}\label{app:prop_algo_search_correct}
For any $1\le \countschools\le \length$, if $A_\countschools\ge 0$, we have $\utility(\countschools; \distdelta_\countschools) - \utility(\countschools-1; \distdelta_\countschools)\ge 0$,
and hence by concavity, we have
\begin{align*}
    \utility(0; \distdelta_\countschools) -\utility(1; \distdelta_\countschools) \le \utility(1; \distdelta_\countschools) - \utility(2; \distdelta_\countschools) \le \ldots \le \utility(\countschools-1; \distdelta_\countschools) - \utility(\countschools; \distdelta_\countschools) \le 0.
\end{align*}
and hence
\begin{align}
    \utility(0; \distdelta_\countschools) \le \cdots \le \utility(\countschools; \distdelta_\countschools).\label{eq:utility_concave_chain}
\end{align}
Hence, if $A_\length\ge 0$, then \Cref{alg:mf} correctly returns equilibrium $\distdelta_\length$ in Line~\ref{line:return_max}. Similarly, for any $0 \le \countschools\le \length-1$, if $B_\countschools\le 0$, we have $\utility(\countschools+1; \distdelta_\countschools) - \utility(\countschools; \distdelta_\countschools) \le 0$, and again by concavity,
\begin{align}
    \utility(\countschools; \distdelta_\countschools) \ge \cdots \ge \utility(\length; \distdelta_\countschools).\label{eq:utility_concave_chain_two}
\end{align}
Hence, if $B_0 \le 0$, then \Cref{alg:mf} correctly returns equilibrium $\distdelta_0$ in  Line~\ref{line:return_zero}.

If \Cref{alg:mf} reaches Line~\ref{line:search}, we have $B_0 > 0$ and $A_\length < 0$. The following lemma shows that there always exists an equilibrium whose support lies within the interval $[\ell, h]$ during the binary search.
\begin{lemma}\label{lem:mf_binary_search}
    Consider any $0 \le \ell < h \le \length$. If $B_\ell \ge 0$ and $A_h \le 0$, then there exists an equilibrium with support in $\{l, \ldots, h\}$.
\end{lemma}
The proof of this lemma is provided at the end of the section. If \Cref{alg:mf} returns strategy $\distdelta_\countschools$ in Line~\ref{line:return_m}, then $A_\countschools\ge 0 \ge B_\countschools$. Combining~\eqref{eq:utility_concave_chain} and~\eqref{eq:utility_concave_chain_two} shows that $\distdelta_\countschools$ is a pure equilibrium. 
Finally, if Algorithm~\ref{alg:mf} reaches Line~\ref{line:root}, then  we have $h = \ell +1$. Applying the argument from the proof of Lemma~\ref{lem:mf_binary_search},  we have that $\probmix^*\distdelta_{\ell} + (1-\probmix^*)\distdelta_{\ell+1}$ is a mixed equilibrium.

\paragraph{Proof of Lemma~\ref{lem:mf_binary_search}.}
  We define value $j^*$ as follows. If there exists $j\in [\ell+1, h-1]$ with $B_j \le 0$, we set
  \begin{align*}
      j^* = \min\{\ell +1 \le j \le h-1: B_j \le 0\}.
  \end{align*}
  Otherwise, we set $j^* = h$. In either case, we have $B_{j^*-1} \ge 0$. Now consider the following two cases:

    \begin{itemize}
       \item If $j^* < h$ and $A_{j^*} \ge 0$, then we have $A_{j^*} \ge 0 \ge B_{j^*}$. By~\eqref{eq:utility_concave_chain} and~\eqref{eq:utility_concave_chain_two}, we have that $\distdelta_{j^*}$ is a pure equilibrium that lies in the interval $[\ell, h]$.

        \item Otherwise, if $j^*< h$, we must have $A_{j^*} < 0$. If $j^* = h$, we have $A_{j^*} = A_h \le 0$ by assumption. In either case, we have $A_{j^*} \le 0$. We show that there must exist an equilibrium with support contained in $\{j^*-1, j^*\}$. Consider the mixed strategy $\strategynum_\probmix\defn \probmix\distdelta_{j^*-1} + (1-\probmix) \distdelta_{j^*}$. Consider
        \begin{align*}
            g(\probmix) \defn \utility(j^*;\strategynum_\probmix) - \utility(j^*-1; \strategynum_\probmix).
        \end{align*}
        We have $g(0) = A_{j^*} \le 0$ and $g(1) = B_{j^*-1} \ge 0$. Since $g(\probmix)$ is continuous in $\probmix$, there must exist a root $\probmix^*\in [0, 1]$ with $g(\probmix^*)=0$. By concavity, we have
        \begin{align*}
            \utility(0; \strategynum_{\probmix^*}) \le \cdots \le \utility(j^*-1; \strategynum_{\probmix^*}) = \utility(j^*; \strategynum_{\probmix^*}) \ge \cdots \ge \utility(\length; \strategynum_{\probmix^*}),
        \end{align*}
        verifying that $\strategynum_{\probmix^*}$ is an equilibrium.
    \end{itemize}

\section{Proofs}\label{app:proofs}
In this appendix, we provide proofs of all theoretical results in Section~\ref{sec:results}.

\subsection{Proof of Theorem~\ref{thm:equivalence}}\label{app:proof_equivalence}

Consider any ordered and contiguous partition $\mathcal{P}_1\union \cdots\union  \mathcal{P}_\numpartitions = [\numstudents]$. Let $\strategy, \strategy'$ be two distinct equilibria. We show that these two equilibria lead to the same match rate and social welfare. 

Recall from Appendix~\ref{app:prelim_partial} that $\flow_{k\idxschool} = \sum_{\idxtype=1}^\numtypes \prior_\idxtype\pi_{t, \partition_k, \idxschool}$ is the probability that a random student from block $\partition_\idxpartition$ applies to school $j$, where $\pi_{t,\mathcal{P}_\idxpartition,j}$ is the probability that a type-$t$ student from this block applies to school $j$. We also refer to $\flow_{\idxpartition\idxschool}$ as the ``flow'' from block $\mathcal{P}_\idxpartition$ to school $j$, and we have  $\sum_{\idxschool=1}^\numschools \flow_{\idxpartition \idxschool} = 1$. Let $\mathbf{q}_\idxpartition =(q_{\idxpartition 1}, \ldots, q_{\idxpartition n})$ be the vector of flows from block $\partition_\idxpartition$ to all schools under equilibrium $\strategy$. We denote by $\mathbf{q}'_k$ the vector of flows under equilibrium $\strategy'$, and likewise use the prime symbol for other quantities under equilibrium $\strategy'$.

The proof consists of two steps: First, we show that the flows $\{\vecflow_\idxpartition\}_\idxpartition$ are identical under the two equilibria $\strategy$ and $\strategy'$. Second, we show that the match rate and social welfare are fully determined by these flows $\{\vecflow_\idxpartition\}_\idxpartition$. 

\paragraph{Step 1: Flows $\{\mathbf{q}_{\idxpartition}\}_\idxpartition$ are identical across equilibria.} 

We prove by induction for every block $\idxpartition \in [\numpartitions]$. We assume that the flows are identical in all preceding blocks, i.e., $\vecflow_\ell = \vecflow'_{\ell}$ for every $\ell < \idxpartition$, and the goal is to show that $\vecflow_\idxpartition = \vecflow'_\idxpartition$. 

Recall from Appendix~\ref{app:prelim_partial} that $\probavail_{\idxpartition j}$ is the admission probability that school $j$ is available to a student in block $\mathcal{P}_\idxpartition$. We have the recursive relation~\eqref{eq:partial_cond_prob_chain}:
\begin{align}
    \probavail_{\idxpartition \idxschool} = \probavail_{\idxpartition\idxschool,\textcond}\cdot \probavailbefore_{\idxpartition\idxschool},\label{eq:partial_cond_prob_chain_recall_equiv}
\end{align}
where $\probavailbefore_{\idxpartition\idxschool}$ is the probability that no student from preceding blocks is matched to school $\idxschool$ (and thus school $\idxschool$ is available to students in block $\partition_\idxpartition$), and $\probavail_{\idxpartition\idxschool, \textcond}$ is the probability that no students in block $\partition_\idxpartition$ with better lottery numbers, from the perspective of an individual student, apply to school $\idxschool$. We write $\boldsymbol{\probavailbefore}_\idxpartition = (\probavailbefore_{\idxpartition 1}, \ldots, \probavailbefore_{\idxpartition\numstudents})$.
Now we decompose the induction into two steps:
\begin{enumerate}[label=(\alph*)]
    \item \textbf{Showing $\boldsymbol{\probavailbefore}_\idxpartition = \boldsymbol{\probavailbefore}'_\idxpartition$.} 
    By~\eqref{eq:partial_probavail_before}, the probability $\probavailbefore_{\idxpartition\idxschool}$ is fully determined by flows $\{\vecflow_{\ell}\}_{\ell < \idxpartition}$. Hence,  we have $\boldsymbol{\probavailbefore}_\idxpartition = \boldsymbol{\probavailbefore}'_\idxpartition$ by the inductive hypothesis that $\vecflow_\ell = \vecflow'_\ell$ for every $\ell < \idxpartition$.

    \item \textbf{Showing $\vecflow_\idxpartition = \vecflow_\idxpartition'$.} 
    If block $\partition_\idxpartition$ is a singleton, then by assumption~\eqref{eq:equivalence_assume_singleton}, we have $\strategy_{t, \partition_k} = \strategy'_{t, \partition_k}$ for every student type $\idxtype$. That is, strategies $\strategy$ and $\strategy'$ are identical for every student type $\idxtype$ in block $\partition_\idxpartition$, and hence they generate the same flows $\vecflow_\idxpartition = \vecflow_\idxpartition'$. It remains to consider non-singleton block $\partition_\idxpartition$ with $\abs*{\partition_\idxpartition} \ge 2$.
        
    Assume for contradiction that $\vecflow_\idxpartition \ne \vecflow'_{\idxpartition}$. Consider the set of schools where students in block $\partition_\idxpartition$ have a positive probability of admission:
    \begin{align*}
        S\defn \{\idxschool\in [\numstudents]: \probavailbefore_{\idxpartition\idxschool} > 0\}.
    \end{align*}
    This set $S$ is non-empty, because the number of schools equals the number of students, so some school must have a positive probability to be available. Consider the set of schools for which students have a higher flow under equilibrium $\strategy$ than equlibrium $\strategy'$:
    \begin{align*}
        S_+ \defn \{\idxschool \in [n] :~\flow_{\idxpartition\idxschool}> \flow'_{\idxpartition\idxschool}\}.
    \end{align*}
    We have $S_+\subseteq S$ because students have a utility of zero when applying to schools outside set $S$, so they only apply to schools within set $S$. Moreover, set $S_+$ is non-empty by the assumption that $\vecflow_\idxpartition \neq \vecflow'_\idxpartition$. By the definition of set $S_+$, we have
    \begin{align*}
        \sum_{\idxschool\in S_+} q_{\idxpartition\idxschool} > \sum_{\idxschool\in S_+} q'_{\idxpartition\idxschool}.
    \end{align*}
    There must exist a student type $\idxtype$ who contributes more flow to schools $S_+$ under equilibrium $\strategy$ than equilibrium $\strategy'$:
    \begin{align*}
        \sum_{\idxschool\in S_+} \pi_{\idxtype, \partition_\idxpartition,\idxschool} > \sum_{\idxschool\in S_+} \pi'_{\idxtype, \partition_\idxpartition,\idxschool}.
    \end{align*}
    and hence
    \begin{align*}
        \sum_{\idxschool\in S\setminus S_+} \pi_{\idxtype, \partition_\idxpartition,\idxschool} < \sum_{\idxschool\in S\setminus S_+} \pi'_{\idxtype, \partition_\idxpartition,\idxschool},
    \end{align*}
    because $\sum_{\idxschool\in S}  \pi_{\idxtype, \partition_\idxpartition,\idxschool}=1$.
    There must exist some school $j \in S_+$ with $\pi_{t, \mathcal{P}_\idxpartition, j}>0$ and some school $\ell \in S\setminus S_+$ with $\pi'_{t, \mathcal{P}_\idxpartition \ell}>0$. By the definition of set $S_+$, we have
    \begin{align*}
        q_{\idxpartition j} &> q'_{\idxpartition j}\\
        q_{\idxpartition \ell} &\le q'_{\idxpartition \ell}.
    \end{align*}
    Recall from~\eqref{eq:partial_cond_prob_chain_recall_equiv} that $ \probavail_{\idxpartition \idxschool} = \probavail_{\idxpartition\idxschool,\textcond}\cdot \probavailbefore_{\idxpartition\idxschool}$, and recall from~\eqref{eq:prob_avail_partial} that the probability $\probavail_{\idxpartition\idxschool,\textcond}$ is strictly decreasing in $\flow_{\idxpartition\idxschool}$ for every school $\idxschool$. Since $\idxschool, \ell\in S$, we have $\beta_{\idxpartition\idxschool}, \beta_{\idxpartition\ell}> 0$. Moreover, since $\boldsymbol{\beta}_\idxpartition=\boldsymbol{\beta}'_\idxpartition$ from Step (a), we have $\probavail_{\idxpartition \idxschool}$ strictly decreasing in $\flow_{\idxpartition\idxschool}$. Therefore,
    \begin{subequations}\label{eq:equiv_flow_inequality_prob_avail}
        \begin{align}
            \probavail_{\idxpartition\idxschool} &< \probavail'_{\idxpartition\idxschool}\label{eq:equiv_flow_inequality_prob_avail_j}\\
            \probavail_{\idxpartition\ell} &\ge \probavail'_{\idxpartition\ell}.\label{eq:equiv_flow_inequality_prob_avail_l}
    \end{align}
    \end{subequations}
By the definition of the equilibria $\strategy$ and $\strategy'$, we have 
\begin{subequations}\label{eq:equiv_flow_inequality_util}
    \begin{align}
    v_{tj}{\alpha}_{\idxpartition j} &\ge v_{t\ell}{\alpha}_{\idxpartition \ell}\label{eq:equiv_flow_inequality_util_equil}\\
    v_{t\ell} {\alpha}'_{\idxpartition\ell} &\ge v_{t\idxschool} {\alpha}'_{\idxpartition\idxschool}.\label{eq:equiv_flow_inequality_util_equil_alt}
    \end{align}
\end{subequations}
Combining~\eqref{eq:equiv_flow_inequality_prob_avail} and~\eqref{eq:equiv_flow_inequality_util}, we have: 
\begin{align*}
    v_{t\ell} \alpha_{\idxpartition \ell} 
    \geq v_{t\ell} \alpha'_{\idxpartition\ell} 
    \geq v_{tj}\alpha'_{\idxpartition j} 
    > v_{tj} \alpha_{\idxpartition j} 
    \geq  v_{t\ell} \alpha_{\idxpartition\ell},
\end{align*}
where the four inequalities are due to~\eqref{eq:equiv_flow_inequality_prob_avail_l}, \eqref{eq:equiv_flow_inequality_util_equil_alt}, \eqref{eq:equiv_flow_inequality_prob_avail_j}, and~\eqref{eq:equiv_flow_inequality_util_equil}, respectively, along with the fact that $v_{\idxtype\idxschool} > 0$, leading to a contradiction. Therefore, we conclude that $\vecflow_\idxpartition =\vecflow'_\idxpartition$. 
\end{enumerate}
Finally, we remark that for the base case, we have $\boldsymbol{\probavailbefore}_1 = \boldsymbol{\probavailbefore}'_1 = 1$, so Step (a) holds trivially as every school is available before the matching process starts. This completes the induction that $\vecflow_\idxpartition =\vecflow'_\idxpartition$ for every $\idxpartition\in [\numpartitions]$.

\paragraph{Step 2: Match rate and social welfare are solely determined by flows $\{\vecflow_\idxpartition\}_\idxpartition$.} 

The match number~\eqref{eq:mn_partial} is
\begin{align*}
    \mn = \sum_{j=1}^n \left(1-\prod_{\idxpartition=1}^K (1-q_{\idxpartition j})^{|\mathcal{P}_\idxpartition|}\right)~,
\end{align*}
which is solely determined by flows $\{\vecflow_\idxpartition\}_\idxpartition$.

The social welfare~\eqref{eq:sw_partial} is
\begin{align}
    \sw = \sum_{\idxpartition=1}^K |\mathcal{P}_\idxpartition| \sum_{t=1}^T p_t \utility_{t,\partition_\idxpartition}\label{eq:equiv_sw_expression_recall}
\end{align}
where $\utility_{t,\partition_\idxpartition}$ is the expected utility of a type-$t$ student from block $\partition_\idxpartition$ at the equilibrium, and we have 
\begin{align}
    \utility_{t,\partition_\idxpartition} = \max_{j\in [\numstudents]} \alpha_{\idxpartition j}\pref_{tj}.\label{eq:util_expression_recall}
\end{align}
Recall from~\eqref{eq:partial_cond_prob_chain_recall_equiv} that $ \probavail_{\idxpartition \idxschool} =\probavail_{\idxpartition\idxschool,\textcond}\cdot \probavailbefore_{\idxpartition\idxschool}$, and the probabilities $\probavail_{\idxpartition\idxschool, \textcond}$ and $\probavailbefore_{\idxpartition\idxschool}$ are determined by flows $\{\vecflow_\idxpartition\}_\idxpartition$ due to~\eqref{eq:prob_avail_partial} and~\eqref{eq:partial_probavail_before}. Combining~\eqref{eq:equiv_sw_expression_recall} and~\eqref{eq:util_expression_recall}, the social welfare is determined by flows $\{\vecflow_\idxpartition\}_\idxpartition$. 

\subsection{Proof of Corollary~\ref{cor:equilbria_convex} }\label{app:proof_cor_convevx}

Consider the partial information policy with any partition $\partition_1\union \ldots \union \partition_\numpartitions=[\numstudents]$, since it strictly generalizes revealed and hidden lotteries. Suppose that $\strategy$ and $\strategy'$ are distinct two equilibria. We show that there must be an infinite number of equilibria. We construct 
\begin{align*}
        \strategy''\defn \lambda \strategy + (1-\lambda)\strategy', \quad \text{where }\lambda\in [0, 1].
\end{align*}
Fix any $\lambda\in (0, 1)$, and we show that $\strategy''$ is also an equilibrium. Consider the earliest block, denoted by $\partition_\idxpartition$, in which  equilibria $\strategy$ and $\strategy'$ differ. We show that there exists a new equilibrium where students in block $\partition_\idxpartition$ follow $\strategy''$, and students in all preceding blocks follow $\strategy$. Given these strategies, students in all subsequent blocks best-respond. To show that $\pi''$ is a best response for students in block $\partition_\idxpartition$, we consider the following two cases:

\begin{itemize}
    \item If $\partition_\idxpartition$ is a singleton, then consider any student type $\idxtype$. Every school $\idxschool$ with $\pi_{\idxtype,\partition_\idxpartition, \idxschool}>0$ or $\pi'_{\idxtype,\partition_\idxpartition, \idxschool}>0$ is utility maximizing, since strategies $\strategy$ and $\strategy'$ are identical in all blocks preceding block $\partition_\idxpartition$. Hence, their linear combination $\strategy''$ is utility maximizing in block $\partition_\idxpartition$.

    \item If $\partition_\idxpartition$ is not a singleton, then by Step 1 in the proof of \Cref{thm:equivalence} from Appendix~\ref{app:proof_equivalence}, the two equilibria have identical flows $\vecflow_\idxpartition=\vecflow'_\idxpartition$. For every school $\idxschool\in [\numstudents]$, we have
\begin{align*}
    q_{k\idxschool} &= q_{k\idxschool}'\\
    \sum_{\idxtype\in [\numtypes]} \prior_\idxtype\pi_{\idxtype, \partition_\idxpartition, \idxschool} &=  \sum_{\idxtype\in [\numtypes]} \prior_\idxtype\pi'_{\idxtype, \partition_\idxpartition, \idxschool}.
\end{align*}
Hence, for strategy $\strategy''$, we have
\begin{align*}
    \sum_{\idxtype\in [\numtypes]} \prior_\idxtype\pi''_{\idxtype, \partition_\idxpartition, \idxschool} = \sum_{\idxtype\in [\numtypes]} \prior_\idxtype(\lambda\pi_{\idxtype, \partition_\idxpartition, \idxschool} + (1-\lambda)\pi_{\idxtype, \partition_\idxpartition, \idxschool}') = q_{\idxpartition\idxschool}.
\end{align*}
That is, strategy $\strategy''$ in block $\partition_\idxpartition$ induces the same flows $\vecflow_\idxpartition$, and hence the same admission probability $\probavail_{\idxpartition\idxschool}$ for every school $\idxschool$. To show that $\strategy''$ is an equilibrium, we want to show that for every type $\idxtype$ and every school $\idxschool^*$ such that $\pi''_{\idxtype, \partition_\idxpartition, \idxschool^*} >0$, we have
\begin{align}
    \idxschool^*\in \argmax_{\idxschool\in [\numstudents]} \probavail_{\idxpartition\idxschool} \pref_{\idxtype\idxschool}.\label{eq:cor_convex_goal_no_deviate}
\end{align}
Note that $\pi''_{\idxtype, \partition_\idxpartition, \idxschool^*} >0$ implies that $\pi_{\idxtype, \partition_\idxpartition, \idxschool^*} >0$ or $\pi'_{\idxtype, \partition_\idxpartition, \idxschool^*} >0$. The assumption that $\strategy$ and $\strategy'$ are both equilibria yields~\eqref{eq:cor_convex_goal_no_deviate}, as desired.
\end{itemize}
Combining the two cases, strategy $\strategy''$ is utility maximizing in block $\partition_\idxpartition$ when students from all preceding blocks follow $\strategy$, completing the proof of a new equilibrium.

\subsection{Proof of Corollary~\ref{cor:unique_equil_two_schools}}\label{app:proof_cor_two_schools}

We index the student types in decreasing order of their valuation for school $1$, so we have $v_{1 1} > v_{21} > \cdots > v_{\numtypes, 1}$. Let $\strategy$ be an equilibrium under the hidden lottery. Consider any student type $\idxtype$ with $\pi_{\idxtype 1} > 0$. We have 
\begin{align*}
    \probavail_{1} \pref_{\idxtype 1} \ge  \probavail_{2} \pref_{\idxtype2}.
\end{align*}
Hence,
\begin{align*}
    \probavail_1\pref_{\ell 1} > \probavail_{2} \pref_{\ell 2}\quad \text{for every }\ell < \idxtype,
\end{align*}
and hence $\pi_{\ell 1} = 1$ for every $\ell < \idxtype$. Equivalently, the strategy profile $(\pi_{11},\ldots, \pi_{\numtypes, 1})$ can be either of the form
\begin{align*}
   (1, \ldots, 1, 0, 0, \ldots, 0),
\end{align*}
including the vector with all ones and that with all zeros, or
\begin{align*}
    (1, \ldots, 1, x, 0, \ldots, 0) \quad \text{for some } x\in (0, 1),
\end{align*}
where $x$ may appear at the first or last index. There is thus a one-to-one correspondence between the strategy profile $(\pi_{11},\ldots, \pi_{\numtypes, 1})$ and flow $\flow_1$. In the proof of \Cref{thm:equivalence}, we show that the flow is unique at equilibrium, so the equilibrium strategy profile is also unique.

\subsection{Proof of Lemma~\ref{lem:reveal_same_ranking_always_one}}\label{app:proof_lem_reveal_homo_always_one}

We show that the equilibrium under the lexicographic tie-breaking rule is that the student with \lnum $\lottery$ applies to school $r$. First, observe that the student with lottery number $1$ always applies to school $1$, because school $1$ is preferred by all types of students by assumption, with ties broken in the lexicographic order. Then by an inductive argument, it is optimal for the student with lottery number $r$, knowing that students with better lottery numbers apply to schools $1$ through $(r-1)$, to apply to school $r$, verifying the equilibrium.

Under this equilibrium, the match rate is $1$ because all students apply to different schools. The social welfare~\eqref{eq:sw_reveal} is:
\begin{align*}
    \swr = \sum_{\lottery=1}^n \sum_{\idxtype=1}^\numtypes p_\idxtype v_{\idxtype\lottery} =  \sum_{\idxtype=1}^\numtypes p_\idxtype \sum_{\lottery=1}^n v_{\idxtype\lottery} = 1,
\end{align*}
completing the proof.

\subsection{Proof of Theorem~\ref{thm:MR-gaps}}\label{app:proof_MR-gaps}

We separately present the proofs for bounds~\eqref{eq:ratio_mr_hr_homo}-\eqref{eq:ratio_mr_rh_hetero}.

\subsubsection{Proof of bound~\texorpdfstring{\eqref{eq:ratio_mr_hr_homo}}{(\ref{eq:ratio_mr_hr_homo})}}

We prove that $\ratiomrhrhomo=1-(1-\frac{1}{n})^n$ under homogeneous preference rankings. 

For the upper bound, recall from \Cref{lem:reveal_same_ranking_always_one} that $\mrr = 1$ under the revealed lottery. Under the hidden lottery, The match rate~\eqref{eq:mn_hidden} is
\begin{align*}
    \mrh = \frac{1}{\numschools} \sum_{\idxschool=1}^\numschools \left(1 - \big(1-\flowh_j\big)^\numschools\right),
\end{align*}
where $\flowh_\idxschool$ is the probability that a random student applies to school $\idxschool$. Since the function $f(q) \defn 1-(1-q)^n$ is concave in $q$, by Jensen's inequality, we have
\begin{align*}
    \mrh = \frac{1}{\numschools}\sum_{\idxschool=1}^\numschools f(\flowh_j) \le f\left(\frac{1}{\numschools}\sum_{\idxschool=1}^\numschools \flowh_\idxschool\right) = f\left(\frac{1}{n}\right) = 1-\left(1-\frac{1}{n}\right)^n,
\end{align*}
proving that $\ratiomrhrhomo\le 1-(1-\frac{1}{n})^n$.

For the lower bound, consider the uniform-prefix economy defined in Appendix~\ref{app:prelim_common_rare}. In this economy, there are $\numtypes=\numstudents$ types with uniform prior $\prior_1 = \cdots = \prior_\numstudents =\frac{1}{n}$. Recall from \Cref{lem:uniform_prefix_equilibrium} that the equilibrium under the hidden lottery is that type-$\idxtype$ students apply to school $\idxtype$, so we have $\flowh_\idxschool = \prior_\idxschool$. The match rate~\eqref{eq:mn_hidden} is:
\begin{align*}
    \mrh =\frac{1}{n}\sum_{\idxschool=1}^\numschools \left(1-\left(1-\frac{1}{n}\right)^n\right) = 1-\left(1-\frac{1}{n}\right)^n,
\end{align*}
proving that $\ratiomrhrhomo\ge 1-(1-\frac{1}{n})^n$.

\subsubsection{Proof of bounds~\texorpdfstring{\eqref{eq:ratio_mr_rh_homo}}{(\ref{eq:ratio_mr_rh_homo})} and~\texorpdfstring{\eqref{eq:ratio_mr_rh_hetero}}{(\ref{eq:ratio_mr_rh_hetero})}}

We prove that $\ratiomrrhhomo=\numstudents$ under homogeneous preference rankings. The proof  for showing $\ratiomrrh=\numstudents$ under heterogeneous preferences is identical.

The upper bound $\ratiomrrhhomo\le\numstudents$ holds trivially because the match rate is always between $\frac{1}{n}$ and $1$.

For the lower bound, consider an economy with one type of students who only value school 1:  We omit the index  $\idxtype=1$ in $v_{tj}$ and write the valuations of all students by:
\begin{align*}
    v_{1} = 1, \quad\text{and } v_{j} = \zeropos \qquad \text{for every } 2\le j \le n.
\end{align*}
Under the hidden lottery, it can be verified that it is an equilibrium for all students to apply to school $1$, leading to match rate
\[\mrh = \frac{1}{n},\]
proving that $\ratiomrrhhomo\ge\numstudents$.

\subsubsection{Proof of bound~\texorpdfstring{\eqref{eq:ratio_mr_hr_hetero}}{(\ref{eq:ratio_mr_hr_hetero})}}

We prove the upper bound $\ratiomrhr=O(\log n)$ and lower bound $\lim_{\numstudents\rightarrow \infty} \ratiomrhr > 1.038$ under heterogeneous preference rankings.

\paragraph{Showing $\ratiomrhr =O(\log\numstudents)$.}

At a high level, we show that $\mrh$ cannot be much higher than $\mrr$. To do so, we consider the ratio $\big\{\frac{\probavailr_{\lottery\idxschool}}{\probavailh_{\idxschool}}\big\}$, and for every lottery number $\lottery\in [\numstudents]$, we define
\begin{align*}
    J_r\defn \left\{\idxschool\in [\numstudents]: \frac{\probavailr_{\lottery\idxschool}}{\probavailh_{\idxschool}} > \frac{1}{2}\right\}
\end{align*}
to be the set of schools whose admission probability under the revealed lottery is relatively higher than the hidden lottery (by at least a factor of $\frac{1}{2}$). We consider the matches contributed by schools in $J_r$ and outside $J_r$ under the hidden lottery. For schools outside $J_r$, they have a relatively low admission probability under the revealed lottery, which means they are likely matched under the revealed lottery too. For schools in $J_r$, their admission probability under the revealed lottery is relatively high, which means students with subsequent lottery numbers find these schools attractive and apply to these schools. In both cases, the matches under the hidden lottery are likely to appear under the revealed lottery as well, providing an upper bound on their ratio.

Specifically, we decompose the match number~\eqref{eq:mn_hidden} under the hidden lottery as:
\begin{align}
    \mnh = \numstudents\sum_{\idxschool=1}^\numschools\flowh_\idxschool\probavailh_\idxschool &= \sum_{\lottery=1}^\numstudents \sum_{\idxschool=1}^\numschools\flowh_\idxschool\probavailh_\idxschool\nonumber\\
    &= \underbrace{\sum_{\lottery=1}^\numschools \sum_{\idxschool\in J_r} \flowh_\idxschool\probavailh_\idxschool}_{\term_1} + \underbrace{\sum_{\lottery=1}^\numschools \sum_{\idxschool\not\in J_r} \flowh_\idxschool\probavailh_\idxschool}_{\term_2},\label{eq:bound_mn_hidden_by_reveal_decompose}
\end{align}
where term $\term_1$ represents matches by schools in $J_r$, and term $\term_2$ represents matches by schools outside $J_r$. We analyze the two terms separately.

\paragraph{Term $\term_2$.} We fix any $\lottery\in [\numstudents]$, and consider any $\idxschool\not\in J_r$. We have $\frac{\probavailr_{\lottery\idxschool}}{\probavailh_\idxschool} \le \frac{1}{2}$, and hence
\begin{align*}
    \probavailr_{\lottery\idxschool} \le \frac{\probavailh_\idxschool}{2} \le \frac{1}{2}.
\end{align*}
We thus have $\probavailr_{\numstudents+1,\idxschool} \le \frac{1}{2}$, since the admission probability $\probavailr_{\lottery\idxschool}$ is non-increasing in the lottery number $\lottery$. The match number~\eqref{eq:mn_reveal} under the revealed lottery is $\mnr = \sum_{\idxschool=1}^\numschools (1-\probavailr_{\numstudents+1, \idxschool})$. We bound term $\term_2$ as
\begin{align}
    \term_2 &\stackrel{\stepone}{\le} \sum_{\lottery=1}^\numstudents\sum_{\idxschool\not \in J_r} \frac{1}{\numstudents} \nonumber\\
    & \stackrel{\steptwo}{\le} \frac{1}{n}\sum_{\lottery=1}^\numstudents\sum_{\idxschool\not\in J_r} 2(1-\probavailr_{\numstudents+1, \idxschool}) 
    \le 2\mnr,\label{eq:mn_term_two_bound}
\end{align}
where step~\stepone is true because $\flowh_\idxschool\probavailh_\idxschool = \frac{1-(1-\flowh_\idxschool)^\numstudents}{\numstudents} \le \frac{1}{n}$, and step~\steptwo is true because $\probavailr_{\numstudents+1, \idxschool}\le \frac{1}{2}$.

\paragraph{Term $\term_1$.}

We bound term $\term_1$ as
\begin{align*}
    \term_1 \le \sum_{\lottery=1}^\numstudents\sum_{\idxschool\in J_r} \flowh_\idxschool 
    \stackrel{\stepone}{\le} \sum_{\lottery=1}^\numstudents \sum_{\idxschool\in J_r} \flowr_{\lottery\idxschool},
\end{align*}
where step~\stepone is true by invoking \Cref{lem:flow_set_compare} with $c = \frac{1}{2}$. It remains to bound $\sum_{\lottery=1}^\numstudents \sum_{\idxschool\in J_r} \flowr_{\lottery\idxschool}$.

Note that $J_1 \supseteq J_2\cdots \supseteq J_n$, because $\probavailr_{\lottery\idxschool}$ is non-increasing in lottery number $\lottery$. Fix any school $\idxschool$, we have $j\in J_1$, because $\probavailr_{1\idxschool} = 1$ and hence $\frac{\probavailr_{1\idxschool}}{\probavailh_\idxschool} \ge 1 > \frac{1}{2}$. Denote by $r_j$ the last lottery number $\lottery$ such that school $\idxschool$ remains in set $J_r$:
 \begin{align*}
     r_{\idxschool}\defn \max \{\lottery\in [\numstudents]: \idxschool\in J_r\}.
 \end{align*}
We have that school $\idxschool$ belongs to sets $J_1, \ldots, J_{r_j}$, and
 \begin{align}
     \term_1 \le \sum_{\idxschool=1}^\numschools\sum_{r=1}^{r_j}\flowr_{\lottery\idxschool}.\label{eq:mr_term_one}
\end{align}
We now bound~\eqref{eq:mr_term_one} for each school $\idxschool\in [\numstudents]$.
\begin{itemize}
    \item \textbf{School $\idxschool$ with $\probavailr_{n+1, \idxschool} \le \frac{1}{2}$.}
Since $\idxschool\in J_{r_\idxschool}$, we have
$\frac{\probavailr_{\lottery_\idxschool, \idxschool}}{\probavailh_\idxschool} \ge \frac{1}{2}$, and hence
\begin{align*}
    \prod_{\lottery=1}^{r_j-1} \big(1-\flowr_{\lottery\idxschool}\big) = \probavailr_{\lottery_\idxschool, \idxschool}
    \ge \frac{1}{2}\probavailh_\idxschool\ge \frac{1}{2\numstudents}.
\end{align*}
Invoking \Cref{lem:prod_sum} with $c=\frac{1}{2n}$, we have
\begin{align*}
    \sum_{r=1}^{r_j-1} \flowr_{\lottery\idxschool} \le \log(2n).
\end{align*}
and hence
\begin{align}
    \sum_{r=1}^{r_j} \flowr_{\lottery\idxschool} \le 1+\log(2n)  \le \left(1 + \log(2n)\right) \cdot 2(1-\probavailr_{n+1, \idxschool}) .\label{eq:sum_flow_one}
\end{align}

    \item \textbf{School $\idxschool$ with $\probavailr_{n+1, \idxschool} > \frac{1}{2}$.}
We have \begin{align*}
    \probavailr_{n+1, \idxschool} = \prod_{\lottery=1}^\numstudents(1-\flowr_{\lottery\idxschool}).
\end{align*}
Invoking \Cref{lem:prod_sum} with $c = \probavailr_{n+1, \idxschool}>\frac{1}{2}$ yields
\begin{align}
    \sum_{\lottery=1}^\numstudents\flowr_{\lottery\idxschool} \le \log \left(\frac{1}{\probavailr_{n+1, \idxschool}}\right) \stackrel{\stepone}{\le} \frac{1-\probavailr_{n+1, \idxschool}}{\probavailr_{n+1, \idxschool}} 
    \le 2(1-\probavailr_{n+1, \idxschool}),\label{eq:sum_flow_two}
\end{align}
where step~\stepone is true because $\log(u) \le u-1$ for every $u>0$.
\end{itemize}

Combining both cases by plugging~\eqref{eq:sum_flow_one} and~\eqref{eq:sum_flow_two} into~\eqref{eq:mr_term_one}, we have
\begin{align}
    \term_1 &\le \sum_{\idxschool=1}^\numschools \sum_{\lottery=1}^{r_\idxschool}\flowr_{\lottery\idxschool} \nonumber\\
    &\lessorder \log n\cdot \sum_{\idxschool=1}^\numschools (1-\probavailr_{n+1, \idxschool}) \nonumber\\
    &\eqorder \log n\cdot \mnr.\label{eq:mn_term_one_bound}
\end{align}
Finally, plugging term $\term_1$ from~\eqref{eq:mn_term_one_bound} and term $\term_2$ from~\eqref{eq:mn_term_two_bound} into~\eqref{eq:bound_mn_hidden_by_reveal_decompose}, we have
\begin{align*}
    \mnh\lessorder \log n\cdot \mnr,
\end{align*}
proving $\frac{\mnh}{\mnr} = O(\log n)$.

\paragraph{Showing $\lim_{\numstudents\rightarrow \infty}\ratiomrhr>1.038$.}

Consider an economy with $\numstudents$ types. For readability, we assume that $\numstudents$ is even. Each odd type $\idxtype$ has valuations
\begin{align*}
    \pref_{\idxtype \idxschool} = \begin{cases}
        1 & \text{if } \idxschool=\idxtype\\
        0^+ & \text{otherwise},
    \end{cases}
\end{align*}
and each even type $\idxtype$ has valuations
\begin{align*}
    \pref_{\idxtype\idxschool} = \begin{cases}
        0.6 & \text{if }\idxschool=\idxtype-1\\
        0.4 & \text{if }\idxschool=\idxtype\\
        0^+ & \text{otherwise}.
    \end{cases}
\end{align*}
with prior $\prior_1 = \ldots = \prior_{\numstudents-1} = \frac{1.5}{n}$ for odd types and $\prior_2 = \ldots = \prior_{\numstudents}=\frac{0.5}{n}$ for even types.
Equivalently, we write $\{\pref_{\idxtype\idxschool}\}$ as a matrix
\begin{align*}
    \begin{bmatrix}
        1 & 0^+ & \cdots & 0^+ & 0^+\\
        0.6 & 0.4 & \cdots & 0^+ & 0^+\\
        \vdots & \vdots &\ddots & \vdots & \vdots \\
        0^+ & 0^+ & \cdots & 1 & 0^+\\
        0^+ & 0^+ & \cdots & 0.6 & 0.4
    \end{bmatrix}.
\end{align*}
Under either the hidden or revealed lottery, it can be verified that students only apply to the schools that they have a positive valuation for. It suffices to analyze students of types $1$ and $2$ who apply to schools $1$ and $2$. Moreover, type-1 students always apply to school 1. We separately analyze the match rate under the hidden and revealed lotteries.

\paragraph{Hidden lottery.} 
We show that it is an equilibrium for type-$2$ students to apply to school $2$. To verify this equilibrium, we have $\flowh_1 = p_1 =\frac{1.5}{n}$ and $\flowh_2 = p_2 = \frac{0.5}{\numstudents}$. The admission probabilities of the two schools are
\begin{align*}
    \probavailh_1 &= \frac{1-(1-\flowh_1)^\numstudents}{\numstudents\flowh_1} = \frac{1-(1-\frac{1.5}{n})^\numstudents}{1.5}\\
    \probavailh_2 &= \frac{1-(1-\flowh_2)^\numstudents}{\numstudents\flowh_2} = \frac{1-(1-\frac{0.5}{n})^\numstudents}{0.5}.
\end{align*}
It can be verified that type-2 students have a higher expected utility of applying to school 2 than school 1:
\begin{align*}
     \utility^\parenh_2(2) &= v_{22}\probavailh_2  > v_{21}\probavailh_1 =\utility^\parenh_2(1) \quad \text{for every }n\ge 11,
\end{align*}
which verifies the equilibrium. Extending the same argument to student types $4, 6, \ldots, \numstudents$, the match number is
\begin{align*}
    \mnh = \frac{n}{2}\cdot \left(2-\left(1-\frac{1.5}{n}\right)^n - \left(1-\frac{0.5}{n}\right)^\numstudents\right)
\end{align*}
and thus
\begin{align}\label{eq:mr_hr_hetero_hidden}
    \lim_{\numstudents\rightarrow \infty} \mrh = \frac{1}{2}\lim_{\numstudents\rightarrow \infty} \left(2-\left(1-\frac{1.5}{n}\right)^n - \left(1-\frac{0.5}{n}\right)^\numstudents\right) = \frac{1}{2}\left(2-e^{-1.5}-e^{-0.5}\right)> 0.58.
\end{align}

\paragraph{Revealed lottery.}
Note that a type-$2$ student with lottery number $1$ applies to school $1$. Let $\lottery_0$ be the earliest lottery number with which a type-$2$ student would apply to school 2 with positive probability. We have (with strict inequality if type-$2$ students break ties in favor of school 1):
\begin{align}
    \utility^\parenr_{2,\lottery_0, 2} &\ge \utility^\parenr_{2,\lottery_0, 1}\nonumber\\
    \pref_{2, 2} \cdot \probavailr_{\lottery_0,2} &\ge \pref_{2, 1} \cdot \probavailr_{\lottery_0, 1}\nonumber\\
    0.4 &\ge0.6 \cdot \left(1-\frac{2}{n}\right)^{r_0-1}\label{eq:revealed_smallest_r}\\
    r_0 &\ge 1+\frac{\log(1.5)}{-\log(1-\frac{2}{n})}.\nonumber
\end{align}
Hence, we have $r_0 = 1+\ceil*{\frac{\log(1.5)}{-\log(1-\frac{2}{n})}}$. We now show by induction that type-$2$ students with $\lottery\ge \lottery_0$ all apply to school 2. Suppose that type-$2$ students with lottery numbers $\lottery_0$ through $\lottery_0 + \ell$ apply to school 2, then for lottery number $\lottery=\lottery_0+\ell+1$, we have
\begin{align*}
    \utility^\parenr_{2, \lottery, 2} = \pref_{22}\probavailr_{\lottery2} = 0.4\cdot \left(1-\frac{0.5}{n}\right)^{\ell+1}
\end{align*}
and
\begin{align*}
    \utility^\parenr_{2, \lottery, 1} = \pref_{21}\probavailr_{\lottery1} = 0.6\cdot \left(1-\frac{2}{n}\right)^{r_0-1}\left(1-\frac{1.5}{n}\right)^{\ell+1}.
\end{align*}
Plugging in~\eqref{eq:revealed_smallest_r} verifies that $\utility^\parenr_{2, \lottery, 2} > \utility^\parenr_{2, \lottery, 1}$, so that a type-$2$ student with lottery number $(\lottery_0 +\ell +1)$ applies to school 2, completing the inductive argument.

Extending the same argument to student types $4, 6, \cdots, \numstudents$, we have
\begin{align*}
    \mnr = \frac{\numstudents}{2}\left(2-\left(1-\frac{2}{n}\right)^{\lottery_0-1}\left(1-\frac{1.5}{n}\right)^{n-\lottery_0+1} - \left(1-\frac{0.5}{n}\right)^{\numstudents-\lottery_0+1}\right)
\end{align*}
With $\lim_{\numstudents\rightarrow \infty} \frac{\lottery_0}{\numstudents} = \lim_{\numstudents\rightarrow \infty}\frac{1}{n}\cdot \frac{\log(1.5)}{-\log(1-\frac{2}{n})}=\frac{\log(1.5)}{2}$, we have
\begin{align}
    \lim_{\numstudents\rightarrow \infty} \mrr &= \frac{1}{2}\lim_{\numstudents\rightarrow \infty}\left(2-\left(1-\frac{2}{n}\right)^{\frac{\log(1.5)}{2}n} \left(1-\frac{1.5}{n}\right)^{\left(1-\frac{\log(1.5)}{2}\right)n} - \left(1-\frac{0.5}{n}\right)^{\left(1-\frac{\log(1.5)}{2}\right)n}\right)\nonumber\\
    &= \frac{1}{2} \cdot \left(2-e^{-\log(1.5)} \cdot e^{-1.5\big(1-\frac{\log(1.5)}{2}\big)} - e^{-0.5\big(1-\frac{\log(1.5)}{2}\big)}\right)<0.57.\label{eq:mr_hr_hetero_revealed}
\end{align}
Combining~\eqref{eq:mr_hr_hetero_hidden} and~\eqref{eq:mr_hr_hetero_revealed}, we have
\begin{align*}
    \lim_{\numstudents\rightarrow \infty}\frac{\mrh}{\mrr} >1.038.
\end{align*}

\subsection{Proof of Theorem~\ref{thm:SW-gaps}}\label{app:proof_SW-gaps}

We separately present the proofs for bounds~\eqref{eq:ratio_sw_hr_homo}-\eqref{eq:ratio_sw_rh_hetero}.

\subsubsection{Proof of bound~\texorpdfstring{\eqref{eq:ratio_sw_hr_homo}}{(\ref{eq:ratio_sw_hr_homo})}}\label{app:sw_ratio_hr_same_ranking}

We prove that $\ratioswhrhomo=\Theta(\log n)$ under homogeneous rankings.
Recall from \Cref{lem:reveal_same_ranking_always_one} that $\swr = 1$, so it remains to show that $\sup_\econ \swh = \Theta(\log n)$ under the hidden lottery.

\paragraph{Showing $\sup_\econ \swh = O(\log n)$.}
Recall that under homogeneous preference rankings, schools are indexed such that $v_{\idxtype1} \ge \ldots \ge v_{\idxtype \numstudents}$ for every student type $\idxtype\in [\numtypes]$, so we have $v_{\idxtype j} \le \frac{1}{j}$ for every school $j\in [\numschools]$. Hence, 
\begin{align*}
    \swh \le \sum_{j=1}^n \frac{1}{j} \eqorder \log n.
\end{align*}

\paragraph{Showing $\sup_\econ \swh = \Omega(\log n)$.}

We consider the uniform-prefix economy from Appendix~\ref{app:prelim_common_rare}. Recall from \Cref{lem:uniform_prefix_equilibrium} that it is an equilibrium for type-$\idxtype$ students to apply to school $\idxtype$, so we have $\flowh_\idxschool = \prior_\idxschool = \frac{1}{\numstudents}$. The admission probability~\eqref{eq:prob_avail_expression} for each school $\idxschool$ is
\begin{align*}
    \probavailh_\idxschool = \frac{1 - \left(1-\flowh_\idxschool\right)^\numstudents}{n\flowh_\idxschool} = 1-\left(1-\frac{1}{n}\right)^n.
\end{align*}
The social welfare~\eqref{eq:sw_hidden} is
\begin{align*}
    \swh = \sum_{\idxtype=1}^\numstudents \utility^\parenh_\idxtype &= \sum_{\idxtype=1}^n v_{\idxtype\idxtype} \probavailh_\idxtype\\
    & =\sum_{\idxtype=1}^n \frac{1- \left(1-\frac{1}{n}\right)^n}{\idxtype} \\
    & \stackrel{\stepone}{>} \left(1-\frac{1}{e}\right)\sum_{\idxtype=1}^n \frac{1}{\idxtype} \eqorder \log n,
\end{align*}
where step~\stepone is true due to inequality~\eqref{eq:inequality}.

\subsubsection{Proof of bound~\texorpdfstring{\eqref{eq:ratio_sw_rh_homo}}{(\ref{eq:ratio_sw_rh_homo})}}\label{app:sw_ratio_rh_same_ranking}

We prove that $\ratioswrhhomo = 2-\frac{1}{n}$ under homogeneous preference rankings.
Recall from \Cref{lem:reveal_same_ranking_always_one} that $\swr = 1$, so it remains to show that $\inf_\econ \swh = \frac{n}{2n-1}$ under the hidden lottery.

\paragraph{Showing $\inf_\econ \swh \le \frac{\numstudents}{2n-1}$.}
Consider a single type of students. We omit the index $\idxtype=1$ in $\pref_{\idxtype\idxschool}$ and write the valuations of all students:
\begin{align*}
    v_1 = v, \quad\text{and} \quad v_2 = \cdots = v_n = \frac{1-v}{n-1},
\end{align*}
where the value of $v\in (0, 1)$ is specified later.
We derive the condition on the value of $v$ such that it is an equilibrium for all students to apply to school 1 under the hidden lottery. Under this strategy profile, we have $\flow_1 = 1$ and $\flow_\idxschool = 0$ for $2\le \idxschool \le \numstudents$. The admission probabilities are
\begin{align*}
    \probavail_1 = \frac{1}{n},\quad \text{and}\quad \probavail_\idxschool = 1 \quad \text{for }2\le \idxschool \le \numstudents.
\end{align*}
A strict equilibrium requires
\begin{align*}
    \utility(1) &> \utility(\idxschool)\\
    v\cdot \probavail_1 &> \frac{1-v}{\numstudents-1}\cdot \probavail_\idxschool\\
    \frac{v}{n} & > \frac{1-v}{n-1}\\
    v & > \frac{n}{2n-1}.
\end{align*}
At this equilibrium, all students apply to school $1$, and therefore $\swh = v$. Taking $v\rightarrow \big(\frac{n}{2n-1}\big)_+$ yields $\swh \rightarrow \frac{n}{2n-1}$. 

\paragraph{Showing $\inf_\econ \swh \ge \frac{\numstudents}{2n-1}$.} 
Consider any economy $\econ$.
Recall from~\eqref{eq:prelim_sum_inv_alpha_ub} that the admission probabilities $\{\probavailh_j\}$ satisfy: 
\begin{align}
    \sum_{j}\frac{1}{\probavailh_j} \le 2n-1.\label{eq:ratio_rh_sum_inv_alpha}
\end{align}
Consider any type $t\in [\numtypes]$. For a type-$t$ student, their expected utility at the hidden equilibrium is\begin{align*}
    \utility_t^\parenh = \max_j v_{\idxtype j} \probavailh_j.
\end{align*}
Thus, we have $v_{\idxtype j} \le \frac{u_t^\parenh}{\probavailh_j}$ for every school $j$, and
\begin{align}
     \sum_{j=1}^n \frac{1}{\alpha^\parenh_j} = \frac{1}{\utility^\parenh_\idxtype}\sum_{j=1}^n \frac{\utility^\parenh_\idxtype}{\alpha^\parenh_j} \nonumber\\
     \ge \frac{1}{\utility_\idxtype^\parenh}\sum_{j=1}^n v_{\idxtype j}=\frac{1}{\utility_\idxtype^\parenh},\label{eq:ratio_rh_sum_inv_alpha_two}
\end{align}
Combining~\eqref{eq:ratio_rh_sum_inv_alpha} and~\eqref{eq:ratio_rh_sum_inv_alpha_two} yields
\begin{align*}
    \utility_t^\parenh \ge \frac{1}{2n-1}.
\end{align*}
We thus have 
\begin{align*}
    \swh =n\sum_t p_t \utility^\parenh_t \ge \frac{n}{2n-1},
\end{align*}
as desired.

\subsubsection{Proof of bound~\texorpdfstring{\eqref{eq:ratio_sw_hr_hetero}}{(\ref{eq:ratio_sw_hr_hetero})}}\label{app:proof_sw_ratio_hr_different_rankings}

We prove that $\ratioswhr= \Theta(\sqrt{n})$ under heterogeneous rankings.

\paragraph{Showing $\ratioswhr=\Omega(\sqrt{n})$.}
We consider the common-rare economy defined in~\eqref{eq:construction_common_rare} from Appendix~\ref{app:prelim_common_rare}, with $m=\floor{\frac{\sqrt{n}}{4}}$ rare types. For readability, we assume that $\frac{\sqrt{n}}{4}$ is an integer and hence $m = \frac{\sqrt{n}}{4}$. Recall that we re-index schools and types to start from $0$, where type $0$ is the common type with valuations:
\begin{align*}
    \pref_{0\idxschool} & = \begin{cases}
        1 - \frac{m}{n} & \text{ if } \idxschool = 0\\
        \frac{1}{n}  & \text{ if } 1\le \idxschool \le \numminority\\
        0 & \text{otherwise,}
    \end{cases}
\end{align*}
and each rare type $t\in [m]$ has valuations
\begin{align*}
    \pref_{\idxtype\idxschool} = \begin{cases}
        1 & \text{if }\idxschool=\idxtype\\
        0& \text{if }\idxschool\ne \idxtype.
    \end{cases}
\end{align*}
The prior is
\begin{align*}
    \prior_0 = 1 - \frac{\numminority}{\numschools},\quad\text{ and }\quad \prior_\idxtype = \frac{1}{\numschools} \quad\text{for every } \idxtype\in [\numminority].
\end{align*}
Without loss of generality, we assume that the tie-breaking rule $\omega_0$ for common-type students break ties among rare schools in the order of schools $1, 2, \ldots, \numminority$. We now compute the social welfare of this economy under the hidden and revealed lotteries.

\paragraph{Hidden lottery.}
By Lemma~\ref{lem:common_rare_equil_hidden}, the social welfare~\eqref{eq:sw_common_rare} under the hidden lottery is at least
\begin{align}
    \swh \ge \left(1-\frac{1}{e}\right)\cdot \numminority\eqorder \sqrt{n}.\label{eq:ratio_sw_hr_bound_hidden}
\end{align}

\paragraph{Revealed lottery.}

It can be verified that it is an equilibrium for students of each rare type $\idxtype\in [\numminority]$ to apply to their preferred school $\idxtype$. We now show that the common-type students have the following strategies at the revealed equilibrium:
\begin{itemize}
    \item If a common-type student has lottery number $1$ or $2$, then they apply to the common school $0$;
    
    \item If a common-type student has lottery number $\lottery \in \{3, \ldots, m+2\}$, then they apply to rare school $(\lottery - 2)$.
\end{itemize}
Students with lottery numbers worse than $(m+2)$ then best-respond accordingly, and we do not specify their equilibrium strategies explicitly, as they do not affect the students with lottery numbers $1$ through $(\numminority+2)$.

We now verify the claimed equilibrium strategies for students with lottery number $r\le m+2$. Recall that $\utility^\parenr_{t r}(j)$ denotes the expected utility for a type-$t$ student with lottery number $r$ to apply to school $\idxschool$, while all other students follow their equilibrium strategies. We show that the claimed equilibrium strategies are utility maximizing for every student type. First, note that common-type students have a valuation of $0$ for every school $\idxschool > m$, and a positive valuation along with a positive admission probability for every school $0\le \idxschool\le \numminority$ (when no students with better lottery numbers apply to school $\idxschool$). Hence, common-type students only apply to schools $0$ through $m$ at the revealed equilibrium.
\begin{itemize}
    \item When a common-type student has lottery number $1$, we have 
\begin{align*}
    \utility^\parenr_{0 1}(0) = \pref_{00} = 1
    - \frac{m}{n} > \frac{1}{n} = \utility^\parenr_{01}(\idxtype), \quad \text{ for every rare  school } \idxtype\in [\numminority].
\end{align*}

\item When a common-type student has \lnum $2$, we have
\begin{align*}
    \utility^\parenr_{02}(0) =\probavailr_{20}\cdot \pref_{00} &= (1-p_0)\cdot \pref_{00} =  \frac{m}{n}\cdot \left(1-\frac{m}{n}\right) \stackrel{\stepone}{>} \frac{1}{n} \ge u_{02}^\parenr(\idxtype),
\end{align*}
where step~\stepone holds for $m=\frac{\sqrt{n}}{4}$ and $n \ge 64$.

\item When a common-type student has \lnum $\lottery\in \{3,\ldots,  m+2\}$, consider any school $\idxschool$ with $\lottery-2\le \idxschool \le \numminority$. School $\idxschool$ is available to them if and only if there is no type-$\idxschool$ student with a better lottery number. The student has the same expected utility for applying to every school $\idxschool$ with $\lottery-2\le \idxschool\le \numminority$:
\begin{align*}
    u^\parenr_{0r}(\idxschool)=\frac{1}{n}\cdot \left(1-\frac{1}{n}\right)^{r-1} \ge \frac{1}{n}\cdot \left(1-\frac{1}{n}\right)^{m+1} \ge \frac{1}{n}\cdot \left(1-\frac{m+1}{n}\right) \ge\frac{1}{2n}.
\end{align*}
Consider any school $1\le \idxschool< \lottery-2$. School $\idxschool$ is available if and only if there is no type-$\idxschool$ student with a better lottery number \emph{and} the student with lottery number $(j+2)$ is not the common type. This admission probability is thus strictly smaller than the admission probability of any school $r-2\le j\le m$. Since common-type students have the same valuation for all rare schools, they have a lower expected utility for applying to any school $1\le \idxschool < \lottery-2$. 

Now consider school $0$. School $0$ is available if and only if the two students with \lnums $1$ and $2$ are not the common type. The expected utility of applying to school 0 is:
\begin{align*}
    u^\parenr_{0r}(0)=\pref_{00}(1-\prior_0)^2=\left(1-\frac{m}{n}\right)\cdot \left(\frac{m}{n}\right)^2  < \left(\frac{m}{n}\right)^2 = \frac{1}{16n},
\end{align*}
Hence, it is optimal for the student to apply to school $\lottery-2\le j \le \numminority$, and thus the student applies to school $(\lottery-2)$ under the tie-breaking rule that favors small indices among rare schools.
\end{itemize}
Having the equilibrium strategies confirmed for common-type students with lottery numbers $\lottery\in [\numminority+2]$, we now analyze the social welfare, by separately considering common-type and rare-type students. The social welfare over all matched common-type students is at most:
\begin{align}
    \sum_{\idxschool=0}^{\numschools-1} \pref_{0\idxschool}= 1.\label{eq:sw_ratio_hr_different_rankings_common_type}
\end{align}
Now we consider the social welfare over all matched rare-type students.  Consider each rare type $\idxschool\in [\numminority]$, and recall that type-$\idxschool$ students only have a positive valuation for school $\idxschool$. We upper bound the probability that school $\idxschool$ is matched to a type-$\idxschool$ student, by considering the following two cases:

\begin{itemize}
    \item \textbf{Case 1:} School $\idxschool$ is matched to a type-$\idxschool$ student with \lnum between $1$ and $(\idxschool + 1)$. By a union bound, the probability that there exists a type-$\idxschool$ student with \lnum between $1$ and $(\idxschool + 1)$ is upper bounded by $(\idxschool+1)\cdot  p_\idxschool = \frac{\idxschool+1}{n}$.
    
    \item \textbf{Case 2:} School $\idxschool$ is matched to a type-$\idxschool$ student with \lnum between $(\idxschool+2)$ and $n$. Consider the student with \lnum $(\idxschool+2)$. If this student is common type, then they apply to school $\idxschool$ at the revealed equilibrium, contradicting the case assumption that school $j$ is matched to a type-$\idxschool$ student.  Hence, the student \withlnum $(\idxschool +2)$ must not be common type, which happens with probability $1-\prior_0 = \frac{m}{n}$.
\end{itemize}
Combining these two cases, the social welfare over all rare-type students is at most
\begin{align}
    \sum_{\idxtype=1}^\numminority \pref_{\idxtype\idxtype}\cdot \left(\frac{t+1}{n}+\frac{m}{n}\right) &=
    \sum_{t=1}^m \left(\frac{t+1}{n}+\frac{m}{n}\right) \eqorder \frac{m^2}{n} \eqorder 1.\label{eq:sw_ratio_hr_different_rankings_rare_type}
\end{align}
Combining~\eqref{eq:sw_ratio_hr_different_rankings_common_type} and~\eqref{eq:sw_ratio_hr_different_rankings_rare_type}, we have that in this common-rare economy:
\begin{align*}
    \swr = \Theta(1).
\end{align*}
Combining with bound~\eqref{eq:ratio_sw_hr_bound_hidden} that $\swh = \Omega(\sqrt{n})$ in this economy, we have 
\begin{align*}
    \frac{\swh}{\swr} = \Omega(\sqrt{n}),
\end{align*}
completing the proof that $\ratioswhr = \Omega(\sqrt{n})$.

\paragraph{Showing $\ratioswhr=O(\sqrt{n})$.}

We show that $\frac{\swh}{\swr} = O(\sqrt{n})$ in every economy.
At a high level, we show that if $\swh$ is large, then $\swr$ cannot be too small. The proof consists of two steps. In the first step, we show that $\swr = \Omega(1)$. It thus suffices to consider the case where $\swh$ is $\Omega(\sqrt{n})$. In the second step, we show that the schools that contribute significantly to $\swh$ must also contribute significantly to  $\swr$, and thus $\swr$ cannot be too small. This is done by establishing connections between the revealed and hidden equilibria. Specifically, we use the equilibrium condition that, at the revealed equilibrium, following the revealed strategy achieves a higher (or equal) utility for each individual student than following the hidden strategy.

\paragraph{Step 1: Showing $\swr = \Omega(1)$.}
Recall from Appendix~\ref{app:prelim_revealed} that $\flowr_{\lottery\idxschool}=\sum_{\idxtype} \prior_\idxtype \pi^\parenr_{\idxtype\lottery\idxschool}$ is the probability that the student \withlnum $r$ applies to school $j$ under the revealed equilibrium $\strategy^\parenr$, and $\probavailr_{\lottery\idxschool}$ is their admission probability for school $\idxschool$.
We fix any type $t\in [\numtypes]$. At the revealed equilibrium, a type-$\idxtype$ student \withlnum $\lottery$ has expected utility~\eqref{eq:util_revealed_max}:
\begin{align*}
    \utility^\parenr_{\idxtype r}= \max_\idxschool v_{\idxtype\idxschool}\probavailr_{\lottery\idxschool}.
\end{align*}
We define the cumulative expected utility  over lottery numbers $1$ to $\lottery$, contributed by type-$\idxtype$ students: \begin{align*}
    U_{r}(\idxtype) \defn \sum_{\ell=1}^r u_{t\ell}^\parenr,
\end{align*}
where we define $U_0(t) = 0$. Then we have $\swr = \sum_{t\in [T]} p_t\cdot U_{n}(\idxtype)$. We now derive a lower bound on $U_{r}(\idxtype)$ by tracking the contribution of each term $\utility^\parenr_{\idxtype\lottery}$. At a high level, if the student \withlnum $r$  has a low expected utility, the admission probabilities $\{\probavailr_{\lottery\idxschool}\}_{\idxschool}$ must be low in general, and consequently students with better \lnums must have already taken these schools and have received high expected utilities. We formalize this intuition by establishing a recursive relation on $U_r(t)$.

We first define an intermediate quantity that measures the expected utility for a type-$t$ student \withlnum $r$ to apply to a school uniformly at random (while all other students follow their equilibrium strategies):
\begin{align*}
    \utilunif_{r}(\idxtype) \defn \frac{1}{\numschools}\sum_\idxschool \pref_{tj}\probavailr_{\lottery\idxschool},
\end{align*}
and hence $\utilunif_{1}(\idxtype) = \frac{1}{n} \sum_{\idxschool} v_{tj} = \frac{1}{n}$.
By definition, we have $u_{tr}^\parenr \ge \utilunif_{r}(\idxtype)$ for every $\lottery\in [\numstudents]$. 
We also have
\begin{align}
    \utilunif_{r}(\idxtype) - \utilunif_{r+1}(\idxtype) &= \frac{1}{n} \sum_j  v_{tj} (\probavailr_{r\idxschool} - \probavailr_{r+1,j})\nonumber\\
    & \stackrel{\stepone}{=} \frac{1}{n} \sum_\idxschool v_{\idxtype\idxschool} \probavailr_{rj}\flowr_{r\idxschool}\nonumber\\
    & \le \frac{1}{n} u_{tr}^\parenr \cdot \sum_j \flowr_{\lottery j} \nonumber\\
    &\stackrel{\steptwo}{=} \frac{u_{tr}^\parenr}{n},\label{eq:util_unif_diff}
\end{align}
where step~\stepone is true by the recursive relation~\eqref{eq:prob_reveal_recursive}, and step~\steptwo is true because  $\sum_j \flowr_{r\idxschool} = 1$.
Applying~\eqref{eq:util_unif_diff} recursively, we have
\begin{align*}
    \utilunif_{r}(\idxtype) & \ge \utilunif_{r-1}(\idxtype) - \frac{u_{t,r-1}^\parenr}{n}\\
    & \ge \cdots \\
    &\ge \utilunif_{1}(\idxtype) - \frac{1}{n} \sum_{\ell = 1}^{r-1}u_{\ell}(\idxtype) \\
    &= \frac{1}{n} - \frac{U_{r-1}(t)}{n}.
\end{align*}
Therefore,
\begin{align*}
    u_{tr}^\parenr \ge \utilunif_{r}(\idxtype) & \ge \frac{1}{n} - \frac{U_{r-1}(t)}{n},
\end{align*}
In words, if students with lottery numbers $1$ through $(\lottery-1)$ have not extracted a high cumulative utility $U_{r-1}(t)$, then the admission probabilities of schools remain high, from which the student with lottery number $\lottery$ is be able to extract a large utility.
We thus have
\begin{align*}
    U_r(t) &= U_{r-1}(t) + u_r(t) \nonumber\\
    & \ge U_{r-1}(t) + \frac{1}{n} - \frac{U_{r-1}(t)}{n} \nonumber\\
    &= \left(1-\frac{1}{n}\right)U_{r-1}(t) + \frac{1}{n}\\
    & \ge \cdots \\
    &\ge \left(1-\frac{1}{n}\right)^\lottery U_0(t) + \frac{1}{n}\cdot \sum_{\ell=0}^{\lottery-1}\left(1-\frac{1}{n}\right)^{\ell}\\
    & \stackrel{\stepone}{=}\frac{1}{n}\cdot \sum_{\ell=0}^{\lottery-1}\left(1-\frac{1}{n}\right)^{\ell}\\
    & = 1 - \left(1-\frac{1}{n}\right)^\lottery,
\end{align*}
where step~\stepone is true because $U_0(t) = 0$ by definition. Therefore,
\begin{align}
    \swr =\sum_{t\in [T]} p_t \cdot U_n(t) 
    \ge 1 - \left(1-\frac{1}{n}\right)^n > 1-\frac{1}{e},\label{eq:ratio_sw_hr_step_one}
\end{align}
completing the proof that $\swr = \Omega(1)$.

\paragraph{Step 2: Showing $(\swh)^2 \le 4n\cdot \swr$.}

Note that\begin{align*}
    4n\cdot \swr >4n\left(1-\frac{1}{e}\right)  > n,
\end{align*}
so the desired inequality holds trivially if $\swh \le \sqrt{n}$. It remains to consider the case when $\swh > \sqrt{n}$.

As an overview of Step 2, we derive the desired relation between $\swh$ and $\swr$ by applying the equilibrium condition that, under the revealed lottery, following the revealed strategies is better than following the hidden strategies for students. We start by analyzing the hidden social welfare $\swh$.

Recall that $\flowh_j = \sum_t p_t \pi^\parenh_{tj}$ is the probability that a student applies to school $\idxschool$ under the hidden equilibrium $\strategy^\parenh$. We consider the expected utility contributed by school $\idxschool$ to a random student, conditional on school $\idxschool$ being available:
\begin{align*}
    w_j \defn \sum_t p_t \pi^\parenh_{tj} \pref_{tj}.
\end{align*}
The social welfare under the hidden equilibrium can be written as
\begin{align}
    \swh &= \numstudents\sum_t p_t \utility_\idxtype^\parenh \nonumber\\
    &=\numstudents\sum_t p_t \sum_j \pi_{tj}^\parenh \probavailh_j \pref_{tj} \nonumber\\
    &= \numstudents \sum_j \probavailh_j w_j,\label{eq:sw_ratio_hr_step_two_sw_h}
\end{align}
where $\probavailh_j w_j$ is the expected utility contributed by school $j$ to a random student. We have
\begin{align}
   \probavailh_j w_j \stackrel{\stepone}{\le} \frac{w_j}{n\flowh_j} = \frac{1}{n}\cdot \frac{\sum_t p_t \pi_{tj}^\parenh\pref_{tj}}{\sum_{t} p_t \pi_{tj}^\parenh} \le \frac{1}{n},\label{eq:ratio_hr_prodct_alpha_w_bound}
\end{align}
where step~\stepone is because $\probavailh_j \le \frac{1}{n\flowh_j}$ by~\eqref{eq:prob_avail_expression}. In words, each school $\idxschool$ contributes at most $1$ to the social welfare, so it contributes at most $\frac{1}{n}$ in expectation to an individual student. 

We now provide a lower bound on social welfare $\swr$ under the revealed equilibrium.
Recall that $\probavailr_{r\idxschool}$ is the admission probability that school $j$ is available to the student \withlnum $r$ under the revealed equilibrium. We take the mean admission probability of $\{\probavailr_{r\idxschool}\}_{\lottery}$ over the lottery number $\lottery$ and define
\begin{align}
    \probavailrmean_j \defn \frac{1}{n}\sum_{r=1}^n \probavailr_{r\idxschool}.\label{def:beta_defn}
\end{align}
 For a type-$t$ student \withlnum $\lottery$, following their revealed strategy $\strategy^\parenr$ achieves a higher or equal expected utility than following their hidden strategy $\strategy^\parenh$, while all other students follow their revealed strategies:
\begin{align*}
    u^\parenr_{tr}\ge \sum_j \pi_{tj}^\parenh v_{tj} \probavailr_{r\idxschool}
\end{align*}
and hence
\begin{align}
    \swr &= \sum_{r\in [n]} \sum_{t\in [\numtypes]} p_t u^\parenr_{tr} \nonumber\\
    &\ge \sum_r \sum_t p_t \sum_j \pi_{tj}^\parenh v_{tj} \probavailr_{r\idxschool}\nonumber\\
    & = n\sum_t p_t \sum_j \pi_{tj}^\parenh v_{tj}  \probavailrmean_j\nonumber\\
    & = n\sum_j \probavailrmean_j\sum_t p_t  \pi_{tj}^\parenh v_{tj}  \nonumber\\
    & = n\sum_j \probavailrmean_j w_j \nonumber\\
    & \ge \numstudents\sum_j \probavailrmean_j w_j\probavailh_j.\label{eq:ratio_sw_hr_bound_sw_reveal}
\end{align}
At a high level, we want to show that if $\swh$ is large, then $\swr$ cannot be too small. We write $\swh$ as a function of $w_j\probavailh_j$ in~\eqref{eq:sw_ratio_hr_step_two_sw_h}. If  $\swh$ is large, then inequality~\eqref{eq:ratio_hr_prodct_alpha_w_bound} suggests that many schools contribute to $\swh$. In what follows, we show that these schools must also contribute to $\swr$ according to~\eqref{eq:ratio_sw_hr_bound_sw_reveal}. To do this, we lower bound the mean probabilities $\{\probavailrmean_j\}_j$ via the following lower bound on $\{\probavailr_{r\idxschool}\}_{r\idxschool}$.
\begin{lemma}\label{lem:sum_probavailrank}
    For any \lnum $\lottery\in [\numstudents]$, for any set $J\subseteq [\numschools]$, we have
    \begin{align}
        \sum_{\idxschool\in J} \probavailr_{\lottery\idxschool} \ge \abs*{J} -r + 1.\label{eq:prob_reveal_sum_lb}
    \end{align}
\end{lemma}
The proof of this lemma is provided at the end of this section. In words, since each student only takes at most one seat, the admission probabilities do not decrease too fast in the lottery number $\lottery$.

Without loss of generality, we re-index the schools such that $\probavailrmean_1 \le \cdots \le \probavailrmean_\numschools$. From~\eqref{eq:ratio_sw_hr_bound_sw_reveal}, we provide a lower bound on $\swr$:
\begin{align}
    \swr &\ge n \sum_{j=1}^{\numschools} \probavailrmean_{j}w_j \probavailh_\idxschool \nonumber\\
    &\stackrel{\stepone}{\ge} \sum_{j=1}^{\floor{\swh}}  \probavailrmean_j \nonumber\\
    &\stackrel{\steptwo}{=} \frac{1}{n}\sum_{j=1}^{\floor{\swh}} \sum_{r=1}^n \probavailr_{r\idxschool} \nonumber\\
    & =\frac{1}{n} \sum_{r=1}^n \sum_{j=1}^{\floor{\swh}} \probavailr_{r\idxschool} \nonumber\\
    &\stackrel{\stepthree}{\ge} \frac{1}{n}\sum_{r=1}^n \max\left\{\floor{\swh}-r+1, 0\right\} \nonumber\\
    &= \frac{1}{n}\cdot \frac{\floor{\swh}(\floor{\swh}+1)}{2} \nonumber\\
    &\stackrel{\stepfour}{\ge} \frac{(\swh)^2}{4n},\nonumber
\end{align}
where step~\stepone is true due to~\eqref{eq:sw_ratio_hr_step_two_sw_h} and~\eqref{eq:ratio_hr_prodct_alpha_w_bound}; step~\steptwo is true by definition~\eqref{def:beta_defn} of $\probavailrmean_j$; step~\stepthree is true by invoking \Cref{lem:sum_probavailrank}; and step~\stepfour is true by the assumption that $\swh > \sqrt{n} >1$. Therefore,
\begin{align}
    4n\cdot \swr \ge (\swh)^2,\label{eq:ratio_sw_hr_step_two}
\end{align}
completing the proof of Step 2.
~\\

Finally, combining~\eqref{eq:ratio_sw_hr_step_two} from Step 2 with inequality~\eqref{eq:ratio_sw_hr_step_one} from Step 1 yields that in every economy,
\begin{align*}
    (\swh)^2 & \le 4n\cdot \swr \lessorder n \cdot (\swr)^2 \\
    \frac{\swh}{\swr} &\lessorder \sqrt{n},
\end{align*}
completing the proof that $\ratioswhr=O(\sqrt{n})$.

\paragraph{Proof of Lemma~\ref{lem:sum_probavailrank}.}

Fix any subset $J \subseteq [\numschools]$. We prove by induction on the lottery number $\lottery\in [\numstudents]$. For the base case of $\lottery=1$, we have $\probavailr_{j1} = 1$ for every $\idxschool\in [\numschools]$ and hence
\begin{align*}
    \sum_{j\in J} \probavailr_{j1} = \abs*{J},
\end{align*}
satisfying~\eqref{eq:prob_reveal_sum_lb}. Now assume that the desired inequality~\eqref{eq:prob_reveal_sum_lb} holds for some lottery number $\lottery$, that is,
\begin{align}
        \sum_{\idxschool\in J} \probavailr_{\lottery\idxschool} \ge \abs*{J} - r + 1.\label{eq:prob_reveal_sum_lb_repeat}
\end{align}
Recall that $\flowr_{\lottery j}$ is the probability that a student with lottery number $\lottery$ applies to school $j$.  Given the recursive relation~\eqref{eq:prob_reveal_recursive} that $ \probavail^\parenr_{r+1,j} = \probavail^\parenr_{rj} \cdot (1-q^\parenr_{rj})$, we have
\begin{align*}
    \sum_{j\in J} \probavailr_{r+1, j} &= \sum_{j\in J} \probavailr_{r\idxschool} (1-\flowr_{\lottery j}) \\
    &= \sum_{j\in J} \probavailr_{r\idxschool} - \sum_{j\in J} \probavailr_{r\idxschool}\flowr_{r\idxschool}\\
    &\stackrel{\stepone}{\ge} (\abs*{J} -r+1) - \sum_{j\in J} \flowr_{rj} \stackrel{\steptwo}{\ge} \abs*{J} - r,
\end{align*}
where step~\stepone is true by the inductive hypothesis~\eqref{eq:prob_reveal_sum_lb_repeat}, and step~\steptwo is true because $\sum_{\idxschool=1}^\numschools \flowr_{r\idxschool} = 1$, completing the proof of \Cref{lem:sum_probavailrank}.

\subsubsection{Proof of bound~\texorpdfstring{\eqref{eq:ratio_sw_rh_hetero}}{(\ref{eq:ratio_sw_rh_hetero})}}

We prove that $\ratioswrh = O(\log^2 n)$ under heterogeneous rankings. The proof consists of two steps. In the first step, we consider the ratio of admission probabilities $\big\{\frac{\probavailr_{\lottery\idxschool}}{\probavailh_{j}}\big\}$ of each school $\idxschool$ under revealed and hidden lotteries. If a student with lottery number $\lottery$ has a much higher expected utility under the revealed lottery, then the admission probability $\probavailr_{\lottery\idxschool}$ must be higher than $\probavailh_{j}$ for some $\idxschool$. This allows us to express $\frac{\swr}{\swh}$ in terms of ratios $\big\{\frac{\probavailr_{\lottery\idxschool}}{\probavailh_{j}}\big\}$. In the second step, we provide upper bounds for these ratios.

\noindent\textbf{Step 1: Writing  $\frac{\swr}{\swh}$ in terms of probability ratios $\big\{\frac{\probavailr_{\lottery\idxschool}}{\probavailh_{j}}\big\}$.}
Recall from Appendix~\ref{app:prelim_hidden} that under the hidden equilibrium, the expected utility of a type-$t$ student is $u_t^\parenh= \max_j v_{tj}\probavailh_j$ and the social welfare~\eqref{eq:sw_hidden} is
\begin{align*}
    \swh = n\sum_t p_tu_t^\parenh.
\end{align*}
Also recall from Appendix~\ref{app:prelim_revealed} that under the revealed equilibrium, $\probavailr_{r\idxschool}$ is the admission probability of school $j$ for the student \withlnum $\lottery$.   The expected utility of a type-$\idxtype$ student \withlnum $\lottery$ is:
\begin{align}
    u^\parenr_{tr} = \max_j v_{tj} \probavailr_{\lottery\idxschool}.\label{eq:ratio_sw_rh_utility_max}
\end{align}
Denoting $\ratioprob_r\defn \max_{\ell\in [\numstudents]} \frac{\probavailr_{\lottery \ell}}{\probavailh_\ell}$, we have that for every $\idxschool\in [\numstudents]$,
\begin{align}
    v_{tj} \probavailr_{\lottery\idxschool} = (v_{tj}\probavailh_j)\cdot \frac{\probavailr_{\lottery\idxschool}}{\probavailh_j} \le u_t^\parenh \cdot \max_\ell \frac{\probavailr_{r \ell }}{\probavailh_\ell} = \utility_t^\parenh \ratioprob_\lottery.\label{eq:ratio_sw_rh_expected_utility_expression}
\end{align}
The social welfare under the revealed lottery is
\begin{align*}
    \swr &= \sum_r \sum_t p_t\utility_{tr}^\parenr \\
    & \stackrel{\stepone}{=} \sum_r \sum_t p_t\cdot \max_j v_{tj} \probavailr_{r\idxschool}\\
    & \stackrel{\steptwo}{\le} \sum_r \sum_t p_t u_t^\parenh \ratioprob_r\\
    & = \sum_r \ratioprob_r\sum_t p_t u_t^\parenh \\
    & = \left(\sum_{r=1}^n \ratioprob_r\right)\cdot \frac{\swh}{n},
\end{align*}
where step~\stepone is true by~\eqref{eq:ratio_sw_rh_utility_max}, and step~\steptwo is true by~\eqref{eq:ratio_sw_rh_expected_utility_expression}. Hence, we have
\begin{align*}
     \frac{\swr}{\swh} &\le \frac{1}{n} \sum_{r=1}^n \ratioprob_r.
\end{align*}
It remains to show that 
\begin{align}
    \sum_{r=1}^n \ratioprob_r = O(n\log^2 n).\label{eq:ratio_sw_rh_different_rankings_remains}
\end{align}

\paragraph{Step 2: Bounding probability ratios $\{\ratioprob_\lottery\}_\lottery$.}

At a high level, for each lottery number $\lottery$, we consider the set of schools $\idxschool$ whose probability ratio $\frac{\probavailr_{\lottery\idxschool}}{\probavailh_j}$ is ``large''. Then the student with \lnum $\lottery$ is inclined to applying to one (or some) of these schools under the revealed lottery. This decreases the admission probability $\probavailr_{\lottery+1, \idxschool}$ for the next \lnum $(\lottery+1)$, and thus decreases the ratio $\frac{\probavailr_{r+1,j}}{\probavailh_j}$. This argument suggests that probability ratios $\big\{\frac{\probavailr_{\lottery\idxschool}}{\probavailh_j}\big\}_j$ cannot stay large throughout many \lnums, implying an upper bound. The proof formalizes this intuition.

\begin{figure}[tb]
    \centering
    \subfloat[]{\includegraphics[width=0.45\linewidth]{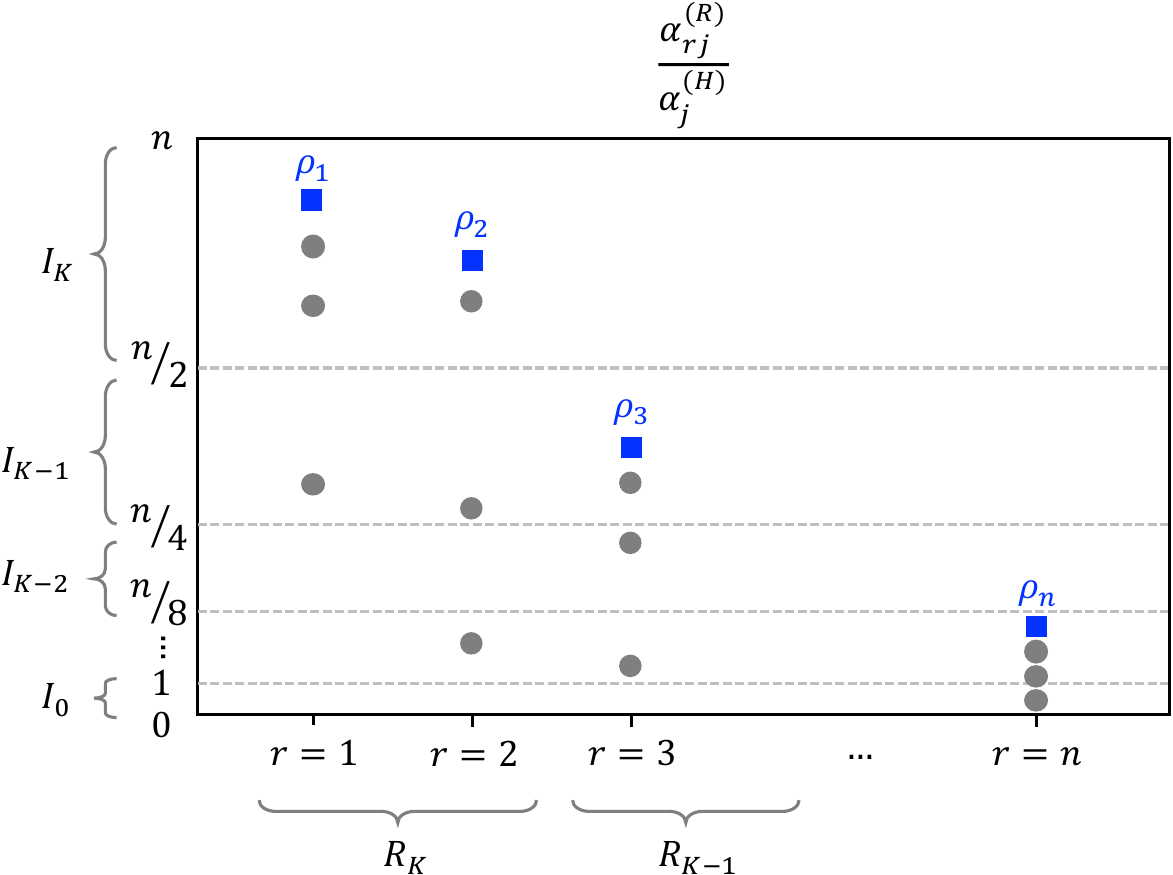}\label{float:proof_interval}
    }~~~~~~~
    \subfloat[]{
        \includegraphics[width=0.45\linewidth]{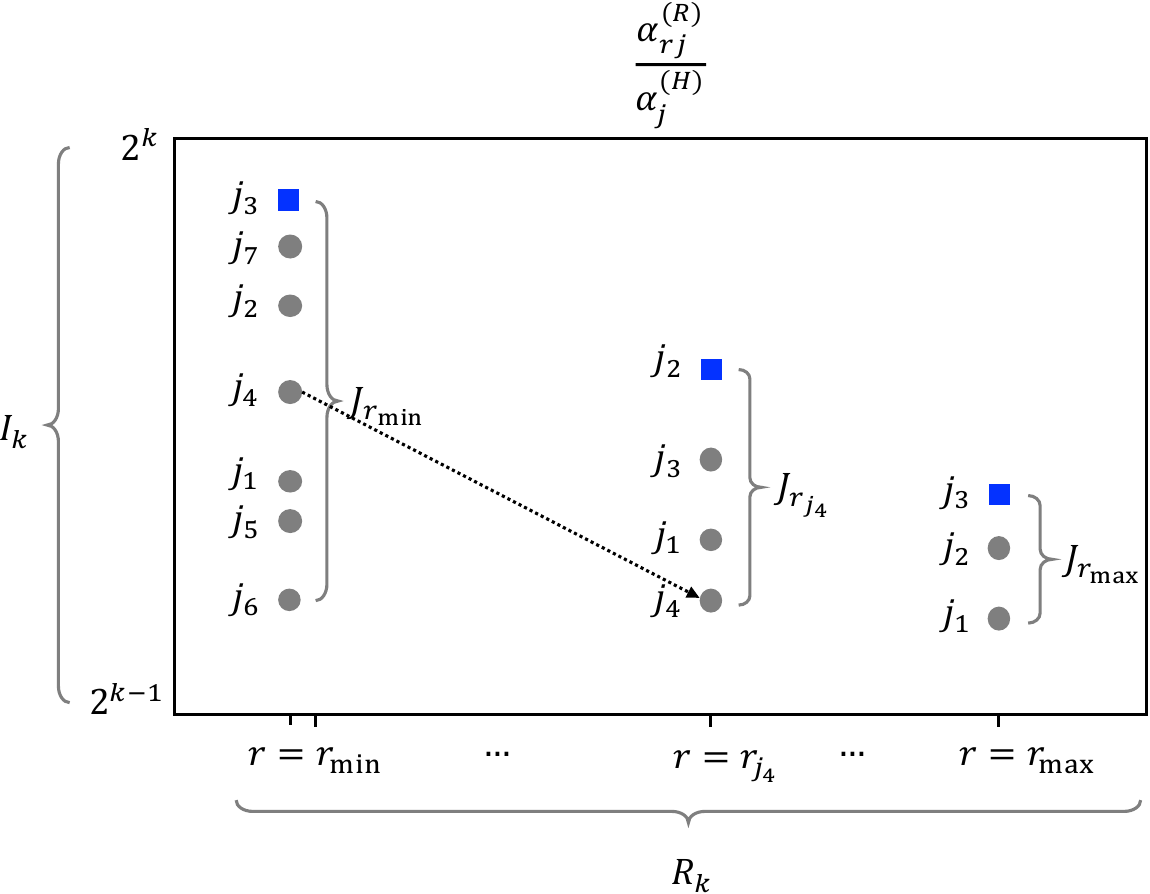}\label{float:proof_interval_block}
    }
    \caption{
        Visualization for the proof of bound~\eqref{eq:ratio_sw_rh_hetero}:
        (a) Gray circles on the same vertical line represent the probability ratios $\big\{\frac{\probavailr_{r\idxschool}}{\probavailh_j}\big\}_{j\in[\numschools]}$ for a fixed \lnum  $r$, where the blue square corresponds to their maximum  $\ratioprob_r=\max_{j} \frac{\probavailr_{r\idxschool}}{\probavailh_j}$. These probability ratios lie in  $[0, n]$, covered by intervals $I_0\union \cdots\union I_K$. Each $R_k$ is the set of \lnums $\lottery$ such that value $\ratioprob_r$ lies in interval $I_k$.
        (b) For a fixed $k$, set $R_k$ covers consecutive lottery numbers from $\lotterymin$ to $\lotterymax$. For each lottery number $r\in \{\lotterymin, \ldots, \lotterymax\}$, set $J_r$ consists of all schools whose gray circle lies in $I_k$. We re-index the schools such that each set $J_r$ contains schools $j_1$ through $j_{\abs*{J_r}}$. We consider how the gray circle corresponding to some school $j$ decreases as a function of $r$, which may eventually go below interval $I_k$. For example, for school $j_4$, it appears from lottery number $\lotterymin$ through $\lottery_{\idxschool_4}$.
    }
    \label{fig:proof_interval}
\end{figure}

Given that $\probavailr_{\lottery\idxschool}$ is non-increasing in $\lottery$, the vector $\big\{\frac{\probavailr_{\lottery\idxschool}}{\probavailh_j}\big\}_j$ dominates the vector  $\big\{\frac{\probavailr_{r+1,j}}{\probavailh_j} \big\}_j$. Hence, we have $\ratioprob_1 \ge \cdots \ge \ratioprob_n$. 
A \naive upper bound for $\ratioprob_r$ is:
\begin{align*}
    \ratioprob_r = \max_{j} \frac{\probavailr_{\lottery\idxschool}}{\probavailh_j} \le \max_j \frac{1}{\probavailh_j}\stackrel{\stepone}{\le} n,
\end{align*}
where step~\stepone is true because $\probavailh_\idxschool \ge \frac{1}{\numstudents}$ for every $\idxschool$ under the hidden lottery. We construct intervals
\begin{align*}
    & I_0 = (0, 1]\\
    & I_k = (2^{k-1}, 2^k] \quad \text{ for }k\in 1, \ldots, \ceil*{\log_2(n)},
\end{align*}
so that the union of these non-overlapping intervals covers $(0, n]$. Each positive-valued $\ratioprob_r$ belongs to exactly one interval $I_k$, and we upper bound the sum $\sum_r \ratioprob_r$ by counting how many values of $\ratioprob_r$ lie in each interval. Formally, we denote by $R_k\defn \{r\in [\numstudents]: \ratioprob_r \in I_k\}$ the set of lottery numbers whose value of $\ratioprob_r$ lies in the interval $I_k$. 
The definitions of interval $I_k$ and set $R_k$ are visualized in \Cref{fig:proof_interval}\subref{float:proof_interval}.
We have
\begin{align}
    \sum_{r=1}^n \ratioprob_r & = \sum_{k=0}^{\ceil{\log_2(n)}} \sum_{r\in R_k} \ratioprob_r\cdot \indicator\{\ratioprob_r \in I_k\}\nonumber\\
    &\le \sum_{k=0}^{\ceil{\log_2(n)}} \sum_{r=1}^n 2^k \cdot \indicator\{\ratioprob_r \in I_k\} = \sum_{k=0}^{\ceil{\log_2(n)}} \underbrace{2^k \cdot \abs*{R_k}}_{\term_k}.\label{eq:ratio_sum_ratio_decompose}
\end{align}
We proceed by bounding each term $T_k$.

\noindent\textbf{Bounding $T_0$ and $T_1$.} With $k=0$ or $1$, we have
\begin{subequations}\label{eq:ratio_T0_bound}
\begin{align}
    \term_0 &= 2^0\cdot \abs*{R_0} \le n\\
    \term_1 &= 2^1 \cdot \abs*{R_1} \le 2n.
\end{align}
\end{subequations}

\noindent\textbf{Bounding $T_k$ for $k\ge 2$.} We fix any $k\ge 2$ and provide an upper bound on $\abs*{R_k}$. Since $\{\ratioprob_\lottery\}_\lottery$ is non-increasing, the set $R_k$ consists of consecutive integers, which we denote by $R_k = \{r_\textmin, \ldots, r_\textmax\}$.
For each \lnum $r\in R_k$, we define the subset of schools whose ratio $\frac{\probavailr_{\lottery \idxschool}}{\probavailh_j}$ falls within interval $I_k$:
\begin{align*}
    J_r = \left\{j\in [\numstudents]: \frac{ \probavailr_{\lottery\idxschool}}{\probavailh_j}\in I_k\right\} 
    = \left\{j\in [\numstudents]: 2^{k-1} < \frac{ \probavailr_{\lottery\idxschool}}{\probavailh_j}\le 2^{k}\right\}.
\end{align*}
In Figure~\ref{fig:proof_interval}, gray circles on the same vertical line represent the ratios $\big\{\frac{\probavailr_{\lottery\idxschool}}{\probavailh_j}\big\}_j$ corresponding to a lottery number $r$; the blue square represents the maximum $\ratioprob_r= \max_j \frac{\probavailr_{\lottery\idxschool}}{\probavailh_j}$. In Figure~\ref{fig:proof_interval}\subref{float:proof_interval_block}, set $R_k$ consists of lottery numbers whose blue square lies in interval $I_k$. For each lottery number $\lottery\in R_k$, set $J_r$ consists of schools $\idxschool$ whose gray circle lies in interval $I_k$. At a high level, gray circles in $J_r$ imply that these schools have a relatively ``high'' admission probability under the revealed lottery compared to the hidden lottery, and thus the students with subsequent lottery numbers are inclined to apply to these schools, decreasing the values of these gray circles. Hence, these schools cannot stay in $J_r$ across many lottery numbers $r$, which implies that $\abs*{R_k}$ cannot be very large.

Recall that $q_j^\parenh$ is the probability that a student applies to school $\idxschool$ under the hidden equilibrium. For each $j\in J_r$, we have
\begin{align*}
    2^{k-1}< \frac{\probavailr_{\lottery\idxschool}}{\probavailh_j} \le \frac{1}{\probavailh_j} \stackrel{\stepone}{\le} 1 + (n-1)q_j^\parenh,
\end{align*}
where step~\stepone is true by~\eqref{eq:prob_avail_single_lb} from Appendix~\ref{app:prelim_hidden}. Therefore, for each $\idxschool\in J_r$,
\begin{align*}
    q_j^\parenh \ge \frac{2^{k-1}-1}{n-1} \ge \frac{2^{k-2}}{n}.
\end{align*}
We define the probability that a student applies to schools $J_r$ under the hidden equilibrium:
\begin{align*}
    \flowh_{J_r} \defn \sum_{j\in J_r} q_j^\parenh \ge \abs*{J_r}\cdot \frac{2^{k-2}}{n}.
\end{align*}
We thus have
\begin{align}
    \abs*{R_k} = \sum_{r\in R_k} 1 \le \frac{n}{2^{k-2}} \sum_{\lottery\in R_k} \frac{\flowh_{J_r}}{\abs*{J_r}}.\label{eq:Rk_flow_hidden_Jr}
\end{align}
We also define
\begin{align*}
    \flowr_{r, J_r} & \defn \sum_{j\in J_r} \flowr_{\lottery\idxschool}.
\end{align*}
Invoking \Cref{lem:flow_set_compare} establishes that 
\begin{align}
        \flowr_{r, J_r} \ge \flowh_{J_r}.\label{eq:flow_subset_inequality}
\end{align}
Specifically, since $r\in R_k$, we have  \begin{align*}
    \frac{\probavailr_{\lottery\idxschool}}{\probavailh_\idxschool} \le 2^k \quad \text{for every }\idxschool\in [n],
\end{align*}
and
\begin{align*}
    2^{k-1} < \frac{\probavailr_{\lottery\idxschool}}{\probavailh_\idxschool} \le 2^k\quad \text{for every }\idxschool\in J_r.
\end{align*}
Invoking \Cref{lem:flow_set_compare} with $c = 2^{k-1}$ yields~\eqref{eq:flow_subset_inequality}.

Plugging~\eqref{eq:flow_subset_inequality} to~\eqref{eq:Rk_flow_hidden_Jr}, we have
\begin{align}
    \abs*{R_k} &\le \frac{n}{2^{k-2}}\sum_{r\in R_k} \frac{\flowr_{ r,J_r}}{\abs*{J_r}}. \label{eq:R_k_expression}
\end{align}
It remains to bound the term $\sum_{r\in R_k}\frac{\flow_{r,J_r}}{\abs*{J_r}}$.  The set $J_r$ shrinks in lottery number $r$. That is, for any $r, r'\in R_k$ with $r < r'$, we have
\begin{align*}
    J_r \supseteq J_{r'},
\end{align*}
as visualized in Figure~\ref{fig:proof_interval}\subref{float:proof_interval_block}. We order and index the schools as
\begin{align*}
    \underbrace{
    \begingroup
    \color{white}
    \underbrace{\color{black}\underbrace{\underbrace{j_1, j_2, \ldots}_{J_{\lotterymax}},  \ldots}_{J_{\lotterymax-1}}}_{\color{black}\vdots}
    \endgroup, \ldots}_{J_\lotterymin}, \ldots, j_n,
\end{align*}
so that for every $r\in R_k$, we have
\begin{align}
    J_r =\{ j_1, \ldots, j_{\abs*{J_r}}\}.\label{eq:reindex_schools}
\end{align}
If more than one indexing satisfies~\eqref{eq:reindex_schools}, then any one of them suffices.
For each $j_\ell$, let 
\begin{align*}
    r_{j_\ell}\defn \max\{r\in [\numstudents]: j_\ell \in J_r\},
\end{align*}
be the latest lottery number such that $j_\ell\in J_r$,
where $\lotterymin\le \lottery_{j_\ell}\le \lotterymax$. Equivalent, we write
\begin{align*}
     \{r\in [\numstudents]: j_\ell \in J_r\}= \{\lotterymin, \ldots, \lottery_{j_\ell}\}.
\end{align*}
We have
\begin{align}
    \sum_{r\in R_k}\frac{\flowr_{r,J_r}}{\abs*{J_r}} &=
    \sum_{r\in R_k}\frac{\sum_{\idxschool\in J_r}\flowr_{\lottery\idxschool}}{\abs*{J_r}}\nonumber\\
    &= \sum_{\ell = 1}^{n} \sum_{r=\lotterymin}^{\lottery_{j_\ell}} \frac{\flowr_{\lottery,j_\ell}}{\abs*{J_r}} \stackrel{\stepone}{\le} \sum_{\ell=1}^n \frac{1}{\ell} \sum_{r=\lotterymin}^{\lottery_{j_\ell}} \flowr_{\lottery,\idxschool_\ell},\label{eq:ratio_fractional_term_expression}
\end{align}
where step~\stepone is true because $j_\ell\in J_r$, and hence by construction~\eqref{eq:reindex_schools}, schools $j_1, \ldots, j_{\ell-1}\in J_r$ and $\abs*{J_r} \ge \ell$.
For every $j_\ell$ and  $\lotterymin\le r\le r_{j_\ell}$, we have 
\begin{align}
    2^{k-1} < \frac{\probavailr_{\lottery,\idxschool_\ell}}{\probavailh_{j_\ell}} \le 2^k.\label{eq:recall_def_set_R_k}
\end{align}
By recursive relation~\eqref{eq:prob_reveal_recursive}, we have
\begin{align*}
    \probavailr_{r_{j_\ell}, j_\ell} &= \probavailr_{r_{j_\ell}-1, j_\ell}\cdot \left(1-\flowr_{r_{j_\ell}-1, j_\ell }\right)\\
    &= \cdots \\
    &= \probavailr_{\lotterymin,j_\ell} \prod_{r=\lotterymin}^{\lottery_{j_\ell}-1} \left(1-\flowr_{\lottery,\idxschool_\ell}\right)
\end{align*}
Therefore,
\begin{align*}
    \prod_{r=\lotterymin}^{\lottery_{j_\ell}-1}(1-\flowr_{\lottery,j_\ell}) = \frac{\probavailr_{ \lottery_{j_\ell},j_\ell}}{\probavailr_{\lotterymin,j_\ell}}\stackrel{\stepone}{\ge} \frac{1}{2},
\end{align*}
where step~\stepone is true by plugging $r=\lotterymin$ and $r = r_{j_\ell}$ to~\eqref{eq:recall_def_set_R_k}.
Invoking \Cref{lem:prod_sum} with $c =\frac{1}{2}$, we have
\begin{align*}
    \sum_{r=\lotterymin}^{r_{j_\ell}-1}{\flowr_{\lottery\idxschool}}\le \log 2,
\end{align*}
and hence
\begin{align}
    \sum_{r=\lotterymin}^{r_{j_\ell}}{\flowr_{\lottery\idxschool}}\le \log 2 +1.\label{eq:ratio_sum_q_revealed_ub}
\end{align}
Plugging~\eqref{eq:ratio_sum_q_revealed_ub} into~\eqref{eq:ratio_fractional_term_expression}, we have
\begin{align}
    \sum_{r\in R_k}\frac{\flowr_{r,J_r}}{\abs*{J_r}} \le \sum_{\ell=1}^n \frac{\log 2 + 1}{\ell} \eqorder \log n.\label{eq:ratio_fractional_expression_ub}
\end{align}
Plugging~\eqref{eq:ratio_fractional_expression_ub} back to~\eqref{eq:R_k_expression}, we have
\begin{align*}
    \abs*{R_k} \lessorder \frac{n\log n}{2^{k-2}}.
\end{align*}
and
\begin{align}
\term_k = 2^k \cdot \abs*{R_k} \lessorder n\log n.\label{eq:ratio_Tk_bound}
\end{align}~\\
Finally, plugging terms $T_0$ and $T_1$ from~\eqref{eq:ratio_T0_bound} and term $T_k$ from~\eqref{eq:ratio_Tk_bound} to~\eqref{eq:ratio_sum_ratio_decompose}, we have
\begin{align*}
    \sum_{r=1}^n \ratioprob_r \lessorder n\log^2 n,
\end{align*}
completing the proof of~\eqref{eq:ratio_sw_rh_different_rankings_remains}.

\subsection{Proof of Theorem~\ref{thm:SW-diff-gaps}}\label{app:proof_SW-diff-gaps}

For bounds~\eqref{eq:diff_sw_hr_homo} and~\eqref{eq:diff_sw_rh_homo} under homogeneous rankings, note from \Cref{lem:reveal_same_ranking_always_one} that $\swr = 1$. Hence, maximizing the social welfare difference $\diffswhr$ (resp. $\diffswrh$) is equivalent to maximizing the ratio $\ratioswhr$ (resp. $\ratioswrh$). From \Cref{thm:SW-gaps}\ref{item:sw_ratio_homo}, we have
\begin{align*}
    \ratioswhrhomo &= \Theta(\log n)\\
    \ratioswrhhomo &= 2-\frac{1}{n} = \frac{2n-1}{n}.
\end{align*}
Therefore,
\begin{align*}
    \diffswhrhomo &= \Theta(\log n)\\
        \diffswrhhomo &= 1-\frac{n}{2n-1}= \frac{n-1}{2n-1}.
\end{align*}
It remains to prove bounds~\eqref{eq:diff_sw_hr_hetero} and~\eqref{eq:diff_sw_rh_hetero} under heterogeneous rankings.

\subsubsection{Proof of bound~\texorpdfstring{\eqref{eq:diff_sw_hr_hetero}}{(\ref{eq:diff_sw_hr_hetero})}}\label{app:proof_sw_diff_hr_different_rankings}

We prove that $\diffswhr=\Theta(n)$ under heterogeneous rankings.
Given the trivial upper bound $\diffswhr = O(n)$, we prove the lower bound $\diffswhr = \Omega(n)$. 

Consider the common-rare economy constructed in Appendix~\ref{app:prelim_common_rare} with $\numminority=\ceil*{\frac{\numstudents}{2}}$ rare types. 
For readability, we assume $n$ is even and hence $m=\frac{n}{2}$. The proof for odd $n$ is similar up to minor modifications.

Recall that we re-index schools and types to start from 0, where the common-type students have valuations:
\begin{align*}
    v_{0\idxschool} = \begin{cases}
        1-\frac{m}{n}=\frac{1}{2} & \text{ if }\idxschool = 0\\
        \frac{1}{\numschools} & \text{ if } 1 \le \idxschool \le \numminority\\
        0^+ & \text{ if }\numminority < \idxschool < \numschools.
    \end{cases}
\end{align*}
Students of each rare type $\idxtype\in [\numminority]$ have valuations
\begin{align*}
    v_{\idxtype j} = \begin{cases}
        1 & \text{if } j=\idxtype\\
        0^+ & \text{if } j \ne \idxtype
    \end{cases}
\end{align*}
with prior
\begin{align*}
    p_0 = 1-\frac{m}{n}=\frac{1}{2}, \quad \text{ and }\quad p_\idxtype = \frac{1}{n} \quad \text{ for every }\idxtype\in [\numminority].
\end{align*}
By Lemma~\ref{lem:common_rare_equil_hidden}, under the hidden lottery, it is an equilibrium for each type to apply to their preferred school. The remaining proof consists of three steps. First, we characterize the equilibrium strategies under the revealed equilibrium. Second, we show that under the revealed lottery, some common-type students with worse \lnums apply to rare schools, because they know that their preferred common school is likely taken by a common-type student with a better lottery number. These common-type students ``block'' rare-type students from being admitted to these rare schools. Finally, we show that the number of ``blocking'' common-type students is $\Theta(n)$ in expectation, hence incurring a loss of $\Theta(n)$ in social welfare under the revealed lottery compared to the hidden lottery.

\paragraph{Step 1: Characterizing the strategies under the revealed lottery.}

We first observe that students of each rare type $\idxtype$ apply to their preferred school $\idxtype$ at equilibrium. For common-type students, we iteratively compute their strategies with each \lnum $\lottery=1, 2, \ldots, n$ in Algorithm~\ref{algo:compute_revealed_strategies_cmmon_type}, where we assume students break ties arbitrarily.
\begin{algorithm}[htb]
    \KwOut{Equilibrium strategies $\{a_\lottery\}_{\lottery\in [\numstudents]}$}
    \DontPrintSemicolon
    Set $n_0 = n_1 = \ldots = n_\numminority = 0$.\;
    \For{$r=1, \ldots, n$}{
        Compute $\utility(0) = v_{00}\cdot \frac{1}{2^{n_0}}$\;
        Compute $\utility(\idxtype) = v_{0t}\cdot\left(\frac{1}{2} - \frac{1}{n}\right)^{n_\idxtype} \cdot \left(1-\frac{1}{n}\right)^{r-1-n_\idxtype}$ for each $\idxtype\in [\numminority]$\;
        Compute $a_r = \argmax_{0 \le \idxschool \le \numminority} \utility(\idxschool)$, where ties are broken arbitrarily\;
        Set $n_{a_\lottery}\leftarrow n_{a_\lottery} + 1$\;
    }
    \caption{Determine the school $a_r$ that a common-type student with each \lnum $\lottery \in [\numstudents]$ applies to under the revealed equilibrium.\label{algo:compute_revealed_strategies_cmmon_type}}
\end{algorithm}

In words, for each \lnum $\lottery$, Algorithm~\ref{algo:compute_revealed_strategies_cmmon_type} keeps track of counts $\{n_t\}_{0\le t \le \numminority}$, denoting the count of better lottery numbers $1$ through $(\lottery-1)$ with which a common-type student would apply to school $\idxtype$. Note that the realized number of students with lottery numbers between $1$ and $(\lottery-1)$ who apply to school $\idxtype$ can be smaller than $n_t$, because some students may be rare types.
Based on these counts $\{n_\idxtype\}_\idxtype$, the algorithm computes the expected utility $\utility(\idxtype)$ for the student with \lnum $\lottery$ to apply to each school $\idxtype$. Specifically, school $0$ is available if and only if the $n_0$ lottery numbers, with which a common-type student would apply to school $0$, are taken by rare-type students. Thus, the expected utility for applying to school $0$ is:
\begin{align*}
    u(0) =v_{00}\cdot (1-p_0)^{n_0} = \pref_{00} \cdot \frac{1}{2^{n_0}}.
\end{align*}
A rare school $\idxtype\in [\numminority]$ is available if and only if there exist no type-$\idxtype$ students with better lottery numbers, and the $n_t$ lottery numbers, with which a common-type student would apply to school $t$, are taken by rare-type students that are not type $\idxtype$. Thus, the expected utility for applying to school $\idxtype\in [\numminority]$ is:
\begin{align*}
    \utility(\idxtype) &= \pref_{0\idxtype}\cdot (1-p_0-p_t)^{n_t}\cdot (1-p_t)^{r-1-n_t}\\
    &=v_{0t}\cdot \left(\frac{1}{2} - \frac{1}{n}\right)^{n_\idxtype} \cdot \left(1-\frac{1}{n}\right)^{\lottery-1-n_\idxtype}.
\end{align*}
Then the algorithm chooses school $a_r\defn \argmax_{0\le \idxschool \le \numminority} u(\idxschool)$ that maximizes this expected utility, under any arbitrary tie-breaking rule. By construction, under the revealed lottery, it is an equilibrium for the common-type student with \lnum $\lottery\in [\numstudents]$ to apply to school $a_r$.

\paragraph{Step 2: Showing that common-type students apply to rare schools.}
Now we show that $n_\idxtype \ge 1$ at the end of Algorithm~\ref{algo:compute_revealed_strategies_cmmon_type} for each rare school $\idxtype \in [\numminority]$. That is, there exists at least one lottery number with which a common-type student would apply to school $\idxtype$. Assume for contradiction that there exists a rare school $\idxtype^*$ with $n_{\idxtype^*}= 0$. Then each rare school $\idxtype\ne \idxtype^*$ can be selected at most once by \Cref{algo:compute_revealed_strategies_cmmon_type}. This is because if any rare school is selected twice, then consider the lottery number when \Cref{algo:compute_revealed_strategies_cmmon_type} selects this rare school the second time. With this lottery number, the admission probability of school $\idxtype^*$ is higher than school $\idxtype$, and \Cref{algo:compute_revealed_strategies_cmmon_type} would have strictly preferred school $\idxtype^*$ over school $\idxtype$. Hence, under the assumption that rare school $\idxtype^*$ is never selected, rare schools in total are selected by \Cref{algo:compute_revealed_strategies_cmmon_type} at most $(\numminority-1) = \frac{n}{2}-1$ times, and the common school is selected at least $(\frac{n}{2}+1)$ times. Consider the worst \lnum $r_\textmax$ with which \Cref{algo:compute_revealed_strategies_cmmon_type} selects the common school for a common-type student. We have $r_\textmax \ge \frac{n}{2}+1$. 

Now we consider the expected utility for a common-type student with \lnum $r_\textmax$ to apply to school $0$ or school $\idxtype^*$. We have
\begin{align*}
    \utility(0) &= v_{00}\cdot (1-p_0)^{n_0} \\
    &\le  v_{00}\cdot (1-p_0)^{n/2} = \frac{1}{2^{n/2+1}}.
\end{align*}
For school $\idxtype^*$, we have $n_{t^*}=0$ by assumption, and hence
\begin{align*}
    \utility(\idxtype^*) &= v_{0\idxtype^*} \cdot \left(1-p_{\idxtype^*}\right)^{r_\textmax -1} \ge \frac{1}{n} \cdot \left(1-\frac{1}{n}\right)^{n}\ge \frac{1}{4\numstudents} \stackrel{\stepone}{>}  \frac{1}{2^{n/2+1}} \ge u(0),
\end{align*}
where step~\stepone holds for any $n > 8$.
Therefore, we have $\utility(\idxtype^*) > u(0)$, contradicting the assumption that \Cref{algo:compute_revealed_strategies_cmmon_type} chooses school $0$ over school $\idxtype^*$ for the common-type student with lottery number $\lottery_\textmax$. Hence, \Cref{algo:compute_revealed_strategies_cmmon_type} selects each rare school $\idxtype$ at least once.

\paragraph{Step 3: Computing the social welfare difference under hidden vs. revealed lotteries.}

We analyze the utility difference at each school, under the hidden and revealed lotteries. First, the utility difference incurred at the common school $0$ is at most $1$. It remains to look at each rare school $\idxtype\in [\numminority]$. Note that under the hidden lottery, school $\idxtype$ can only be matched to a rare type-$\idxtype$ student. We consider the following two cases, depending on whether school $\idxtype$ is matched under the hidden lottery.

\paragraph{Case 1: School $\idxtype$ is unmatched under the hidden lottery. }  There must exist no type-$\idxtype$ students. If the utility differs at school $\idxtype$ under the two lotteries, then school $\idxtype$ must be matched to a common-type student under the revealed lottery. In this case, the utility differs by $v_{0\idxtype} =\frac{1}{n}$. Summing over all rare schools $\idxtype\in [\numminority]$ contributes at most $\numminority\cdot \frac{1}{n} =\frac{1}{2}$ more social welfare under the revealed lottery under Case 1.

\paragraph{Case 2: School $\idxtype$ is matched to a type-$\idxtype$ student under the hidden lottery.} Since there exists at least one type-$\idxtype$ student, school $\idxtype$ must be matched under the revealed lottery too, to either a type-$\idxtype$ student or a common-type student. The utility at school $\idxtype$ differs if school $\idxtype$ is matched to a common-type student under the revealed lottery. In this case, school $\idxtype$ achieves a utility $1$ under the hidden lottery, and $\frac{1}{n}$ under the revealed lottery, which causes the social welfare to be higher by $(1-\frac{1}{n})$ under the hidden lottery. Our goal is to show that this is incurred at a linear number of rare schools.

Recall from Step 2 that there exists at least one \lnum with which a common-type student would apply to school $\idxtype$. Denote by $\lottery_\idxtype$ the earliest such \lnum. A sufficient condition for the utility to differ under Case 2 is that the following events hold simultaneously:
\begin{enumerate}[label=($E_{\arabic*}$)]
    \item The student \withlnum $\lottery_\idxtype$ is a common-type student;
    \item There exists no type-$\idxtype$ student \withlnum strictly better than $r_\idxtype$;
    \item There exists at least one type-$\idxtype$ student with \lnum strictly worse than $r_\idxtype$.
\end{enumerate}
The first two events guarantee that school $\idxtype$ is matched to a common-type student under the revealed lottery, and the third event guarantees that school $\idxtype$ is matched to a type-$\idxtype$ student under the hidden lottery. Since the types are independent across students, the probability that these events hold simultaneously is:
\begin{align*}
    \Prob(E_1)\cdot \Prob(E_2) \cdot \Prob(E_3) &= p_0\cdot (1-p_t)^{r_\idxtype-1}\cdot [1-(1-p_t)^{n-r_\idxtype}]\\
    &= \frac{1}{2}\cdot \left(1-\frac{1}{n}\right)^{r_\idxtype-1} \cdot \left[1 -\left(1-\frac{1}{n}\right)^{n-r_\idxtype}\right] \\
    &\stackrel{\stepone}{>} \frac{1}{8}\cdot \left[1 -\left(1-\frac{1}{n}\right)^{n-r_\idxtype}\right],
\end{align*}
where step~\stepone is true due to inequality~\eqref{eq:inequality}.
Summing over all rare schools $\idxtype\in [\numminority]$ yields that Case 2 causes the social welfare to be higher in expectation under the hidden lottery than the revealed lottery by at least:
\begin{align*}
    \left(1-\frac{1}{\numschools}\right) \cdot \sum_{\idxtype=1}^{\numminority}  \frac{1}{8}\cdot \left[1 -\left(1-\frac{1}{n}\right)^{n-r_\idxtype}\right]
    &> \frac{1}{16}\cdot \sum_{\idxtype=1}^{\numminority} \left[1 -\left(1-\frac{1}{n}\right)^{n-r_\idxtype}\right]\\
    & \stackrel{\stepone}{\ge} \frac{1}{16}\cdot \sum_{\idxtype=0}^{\numminority-1} \left[1 -\left(1-\frac{1}{n}\right)^{\idxtype}\right]\\
    & \stackrel{\steptwo}{\ge} \frac{1}{16}\cdot \sum_{\idxtype=\ceil{n/4}}^{n/2-1} \left[1 -\left(1-\frac{1}{n}\right)^{\idxtype}\right]\\
    & \ge \frac{1}{16}\cdot \left(\frac{n}{2} - \ceil*{\frac{n}{4}}\right) \left[1 -\left(1-\frac{1}{n}\right)^{\frac{n}{4}}\right]\\
    & \stackrel{\stepthree}{\gtorder} n\cdot \left(1- \left(\frac{1}{e}\right)^{\frac{1}{4}}\right)\gtorder n,
\end{align*}
where step~\stepone is true because all values $\{r_\idxtype\}_\idxtype$ are distinct between $1$ and $n$; step~\steptwo is true by plugging in $\numminority=\frac{n}{2}$; and step~\stepthree is true by inequality~\eqref{eq:inequality}.

Combining the two cases, we have
\begin{align*}
    \diffswhr\gtorder n,
\end{align*}
completing the proof.

\subsubsection{Proof of bound~\texorpdfstring{\eqref{eq:diff_sw_rh_hetero}}{(\ref{eq:diff_sw_rh_hetero})}}

We prove that $\diffswrh = \Theta(n)$ under heterogeneous rankings. Given the trivial upper bound $\diffswrh = O(n)$, we prove the lower bound $\diffswrh = \Omega(\numstudents)$.

Consider an economy with $m=\floor*{\frac{n}{2}}$ student types. For readability, we assume $n$ is even and hence $m=\frac{n}{2}$. The proof for odd $\numstudents$ is similar up to minor modifications.
 Each type $t\in [m]$ has valuations
\begin{align*}
    & v_{t\idxschool} = \begin{cases}
        \frac{3}{4} & \text{if }\idxschool = 2t-1\\
        \frac{1}{4} & \text{if }\idxschool = 2\idxtype\\
        0^+ & \text{otherwise},
    \end{cases}
\end{align*}
with uniform prior $p_1 =\cdots = p_m = \frac{1}{m}$. Equivalently, we write $\{\pref_{\idxtype\idxschool}\}$ as a matrix
\begin{align*}
    \begin{bmatrix}
        3/4 & 1/4 & 0^+ & 0^+ & \cdots & 0^+ & 0^+\\
        0^+ & 0^+ & 3/4 & 1/4 & \cdots & 0^+ & 0^+\\
        \vdots & \vdots & \vdots & \vdots & \ddots & \vdots & \vdots\\
        0^+ & 0^+ & 0^+ & 0^+ & \cdots & 3/4 & 1/4
    \end{bmatrix}.
\end{align*}
Under either the hidden or revealed lottery, it can be verified that type-$t$ students only apply to the two schools of which they have a positive valuation. It suffices to consider type-1 students applying to schools 1 and 2, and we separately analyze the social welfare under the hidden and revealed lotteries.

\paragraph{Hidden lottery.}
We show that it is an equilibrium for type-$1$ students to apply to school $1$. To verify this equilibrium, we have $\flowh_1 = p_1 = \frac{1}{m}$. The admission probability of school $1$ is:
\begin{align*}
    \probavailh_{1} = \frac{1- \left(1-\flowh_1\right)^n}{n \flowh_1} = \frac{1-(1-\frac{1}{m})^{2m}}{2} \stackrel{\stepone}{>} \frac{1-e^{-2}}{2},
\end{align*}
where step~\stepone is true by inequality~\eqref{eq:inequality}.
Hence, the expected utility of applying to school $1$ is
\begin{align*}
    \utility^\parenh_1(1) =\pref_{11}\cdot \probavailh_{1}> \frac{3}{4} \cdot \frac{1-e^{-2}}{2} > \frac{1}{4}.
\end{align*}
The expected utility of applying to school $2$ is
\begin{align*}
    \utility^\parenh_1(2) =\pref_{12} \cdot \probavailh_{2}  =  \frac{1}{4}.
\end{align*}
Hence, we have $\utility^\parenh_1(1) > \utility^\parenh_1(2)$. Applying to school $1$ is utility maximizing for type-$1$ students, verifying the equilibrium under the hidden lottery.
School $1$ contributes utility $\frac{3}{4}$ to the social welfare if and only if there exists at least one type-$1$ student, which happens with probability $(1-(1-p_1)^n)$. Extending the same argument to every type $\idxtype\in [m]$ yields social welfare
\begin{align*}
    \swh = m\cdot \frac{3}{4}\cdot \left(1- \left(1-\frac{1}{m}\right)^{2m}\right).
\end{align*}
We thus have
\begin{align}
\lim_{m\rightarrow \infty} \frac{\swh}{m} = \frac{3}{4}\cdot \lim_{m\rightarrow \infty} \left(1-\left(1-\frac{1}{m}\right)^{2m}\right)
= \frac{3}{4}\cdot \left(1-e^{-2}\right) < 0.65.\label{eq:sw_diff_rh_hidden}
\end{align}

\paragraph{Revealed lottery.}

The equilibrium strategy for a type-$1$ student \withlnum $r$ is determined by \Cref{algo:compute_revealed_strategies_high_or_low}. In words, for each \lnum $\lottery$, \Cref{algo:compute_revealed_strategies_high_or_low} keeps track of a count $\counthigh$ (resp. $\countlow$), which is the count of better lottery numbers with which a type-$1$ student applies to school $1$  (resp. school $2$). We have $\counthigh + \countlow = r-1$. The student \withlnum $\lottery$ computes their expected utility for applying to school $1$ or school $2$, and applies to one of the two schools that maximizes their expected utility, with ties broken arbitrarily.
\begin{algorithm}[htb]
    \KwOut{Equilibrium strategies $\{a_\lottery\}_{\lottery\in [\numstudents]}$}
    \DontPrintSemicolon
    Set $\counthigh = \countlow = 0$.\;
    \For{$r=1, \ldots, n$}{
        Compute $\utility(1) = \frac{3}{4}\cdot (1-\frac{1}{m})^{\counthigh}$\;
        Compute $\utility(2) = \frac{1}{4}\cdot (1-\frac{1}{m})^{\countlow}$\;
        Compute $a_r = \argmax_{a \in \{1, 2\}} \utility(a)$, where ties are broken arbitrarily\;
        Set $n_{a_\lottery}\leftarrow n_{a_\lottery} + 1$\;
    }
    \caption{Determine the school $a_r$ that a type-$1$ student with \lnum $\lottery \in [\numstudents]$ applies to under the revealed equilibrium.\label{algo:compute_revealed_strategies_high_or_low}}
\end{algorithm}

Specifically, school $1$ is available to the student \withlnum $\lottery$ if and only if the $\counthigh$ lottery numbers, with which a type-$1$ student would apply to school $1$, are not taken by type-$1$ students. The expected utility for applying to school $1$ is
\begin{align*}
    \utility_1(1) = \frac{3}{4}\cdot (1-p_1)^{\counthigh} = \frac{3}{4}\cdot \left(1-\frac{1}{m}\right)^{\counthigh}.
\end{align*}
Likewise the expected utility for applying to school $2$ is
\begin{align*}
    \utility_1(2)= \frac{1}{4}\cdot \left(1-\frac{1}{m}\right)^{\countlow}.
\end{align*}
Hence, school $1$ has a higher expected utility if
\begin{align*}
    \utility_1(1) &> \utility_1(2)\\
    \frac{3}{4}\cdot \left(1-\frac{1}{m}\right)^{\counthigh} &> \frac{1}{4}\cdot \left(1-\frac{1}{m}\right)^{\countlow}\\
    \left(1-\frac{1}{m}\right)^{\counthigh - \countlow} & > \frac{1}{3}\\
    \counthigh - \countlow &< \frac{-\log 3}{\log(1-\frac{1}{m})}\defnright \thresh_m.
\end{align*}
Hence, in each iteration of \Cref{algo:compute_revealed_strategies_high_or_low}, if the count difference $\counthigh-\countlow$ is smaller than the threshold $\thresh_m$, then count $\counthigh$ increments by $1$, and otherwise count $\countlow$ increments by $1$, with ties broken arbitrarily. Intuitively, this count difference oscillates around the threshold $\thresh_m$.
We consider the final counts at the end of \Cref{algo:compute_revealed_strategies_high_or_low}, denoted by $\counthigh^*$ and $\countlow^*$ with $\counthigh^* + \countlow^* = \numstudents$. Using the approximation that $\log(1-\frac{1}{m}) \approx -\frac{1}{m}$ for large $m$, we have
\begin{align*}
    \counthigh^* - \countlow^* &\approx \frac{-\log 3}{\log(1-1/m)} \approx \log 3 \cdot m,
\end{align*}
and thus
\begin{subequations}\label{eq:counts_approx}
\begin{align}
    \counthigh^* &\approx m\left(1 + \frac{\log 3}{2}\right)\\
    \countlow^* &\approx m\left(1-\frac{\log 3}{2}\right).
\end{align}
\end{subequations}
The following lemma formalizes~\eqref{eq:counts_approx}.
\begin{lemma}\label{lem:counts_ratio_lim}
     Let $n\ge 4$ and thus $m \ge 2$. We have
\begin{align*}
    \lim_{m\rightarrow \infty} \frac{\counthigh^*}{m} &= 1+\frac{\log 3}{2}\\
    \lim_{m\rightarrow \infty} \frac{\countlow^*}{m} &= 1-\frac{\log 3}{2}.
\end{align*}
\end{lemma}
The proof of this lemma is provided at the end of this section. Given this lemma, there exists a universal constant $m_0$, such that for every $m \ge m_0$, we have
\begin{align*}
    \counthigh^* &> 1.5 m\\
    \countlow^* & > 0.4 m.
\end{align*}
To compute the social welfare, note that school $1$ is matched to a student if and only if there exists at least one type-$1$ student with one of the $\counthigh$ lottery numbers. The probability of school $1$ being matched is
\begin{align*}
    1 - \left(1-\frac{1}{m}\right)^{\counthigh^*} > 1 - \left(1-\frac{1}{m}\right)^{1.5 m} > 1 - e^{-1.5}
\end{align*}
and likewise the probability of school $2$ being matched is
\begin{align*}
    1 - \left(1-\frac{1}{m}\right)^{\countlow^*} > 1 - e^{-0.4}.
\end{align*}
Summing over student type $t\in [m]$, the social welfare under the revealed lottery is
\begin{align}
    \swr > m\cdot \left[\frac{3}{4}\left(1-e^{-1.5}\right) + \frac{1}{4}\left(1-e^{-0.4}\right)\right]> 0.66 m \quad \text{ for every }m \ge m_0.\label{eq:sw_diff_rh_reveal}
\end{align}
Combining~\eqref{eq:sw_diff_rh_hidden} and~\eqref{eq:sw_diff_rh_reveal}, there exists a universal constant $m_0'$ such that
\begin{align*}
    \swr - \swh > 0.66m-0.65m = 0.01 m\quad  \text{ for every }m \ge m_0'.
\end{align*}
completing the proof that $\diffswrh = \Omega(\numstudents)$.

\paragraph{Proof of Lemma~\ref{lem:counts_ratio_lim}.}

Let $\thresh_m \defn \frac{-\log 3}{\log(1-\frac{1}{m})}$. The counts $\counthigh^*$ and $\countlow^*$ equal to the final value of $\counthigh$ and $\countlow$ at the end of \Cref{algo:compute_counts_high_or_low}.

\begin{algorithm}[htb]
    \KwOut{Final counts $\counthigh$ and $\countlow$.}
    \DontPrintSemicolon
    Set $\counthigh = \countlow = 0$.\;
    \For{$r=1, \ldots, n$}{
        \uIf{$\counthigh - \countlow < \thresh_m$}{
            Set $\counthigh \leftarrow\counthigh+1$\;
        }
        \uElseIf{$\counthigh - \countlow >\thresh_m$}{
            Set $\countlow \leftarrow\countlow+1$\;
        }
        \Else{
            Pick $a\in \{1, 2\}$ arbitrarily\;
            Set $n_a \leftarrow n_a + 1$\;
        }
    }
    \caption{Computation of $\counthigh^*$ and $\countlow^*$.\label{algo:compute_counts_high_or_low}}
\end{algorithm}

Denote by $\countdiff_r$ the value of $\counthigh - \countlow$ at the end of the $\lottery^\textth$ iteration of the for loop in \Cref{algo:compute_counts_high_or_low}, and define $\countdiff_0 = 0$. In \Cref{algo:compute_counts_high_or_low}, if $\countdiff_r < \thresh_m$, then $\countdiff_r$ increases by 1, and vice versa. Hence, $\countdiff_r$ stays around $\thresh_m$. Formally, if there exists $r_0$ satisfying $\countdiff_{r_0} \in [\thresh_m - 1, \thresh_m + 1]$, then we have $\countdiff_r\in [\thresh_m -1, \thresh_m+1]$ for every $r \ge r_0$.

Assume for contradiction that there does not exist $r_0$ satisfying $\countdiff_{r_0} \in [\thresh_m - 1, \thresh_m + 1]$. Since $\countdiff_0 = 0 < \thresh_m$, we must have $\countdiff_0, \ldots, \countdiff_n < \thresh_m -1$. In \Cref{algo:compute_counts_high_or_low}, we have \begin{align*}
    \countdiff_n &= \countdiff_{n-1} + 1\\
    &= \cdots \\
    &= n > \log 3 \cdot m \stackrel{\stepone}{>} \thresh_m,
\end{align*}
where step~\stepone is true because $\log\big(1-\frac{1}{m}\big) < -\frac{1}{m} < 0$. This contradicts the fact that $\countdiff_n < \thresh_m - 1$.
Hence, $\countdiff_n =\counthigh^* - \countlow^* \in [\thresh_m - 1, \thresh_m + 1]$. Combining with the fact that $\counthigh^* +\countlow^* = n = 2m$, we have
\begin{subequations}\label{eq:counts_lim_range}
\begin{align}
    \counthigh^* - \left(m+\frac{\thresh_m}{2}\right) &\in \left[-\frac{1}{2}, \frac{1}{2}\right]\\
    \countlow^* - \left(m-\frac{\thresh_m}{2}\right) &\in \left[-\frac{1}{2}, \frac{1}{2}\right].   
\end{align}
\end{subequations}
Dividing both sides of~\eqref{eq:counts_lim_range} by $m$, taking the limit of $m\rightarrow \infty$, and using the fact that $\lim_{m\rightarrow \infty} \frac{\thresh_m}{m}= \log 3$ completes the proof of the lemma.

\subsection{Proof of Proposition~\ref{prop:partial-MR}}\label{app:proof_partial-MR}

We construct economies that satisfy orderings~\eqref{eq:ranking_partial_mr_RPH} and~\eqref{eq:ranking_partial_mr_RHP}.

\paragraph{Ordering 1: $\text{H} < \text{P}<\text{R}$.}
In Example~\ref{ex:equilibria}, we have
\begin{align*}
    \mrh \approx 0.493, \quad \mrp \approx 0.666, \quad \mrr =1.
\end{align*}
The derivations are provided in Appendix~\ref{app:derivations}. 

\paragraph{Ordering 2: $\text{P} < \text{H}<\text{R}$.}
Consider the following economy with $n=4$ students of $T=3$ types:
\begin{center}
\begin{tabular}{c|c|c|c|c||c}
     & school 1 & school 2 & school 3 & school 4 & prior \\\hline
     type 1 & 0.9 & 0.08 & 0.01 & 0.01 & 0.6\\
     type 2 & 0.5 & 0.4 & 0.09 & 0.01 & 0.3\\
     type 3 & 0.5 & 0.38 & 0.11 & 0.01 & 0.1
\end{tabular}
\end{center} 
For the partial information lottery, let the partition be $\mathcal{P}_1=\{1,2\}$ and $\mathcal{P}_2=\{3,4\}$. The match numbers are:
\begin{align*}
    \mnp=1.8092, \quad \mnh = 1.8448, \quad \mnr= 4.
\end{align*}

\paragraph{Revealed lottery.} Lemma~\ref{lem:reveal_same_ranking_always_one} yields that $\mrr=1$, and hence $\mnr=4$.

\paragraph{Hidden lottery.}
It can be verified that it is an equilibrium that students of type 1 apply to school $1$; students of types $2$ and $3$ apply to school $2$. At this equilibrium, the flows are
\begin{align*}
    \flowh_1 &= 0.6, \quad \flowh_2 = 0.4.
\end{align*}
The match number~\eqref{eq:mn_hidden} is
\begin{align*}
    \mnh = 1-\big(1-\flowh_1\big)^4 + 1-\big(1-\flowh_2\big)^4 = 1.8448.
\end{align*}

\paragraph{Partial information lottery.} The following equilibrium can be verified:
\begin{itemize}
    \item For block $\partition_1$, students of types 1 and 3 apply to school 1. Students of type 2 apply to school 2. 

    \item For block $\partition_2$, students of type 1 apply to school 1. Students of types 2 and 3 apply to school 2. 
\end{itemize}
At this equilibrium, the flows are
\begin{align*}
    \flowp_{\partition_1,1} &= 0.7, \quad \flowp_{\partition_1,2} = 0.3 \quad \text{for block }\partition_1\\
    \flowp_{\partition_2,1} &= 0.6, \quad \flowp_{\partition_2,2} = 0.4\quad \text{for block }\partition_2. 
\end{align*}
The match number~\eqref{eq:mn_partial} is
\begin{align*}
    \mnp = 1 - \left(1-\flowp_{\partition_1,1}\right)^2\left(1-\flowp_{\partition_2,1}\right)^2 + 1- \left(1-\flowp_{\partition_1,2}\right)^2 \left(1-\flowp_{\partition_2,2}\right)^2 
    = 1.8092.
\end{align*}

\subsection{Proof of Proposition~\ref{prop:partial-MR-hetero}}\label{app:proof_partial_MR_hetero}

We construct economies that satisfy the six possible orderings.

\paragraph{Orderings 1-2: $\text{H} < \text{P} < \text{R}$ and $\text{P} < \text{H} < \text{R}$.}

The same economies constructed in Appendix~\ref{app:proof_partial-MR} hold.

\paragraph{Ordering 3: $\text{H} < \text{R} < \text{P}$.}

Consider the following economy with $n=3$ students of $T=2$ types:
\begin{center}
\begin{tabular}{c|c|c|c||c}
     & school 1 & school 2 & school 3 & prior \\\hline
     type 1 & 0.6 & 0.3 & 0.1 & $0.6$\\\hline
     type 2 & 0.1 & 0.3 & 0.6 & $0.4$
\end{tabular}
\end{center}
For the partial information lottery, let the partition be $\mathcal{P}_1=\{1,2\}$ and $\mathcal{P}_2=\{3\}$. The match numbers are:
\begin{align*}
    \mnh = 1.72, \quad \mnr=2.224, \quad \mnp =2.48.
\end{align*}
Under the hidden lottery, type-1 students apply to school 1 and type-2 students apply to school $3$. The match number is
\begin{align*}
    \mnh = 2- (1-\prior_1)^3 - (1-\prior_2)^3 =1.72.
\end{align*}
Under the revealed lottery, type-1 students with lottery number $1, 2$ and $3$ apply to schools $1$, 2, and 1, respectively. Type-2 students always apply to school $3$.
The match number is
\begin{align*}
    \mnr = 3-(1-p_1)^2 - (1-p_1) - (1-p_2)^3 =2.224.
\end{align*}
Under the partial information lottery, students in block $\partition_1$ apply to their most preferred school, and students in block $\partition_2$ always apply to school $2$. The match number is
\begin{align*}
    \mnp = 3 - (1-p_1)^2-(1-p_2)^2 = 2.48.
\end{align*}

\paragraph{Ordering 4: $\text{R} < \text{H} < \text{P}$.} 
Consider the following economy with $\numstudents=3$ students of $\numtypes=3$ types:
\begin{center}
\begin{tabular}{c|c|c|c||c}
     & school 1 & school 2 & school 3 & prior \\\hline
     type 1 & 0.85 & 0.05 & 0.1 & 0.7\\\hline
     type 2 & 0.1 & 0.3 & 0.6 & $0.25$\\\hline
     type 3 & 0.6 & 0.3 & 0.1 & $0.05$
\end{tabular}
\end{center}
For the partial information lottery, let the partition be $\partition_1 = \{1, 2\}$ and $\partition_2 = \{3\}$. The match numbers are
\begin{align*}
    \mnr\approx 1.653,\quad \mnh\approx 1.694, \quad\mnp \approx 1.959.
\end{align*}
Under the hidden lottery, type-$1$ students apply to school $1$; type-$2$ students apply to school $3$; type-$3$ students apply to school $2$. The match number is
\begin{align*}
    \mnh = 3 - (1-p_1)^3 - (1-p_3)^3 - (1-p_2)^3 \approx 1.694.
\end{align*}
Under the revealed lottery, type-$1$ students always apply to school $1$; type-$2$ students always apply to school $3$; type-$3$ students apply to school $1$ if they have lottery number $1$, and otherwise apply to school $2$. The match number is
\begin{align*}
    \mnr = 3 - (1-p_1-p_3)(1-p_1)^2 - (1-p_3)^2 - (1-p_2)^3 \approx 1.653.
\end{align*}
Under the partial information lottery, type-$1$ students apply to school $1$ in block $\partition_1$, and school $3$ in block $\partition_2$; type-$2$ students always apply to school $3$; type-$3$ students apply to school $1$ in block $\partition_1$, and school $2$ in block $\partition_2$. The match number is
\begin{align*}
    \mnp = 3 - (1-p_1-p_3)^2 - (1-p_3) - (1-p_2)^2(1-p_1-p_2) \approx 1.959.
\end{align*}

\paragraph{Ordering 5: $\text{R} < \text{P} < \text{H}$.}
Consider the following economy with $\numstudents=3$ students of $\numtypes=3$ types:
\begin{center}
\begin{tabular}{c|c|c|c||c}
     & school 1 & school 2 & school 3 & prior \\\hline
     type 1 & 0.6 & 0.3 & 0.1 & 0.5\\\hline
     type 2 & 0.3 & 0.6 & 0.1 & $0.4$\\\hline
     type 3 & 0.45 & 0.25 & 0.3 & $0.1$
\end{tabular}
\end{center}
For the partial information lottery, let the partition be $\partition_1 = \{1, 2\}$ and $\partition_2 = \{3\}$. The match numbers are
\begin{align*}
    \mnr = 1.874, \quad \mnp = 1.904, \quad \mnh = 1.93.
\end{align*}
Under the hidden lottery, type-$\idxtype$ students apply to school $\idxtype$. The match number is
\begin{align*}
    \mnh = 3 - \sum_{\idxschool=1}^3 (1-p_\idxschool)^3 = 1.93.
\end{align*}
Under the revealed lottery, type-$1$ and type-$2$ students always apply to their preferred school. Type-$3$ students apply to school 1 if they have lottery number 1, and apply to school 3 otherwise. The match number is
\begin{align*}
    \mnr = 3 - (1-p_1-p_3)(1-p_1)^2 - (1-p_2)^3 - (1-p_3)^2 = 1.874.
\end{align*}
Under the partial information lottery, all students in block $\partition_1$ apply to their preferred school. For block $\partition_2$, students of type 1 and type 2 apply to school 2, and students of type 3 apply to school 3. The match number is
\begin{align*}
    \mnp  = 3 - (1-p_1-p_3)^2 - (1-p_2)^2 (1-p_1-p_2) - (1-p_3) = 1.904.
\end{align*}

\paragraph{Ordering 6: $\text{P} < \text{R} < \text{H}$.}
Consider the following economy with $\numstudents=3$ students of $\numtypes=3$ types:
\begin{center}
\begin{tabular}{c|c|c|c||c}
     & school 1 & school 2 & school 3 & prior \\\hline
     type 1 & $0.6$ & $0.3$ & $0.1$ & $0.45$\\\hline
     type 2 & 0.3 & 0.6 & 0.1 & $0.45$\\\hline
     type 3 & 0.15 & 0.5 & 0.35 & $0.1$
\end{tabular}
\end{center}
For the partial information lottery, let the partition be $\partition_1 = \{1, 2\}$ and $\partition_2 = \{3\}$. The match numbers are
\begin{align*}
     \mnp \approx 1.822, \quad \mnr = 1.8875, \quad \mnh \approx 1.938.
\end{align*}
Under the hidden lottery, type-$\idxtype$ students apply to school $\idxtype$. The match number is
\begin{align*}
    \mnh = 3 - \sum_{\idxschool=1}^3 (1-p_\idxschool)^3 \approx 1.938.
\end{align*}
Under the revealed lottery, students with lottery number 1 apply to their preferred school. Type-$\idxtype$ students with lottery number 2 or 3 apply to school $\idxtype$. The match number is
\begin{align*}
    \mnr = 3 - (1-p_1)^3 - (1-p_2-p_3)(1-p_2)^2 -(1-p_3)^2 = 1.8875
\end{align*}
Under the partial information lottery, students in block $\partition_1$ apply to their preferred school. Type-$\idxtype$ students in block $\partition_2$ apply to school $\idxtype$. The match number is
\begin{align*}
    \mnp = 3 - (1-p_1)^3 - (1-p_2-p_3)^2(1-p_2) - (1-p_3) \approx 1.822.
\end{align*}

%
\subsection{Proof of Proposition~\ref{prop:partial-SW}}\label{app:proof_partial-SW}

We construct economies that satisfy the six possible orderings.

\paragraph{Ordering 1: $\text{R} < \text{H} < \text{P}$.}
Consider the following economy with $n=4$ students of $T=2$ types:
\begin{center}
\begin{tabular}{c|c|c|c|c||c}
     & school 1 & school 2 & school 3 & school 4 & prior \\\hline
     type 1 & 0.95 & 0.03 & 0.01 & 0.01 & $0.2$\\\hline
     type 2 & 0.6 & 0.38 & 0.01 & 0.01 & 0.8
\end{tabular}
\end{center}
For the partial information lottery, let the partition be $\mathcal{P}_1=\{1,2\}$ and $\mathcal{P}_2=\{3,4\}$. The values of social welfare are:
\begin{align*}
    \swr = 1, \quad \swh \approx 1.006, \quad \swp \approx 1.044.
\end{align*}

\paragraph{Revealed lottery.} 
Lemma~\ref{lem:reveal_same_ranking_always_one} yields that $\swr=1$. 

\paragraph{Hidden lottery.}
It can be verified that an equilibrium is
\begin{align*}
    \strategy_1 = (1, 0, 0, 0), \quad \strategy_2 = (x, 1-x, 0, 0),
\end{align*}
where $x\approx 0.5708$ is the root for the equation
\begin{align*}
    \pref_{21} \probavail_1 = \pref_{22}\probavail_2.
\end{align*}
The flows are
\begin{align*}
    q_1=0.2 + 0.8x,\quad q_2= 0.8(1-x).
\end{align*}
The admission probabilities are
\begin{align*}
    \alpha_1\approx 0.3754,\quad \alpha_2 \approx 0.5928.
\end{align*}
The social welfare is
\begin{align*}
    \swh = 4\cdot (p_1u^\parenh_{1} + p_2u^\parenh_2) =4\cdot (p_1v_{11} + p_2 v_{21} )\cdot \probavail_1 \approx 1.006.
\end{align*}

\paragraph{Partial information lottery.} It can be verified that an equilibrium is
\begin{align*}
    \strategy_{1, \partition_1} &= (1, 0, 0, 0), \quad \strategy_{2, \partition_1} = \left(\frac{39}{49}, \frac{10}{49}, 0, 0\right) \quad\text{for block }\partition_1\\
    \strategy_{1, \partition_2} &= (1, 0, 0, 0), \quad \strategy_{2, \partition_2} = (0, 1, 0, 0) \quad\text{for block }\partition_2.
\end{align*}
The flows are
\begin{align*}
    (\flow_{\partition_1, 1}, \flow_{\partition_1, 2}) &= \left(\frac{41}{49}, \frac{8}{49}\right) \quad \text{for block }\partition_1\\
    (\flow_{\partition_2, 1}, \flow_{\partition_2, 2}) &= (0.2, 0.8)\quad \text{for block }\partition_2.
\end{align*}
The admission probabilities are
\begin{align*}
    (\probavail_{\partition_1, 1}, \probavail_{\partition_1, 2}) &= \left(\frac{57}{98}, \frac{45}{49}\right) \quad \text{for block }\partition_1\\
    (\probavail_{\partition_2, 1}, \probavail_{\partition_2, 2}) &\approx (0.024, 0.420)\quad \text{for block }\partition_2. 
\end{align*}
The social welfare is 
    \begin{align*}
        \swp = 2\cdot (\prior_1 v_{11} \probavail_{\partition_1, 1} + \prior_2 v_{21}\probavail_{\partition_1, 1}) + 2\cdot (\prior_1 v_{11}\probavail_{\partition_2, 1}+ \prior_2 v_{22}\probavail_{\partition_2, 2} ) \approx 1.044.
\end{align*}

\paragraph{Ordering 2: $\text{H}< \text{R} <\text{P}$.} Consider the same economy from Ordering 1, but with a different prior: 
\begin{center}
\begin{tabular}{c|c|c|c|c||c}
     & school 1 & school 2 & school 3 & school 4 & prior \\\hline
     type 1 & 0.95 & 0.03 & 0.01 &  0.01 & $0.1$\\
     type 2 & 0.6 & 0.38 & 0.01 & 0.01 & $0.9$
\end{tabular}
\end{center}
with the same partition $\mathcal{P}_1=\{1,2\}$ and $\mathcal{P}_2=\{3,4\}$.  
The values of social welfare are:
\begin{align*}
    \swh = 0.954, \quad \swr =1, \quad \swp \approx 1.007.
\end{align*}
\paragraph{Hidden lottery.} An equilibrium is
\begin{align*}
    \strategy_1 = (1, 0, 0, 0), \quad \strategy_2 = (x, 1-x, 0, 0),
\end{align*}
with $x\approx0.6185$. The social welfare is
\begin{align*}
    \swh = 4\cdot (p_1 v_{11} + p_2 v_{21})\probavail_1 \approx 0.954.
\end{align*}

\paragraph{Partial information lottery.}
An equilibrium is
\begin{align*}
    \strategy_{1,\partition_1} &= (1, 0, 0, 0), \quad \strategy_{2,\partition_1} = \left(\frac{361}{441}, \frac{80}{441}, 0, 0\right) \quad \text{for block }\partition_1\\
    \strategy_{1,\partition_2} &= (1, 0, 0, 0), \quad \strategy_{2,\partition_2} = (0, 1, 0, 0) \quad \text{for block }\partition_2.
\end{align*}
The admission probabilities are
\begin{align*}
    (\probavail_{\partition_1, 1}, \probavail_{\partition_1, 2}) &= \left(\frac{57}{98}, \frac{45}{49}\right) \quad\text{for block }\partition_1\\
    (\probavail_{\partition_2, 1}, \probavail_{\partition_2, 2}) &= (0.0253, 0.3851)\quad \text{for block }\partition_2. 
\end{align*}
The social welfare is 
    \begin{align*}
        \swp = 2\cdot (\prior_1 v_{11} \probavail_{\partition_1, 1} + \prior_2 v_{21}\probavail_{\partition_1, 1}) + 2\cdot(\prior_1 v_{11}\probavail_{\partition_2, 1}+ \prior_2 v_{22}\probavail_{\partition_2, 2} ) \approx 1.007.
\end{align*}

\paragraph{Ordering 3: $\text{H} < \text{P} <\text{R}$.} 
Consider the following economy with $\numstudents=3$ students of $\numtypes=1$ type:
\begin{center}
\begin{tabular}{c|c|c|c||c}
     & school 1 & school 2 & school 3 & prior \\\hline
     type 1 & 0.75 & 0.2 & 0.05 & $1$
\end{tabular}
\end{center}
For the partial information lottery, let the partition be $\partition_1 = \{1\}$ and $\partition_2 = \{2, 3\}$. The values of social welfare are:
\begin{align*}
    \swh=0.75, \quad \swp=0.95, \quad \swr=1.
\end{align*}
Under the revealed lottery, Lemma~\ref{lem:reveal_same_ranking_always_one} yields that $\swr=1$.

Under the hidden lottery, it is an equilibrium for every student to apply to school 1, yielding social welfare $\swh=0.75$.

Under the partial information lottery, the student in block $\partition_1$ applies to school $1$, and students in block $2$ apply to school $2$, yielding social welfare $\swp=0.95$.

\paragraph{Ordering 4: $\text{P} < \text{H} < \text{R}$.} 
Consider the following economy with $n=4$ students of $T=1$ type:
\begin{center}
\begin{tabular}{c|c|c|c|c||c}
     & school 1 & school 2 & school 3 & school 4 & prior \\\hline
     type 1 & 0.5 & 0.45 & 0.03 & 0.02 & $1$
\end{tabular}
\end{center}
with partition $\partition_1 = \{1,2\}$ and $\partition_2 = \{3, 4\}$.
The values of social welfare are
\begin{align*}
    \swp \approx 0.878, \quad \swh \approx 0.890, \quad \swr = 1.
\end{align*}

\paragraph{Hidden lottery.}
It can be verified that an equilibrium is
\begin{align*}
    \strategy = (x, 1-x, 0, 0) 
\end{align*}
with $x \approx 0.5359$. 
The admission probabilities are
\begin{align*}
    \probavail_1 \approx 0.445, \quad \probavail_2\approx 0.494.
\end{align*}
The social welfare is \begin{align*}
    \swh = 4 \cdot v_{11}\probavail_1 \approx 0.890.
\end{align*}

\paragraph{Partial information lottery.}
An equilibrium is
\begin{align*}
    \strategy_{\partition_1} = \left(\frac{11}{19}, \frac{8}{19}, 0, 0\right), \quad \strategy_{\partition_2} = (0.1105, 0.8895, 0, 0).
\end{align*}
The admission probabilities are
\begin{align*}
    \probavail_{\partition_1, 1} =\frac{27}{38}, \quad \probavail_{\partition_2, 1} = 0.1675.
\end{align*}
The social welfare is
\begin{align*}
    \swp = 2v_1\cdot \left(\probavail_{\partition_1, 1} +  \probavail_{\partition_2, 1}\right) \approx 0.878.
\end{align*}

\paragraph{Ordering 5: $\text{R} < \text{P} < \text{H}$.} 
In Example~\ref{ex:equilibria}, we have:
\begin{align*}
    \swr = 1, \quad\swp \approx 1.012, \quad \swh \approx 1.128.
\end{align*}
The derivations are provided in Appendix~\ref{app:derivations}.

\paragraph{Ordering 6: $\text{P} < \text{R} < \text{H}$.} Consider the economy in Example~\ref{ex:equilibria}, where for the partial lottery, we instead consider  the partition $\mathcal{P}_1=\{1\}$ and $\mathcal{P}_2=\{2,3\}$. The values of social welfare are
\begin{align*}
    \swp =  0.99, \quad\swr = 1, \quad \swh \approx 1.128.
\end{align*}
Under the partial information lottery, the student in block $\partition_1$ always applies to school $1$. Both students in block $\partition_2$ apply to school $2$.
We have
\begin{align*}
\swp = (p_1 v_{11} +p_2 v_{21}) + (p_1 v_{12} + p_2 v_{22}) =  0.99.
\end{align*}

\end{document}